\documentclass[12pt]{article} 

\usepackage{url,fullpage,color}
\usepackage{graphicx}
\usepackage{amsmath, amsthm, amssymb}
\usepackage{algpseudocode}
\usepackage{algorithm}
\usepackage{subfigure}
\usepackage{comment}
\RequirePackage[round]{natbib}
\RequirePackage[colorlinks,citecolor=blue,urlcolor=blue]{hyperref}

\usepackage{setspace}

\usepackage{multirow}

\newtheorem{thm}{Theorem}
\newtheorem{lem}[thm]{Lemma}
\newtheorem{cor}[thm]{Corollary}
\newtheorem{proposition}[thm]{Proposition}
\let\hat\widehat

\theoremstyle{remark}
\newtheorem{remark}{Remark}

\newcommand\E{\mathbb{E}}

\catcode`@=11
\newskip\beforeproofvskip
\newskip\afterproofvskip
\beforeproofvskip=\medskipamount
\afterproofvskip=\bigskipamount

\def\prooftag{Proof}
\def\proofskip{\enspace}

\def\proof{\@ifnextchar[{\@@proof}{\@proof}}  
\def\@startproof{\par\vskip\beforeproofvskip\leavevmode}
\def\@proof{\@startproof{\scshape\prooftag.}\proofskip}
\def\@@proof[#1]{\@startproof {\scshape\prooftag #1.}\proofskip}

\catcode`@=12

\newcommand{\blind}{0}

\usepackage{bbm}
\usepackage{amsmath, amssymb}

\usepackage{algpseudocode,algorithm}
\usepackage{graphicx}
\usepackage{subfigure}
\usepackage{lineno}
\usepackage{multirow}
\usepackage{booktabs}

\usepackage{caption, tabularx}

\usepackage{enumitem}

\usepackage{tikz}
\usetikzlibrary{shadows,arrows,positioning}
\pgfdeclarelayer{background}
\pgfdeclarelayer{foreground}
\pgfsetlayers{background,main,foreground}

\tikzstyle{materia}=[draw, fill=blue!20, text width=6.0em, text centered,
minimum height=1.5em,drop shadow]
\tikzstyle{etape} = [materia, text width=8em, minimum width=10em,
minimum height=3em, rounded corners, drop shadow]
\tikzstyle{texto} = [above, text width=6em, text centered]
\tikzstyle{linepart} = [draw, thick, color=black!50, -latex', dashed]
\tikzstyle{line} = [draw, thick, color=black!50, -latex']
\tikzstyle{ur}=[draw, text centered, minimum height=0.01em]

\newtheorem{thm1}{Theorem}
\newtheorem{example}[thm1]{Example}

\newtheorem{thm2}{Theorem}
\newtheorem{definition}[thm2]{Definition}

\let\hat\widehat
\let\tilde\widetilde

\DeclareMathAlphabet{\mathpzc}{OT1}{pzc}{m}{it}

\newcommand{\M}{\mathcal{M}}

\newcommand{\Corr}{\text{Corr}}

\usepackage{multicol}

\usepackage{xr}

\begin{document}

\def\spacingset#1{\renewcommand{\baselinestretch}%
{#1}\small\normalsize} \spacingset{1}

\if0\blind
{
  \title{\bf Multistage Estimators for Missing Covariates and Incomplete Outcomes}
  \author{ Daniel Suen
  	\hspace{.2cm}\\
    Department of Statistics, University of Washington\\
    and \\
    Yen-Chi Chen
    \\
    Department of Statistics, University of Washington}
  \maketitle
} \fi

\if1\blind
{
  \bigskip
  \bigskip
  \bigskip
  \begin{center}
    {\LARGE\bf 
    	\vspace{1 in}
    Multistage Estimators for Missing Covariates and Incomplete Outcomes}
\end{center}
  \medskip
} \fi

\bigskip

\begin{abstract}
\noindent
We study problems with multiple missing covariates and partially observed responses. 
We develop a new framework to handle complex missing covariate scenarios via inverse probability weighting,
regression adjustment, and a multiply-robust procedure.
We apply our framework to three classical problems: the Cox model from survival analysis, missing response, and binary treatment from causal inference. We also discuss 
how to handle missing covariates in these scenarios, and develop associated identifying theories and asymptotic theories. We apply our procedure to simulations and an Alzheimer's disease dataset
and obtain meaningful results.

\end{abstract}

\noindent%
\vfill

\newpage
\spacingset{1.45} 
\section{Introduction}		\label{sec::intro}

Missing data occurs across many different fields, and as the size of datasets becomes ever larger, it becomes increasingly unrealistic to expect fully observed datasets.  In many statistical problems, we have outcome variables of interest
and a set of associated covariates. However, these outcome variables may not be completely observed.  For example, in a survival  problem, either the censoring time or the time-to-event are observed.
In a missing response problem, a response may be subject to missingness and only be observed for a subset of the sample. 
In the binary treatment problem, only one of the two potential outcomes will be observed.
In a traditional setting, all the covariates are assumed to be completely measured
and there has been a vast amount of literature on how to estimate a parameter of interest. 
However, little is known when the covariates could be missing.

Some early attempts to handle missing covariates rely on ignorability assumptions, utilize estimation procedures in which the assumptions are not well-specified, or assume that there is only one missing pattern (one covariate subject to missing).  For example, many standard assumptions to identify an effect of interest take the form of a conditional independence statement, where the conditioning is done on the covariates $X$.  In the case of missing covariates, strong ignorability assumptions have been proposed by conditioning on the observed covariates $X_R$ and the missing covariate pattern $R$ and using generalized propensity scores with propensity score matching \citep{rosenbaum1984reducing, rosenbaum1985constructing}.  Other approaches include augmenting the support of the covariates with a new category for missing data $\{\texttt{NA}\}$ \citep{Mayer2020,mayer2021transporting}.  One downside to both of these methods is that in some situations, the interpretation of these missing data assumptions may not be so natural as there are effectively different sets of covariates being conditioned on for different individuals.  Various imputation methods have also been proposed in the Cox model such as those in \cite{Paik1997, White2009, Beesley2016}, but those proposed methods rely on specific assumptions that may not extend to more general settings.  The multiple imputation approach discussed in \cite{qu2009propensity,crowe2010comparison}  rely on missing data assumptions that are not clearly stated, which can lead to difficulty in interpretation.  \cite{berg2021} also recently examined estimation under missing-not-at-random (MNAR) but they only consider a single covariate that are subject to missingness.

In the case of multiple missing variables, the missing-at-random (MAR) assumption 
becomes hard to interpret \citep{robins1997non},
so a nonignorable/MNAR missing data assumption may be preferred.
There have been previous developments using nonignorable missing data assumptions in the literature.  For example, \cite{robins1997non} proposed the permutation missingness process as a generalization of MAR processes.  Also, \cite{tchetgen2018discrete} used discrete-choice models as a way to generate a class of MNAR assumptions.  \cite{shpitser2016consistent, sadinle2017itemwise, malinsky2020semiparametric} considered a ``no self-censoring'' or ``itemwise conditionally independent nonresponse'' assumption.  \cite{chen2020pattern} introduced the idea of a pattern graph to generate further MNAR assumptions.  Other approaches include various graphical ones: \cite{mohan2013graphical, mohan2014graphical, nabi2020full, Mohan2021graphical}.

In this paper, we build on this previous work by introducing a general framework for constructing a set of compatible assumptions on  problems with missing covariates and incomplete outcomes.  This framework has several advantages.  First, one can construct identifying restrictions on the missing covariates that are compatible with common identifying restrictions for the outcome variables.  This ensures that the final outcome of interest is nonparametrically identifiable \citep{robins2000sensitivity}.  Also, because we start with a clear set of assumptions and then develop an estimation method based on those assumptions, this framework provides a principled approach for estimating the final outcome of interest.  Previous work on MNAR mechanisms can be utilized in our estimation procedure.  Lastly, our framework can be applied to various settings including but not limited to: survival analysis with censoring, binary treatment from causal inference, and missing response.

\emph{Outline.}
In Section \ref{sect:background}, we introduce our framework and several problems of interest.  In Section \ref{sect:multistage}, we outline how to construct a multistage estimator with the Cox model as a motivating example.  In Sections \ref{sect:simulation} and \ref{sect:data}, we explore simulation results and include a specific data analysis.  As a matter of theoretical interest, we include details on a multistage estimator for the missing response problem in Appendix \ref{appendix:missingresp}. 
We discuss the asymptotic theory of our estimators and demonstrate the $\sqrt{n}$-consistency of our estimators in Appendix \ref{appendix:asymptotic}.  Further simulations, data analysis results, and theoretical details such as sensitivity analyses are provided in Appendices  \ref{appendix:beyondCCMV}, \ref{appendix:sens}, \ref{appendix:sim},  \ref{appendix:data}, and \ref{appendix:proofs}.


\section{Background and framework} \label{sect:background}

In this section, we discuss the  problem setup and introduce notation that will be helpful in describing our framework.  We also discuss the setting of three classical problems from the statistics literature: the Cox model from survival analysis, missing response, and binary treatment from causal inference.

\subsection{Problem Setup} In our problems, there are both covariate and outcome-related variables. Let $Z = (X,R)$ represent the covariate-related variables, where $X\in\mathbb{R}^d$ is a set of covariate variables.  Since $X$ is multivariate, it can have an arbitrary missing pattern.  We use the binary vector $R \in \{0,1\}^d$ to represent the missing response for $X$.  If $R_j=1$, then the variable $X_j$ is observed; otherwise, it is missing.  Let $1_d = (1, 1,\ldots, 1)^\top$ denote the case where $X_j$ is observed for all $j$.  We denote $\bar{r}=1_d-r$ as the flipped version of pattern $r$.  Thus, the notation $X_r = (X_j : r_j = 1)$ and $X_{\bar{r}} = (X_j : r_j = 0)$ refer to observed and missing covariate variables, respectively.

Next, we define $S = (W,A)$ as the outcome-related variables.  Here $W \in \mathbb{R}^k$ is a set of outcome variables, and $A\in\mathcal{A}$ is a random variable over a discrete support $\mathcal{A}$ that represents the observed pattern of the outcome variables.  Note that $A$ has a more general interpretation than the missing pattern $R$ for $X$ because $A$ does not necessarily correspond to a missing pattern for $W$ in the traditional sense.  This idea will be made more precise in the following subsections.  Similar to the previous case with the covariates, we will use the notation $W_A$ to denote the set of outcome variables that are observed under the observed pattern $A$.

Therefore, with a slight abuse of notation, the observed dataset consists of $n$ IID observations of the form $\{(X_{i,R_i}, R_i, W_{i, A_i}, A_i)\}_{i=1}^n$.  Here $R_i$ is the covariate missingness pattern for the $i$th individual.  As $X_i$ consists of the covariates for the $i$th individual, some of its entries may be subject to missingness; thus, we use $X_{i,R_i}$ to denote the observed covariates for the $i$th individual.  Similarly, the observed outcome variables may differ between individuals, so we use $W_{i,A_i}$ to denote the observed outcome variables for the $i$th individual, where $A_i$ is the observed pattern of the outcome variables for the $i$th individual.

To further describe our problem formulation, we provide a brief description of each of the three problems here (Cox model, missing response, and binary treatment).

\subsection{Cox model} \label{3.2:CoxModel}

In survival analysis, we typically wish to determine the effect of covariates on a time-to-event variable.  However, the event time is often not known for some individuals due to censoring.  For example, people participating in a medical trial may dropout prematurely, which may mean that the event of interest is not observed for those individuals.  Thus, the observed data is of the form
\begin{equation*}
	(X_{1,R_1}, R_1, Y_1, \Delta_1), \ldots, (X_{n,R_n}, R_n, Y_n, \Delta_n),
\end{equation*}
where $Y_i = \min\{T_i,C_i\} \in\mathbb{R}$ is the observed time, and $\Delta_i = 1(Y_i=T_i) = 1(C_i\geq T_i)$ is the status indicator for a time-to-event $T_i$ and a censoring time $C_i$.  Thus, the outcome variables are $W=(C,T)$ and the observed outcome pattern corresponds to $A = \Delta$.  We see that $W_0 = C$ and $W_1 = T$, as the observed outcome pattern denotes whether or not the time-to-event variable was censored.  One can see that the observed outcome pattern $A$ is not associated with a missingness pattern.  Importantly, the outcome variables $W=(C,T)$ are never completely observed for any individuals.


The Cox proportional hazards model assumes that the effect of the covariates is multiplicative on $\lambda(t|x)$ through a baseline hazard function
	$\lambda(t|x) = \lambda_0(t)\exp(\beta^\top x)$.
Under the Cox model and an independent censoring assumption, 
the parameter of interest $\beta^*$ satisfies the population estimating equation:
\begin{equation}
\begin{aligned}
	& U(\beta) = \E\left[\Delta \left(X - \frac{s^{(1)}(Y; \beta)}{s^{(0)}(Y;\beta)}\right)\right] = 0, \\
	& s^{(1)}(t;\beta) = \E[ 1(Y\geq t) \exp(\beta^\top X) X], \\
	& s^{(0)}(t;\beta) = \E[ 1(Y\geq t) \exp(\beta^\top X)].
\end{aligned}
\label{eq::Cox::EE}
\end{equation}
Our goal is to estimate $\beta^*$ in this semiparametric setting.  When the covariates $X$ are fully observed, under suitable regularity conditions, it is known that finding a near-solution $\hat \beta$ to the estimating equation
\begin{align*}
	& U_n(\beta) = \frac{1}{n}\sum_{i=1}^n \Delta_i \left(X_i - \frac{\hat{s}^{(1)}(Y_i; \beta)}{\hat{s}^{(0)}(Y_i;\beta)} \right) = o_p(1/\sqrt{n}), \\
	& \hat{s}^{(1)}(t;\beta) = \frac{1}{n} \sum_{i=1}^n 1(Y_i\geq t) \exp(\beta^\top X_i) X_i, \\
	& \hat{s}^{(0)}(t;\beta) = \frac{1}{n} \sum_{i=1}^n 1(Y_i\geq t) \exp(\beta^\top X_i)
\end{align*}
provides a consistent and asymptotically normal estimator of $\beta^*$.  Since $X$ can now be subject to missingness, we need a way to estimate these nuisance functions.

\subsection{Missing response}

In the missing response problem, we assume that we have a single outcome variable of interest $Y\in\mathbb{R}$ that is possibly subject to missingness.  We use the binary random variable $A\in\{0,1\}$, which is $1$ if and only if $Y$ is observed.  The observed data is either in the form of $(X_{i,R_i}, R_i, Y_i, A_i = 1)$ or $(X_{i,R_i}, R_i, A_i = 0)$.
In the second case ($Y_i$ is missing), we can think of the data as $(X_{R_i}, R_i,\texttt{NA},  A_i = 0)$. 
Using our aforementioned notations, $W_0 = \texttt{NA}$ and $W_1= Y$.
Since $Y$ can be subject to missingness, $A$ takes a traditional interpretation as a missing response pattern.  Then, $Y_{A_i}$ is the observed outcome variable for the $i$th individual.  
In a typical case, the parameter of interest is the marginal mean $\E[Y]$ and
we often assume $Y\perp A|X$
to identify the marginal mean.
Under this assumption,
we have
\begin{align}
	\E[Y] &= \E\left[\frac{YA}{\pi_A(X)}\right] \label{eq:miss-ipw} \\
	&= \E[m_1(X)], \label{eq:miss-ra}
\end{align}
where $\pi_A(X) = P(A=1|X)$ is the propensity score and $m_1(X) = \E[Y|A=1,X]$ is the outcome regression.  The expressions \eqref{eq:miss-ipw} and \eqref{eq:miss-ra} are associated with the \textit{inverse probability weighting} and \textit{regression adjustment} estimators, respectively.  When the covariates $X$ are always observed, these quantities can be easily estimated from the observed data, but the situation becomes more complicated when the covariates are multivariate and subject to nonmonotone missingness.


\subsection{Binary treatment}

Under the Neyman-Rubin causal model \citep{rubin1978bayesian, Neyman1990}, we have a binary treatment $A \in \{0,1\}$ that corresponds to potential outcomes $Y(0)$ and $Y(1)$, respectively.  Here the outcome variables are $W=(Y(0),Y(1))$.  Similar to the survival analysis problem with censoring, only one of the outcome variables is observed for a given individual.
The observed outcome variables are $W_0 = Y(0)$ and $W_1 = Y(1)$.

A common parameter of interest is the average treatment effect (ATE)
	$\E[Y(1)] - \E[Y(0)]$
and under the classical assumptions (consistency, SUTVA, weak ignorability; \citep{rubin1974estimating, rubin1978bayesian})
%
the ATE admits the following expression
\begin{align*}
	\E[Y(1)] - \E[Y(0)] &= \E\left[\frac{YA}{\pi_A(X)}\right] - \E\left[\frac{Y(1-A)}{1-\pi_A(X)}\right] \\
	&= \E[m_1(X)] - \E[m_0(X)],
\end{align*}
where $\pi_A(X) = P(A=1|X)$ is the propensity score and $m_a(X) = \E[Y|A=a,X]$ is the outcome regression.  
Thus, this leads to the inverse probability weighting or regression adjustment estimators for inferring the ATE.

\section{A multistage estimator} \label{sect:multistage}

The intermediate quantities that we described in the previous section such as $s^{(1)}(t;\beta)$ and $\pi_A(X)$ are calculated using the covariates $X$ and the outcome-related variables $(W,A)$.  Thus, due to this shared property, before we construct our multistage estimator, we first consider a general real-valued function $h$ of the covariate-related variables $X$ and the outcome-related variables $(W,A)$, as defined below.

\begin{definition}[Observed-decomposable functions] \label{def:ODF}
	Let $h : \mathbb{R}^d \times \mathbb{R}^j \times \mathcal{A} \mapsto \mathbb{R}$ be a function of interest.  
	Call $h$ an observed-decomposable function (ODF) if
	\begin{equation*}
		h(X,W,A) = \sum_{a\in \mathcal{A}} f_a(X,W_a)1(A=a), \label{eq:h-function}
	\end{equation*}
	where $f_a: \mathbb{R}^d \times \mathbb{R}^{\text{dim}(W_a)} \mapsto \mathbb{R}$ is a fixed (non-random) real-valued function for all $a\in\mathcal{A}$.
\end{definition}
The interpretation of an ODF is that for each observed outcome pattern $a$, we have a separate real-valued function $f_a$ that is dependent on all of the covariates $X$ and the observed outcome variables $W_a$ under pattern $a$.  Since $h$ has this decomposition, it is computable in the traditional setting with incomplete outcomes and completely observed covariates.  As we will see in the construction of a multistage estimator, many quantities of interest can be written in this way, so this is a very natural decomposition.  We will either typically be interested in estimating $\theta=\E[h(X,W,A)]$ or $\E[h(X,W,A)]=0$ will be an estimating equation.  We provide an illustrative example of $h$ using the missing response problem.

\begin{example}[Missing response]
	In the traditional missing response problem without missing covariates where we commonly wish to estimate $\E[Y]$, there are 3 traditional estimators: regression adjustment on $Y$, inverse probability weighting on $Y$, and doubly robust on $Y$ estimators.  Because our goal is to estimate $\E[Y]$, the corresponding $h$, $f_0$, and $f_1$ functions take the following forms in these three situations.
	Note that in the case $A= 0 $ ($Y$ is missing), we simplify notation such that $f_0(X,W_0 ) = f_0(X, \texttt{NA}) = f_0(X)$ since $W_0$ 
	can only take one possible outcome \texttt{NA}.
	
	\begin{enumerate}
		\item Regression adjustment on $Y$.  We have
			$h(X,W,A) = \E[Y|A=1,X],$
		which implies that $f_0(X) = f_1(X,Y) = \E[Y|A=1,X]$.
		
		\item Inverse probability weighting on $Y$.  We have
			$h(X,W,A) = \frac{1(A=1)Y}{P(A=1|X)},$
		which implies that $f_0(X) = 0$ and $f_1(X,Y) = \dfrac{Y}{P(A=1|X)}$.
		
		\item Doubly-robust on $Y$.  We have
			$h(X,W,A) = \frac{1(A=1)Y}{P(A=1|X)} + \E[Y|A=1,X]\left(1-\frac{1(A=1)}{P(A=1|X)}\right),$
		which implies that $f_0(X) = \E[Y|A=1,X]$ and $f_1(X,Y) = \dfrac{Y-\E[Y|A=1,X]}{P(A=1|X)}+\E[Y|A=1,X]$.
	\end{enumerate}
\end{example}

\begin{example}[Cox model]
In the Cox model, $\Delta = A$ and $W_0 = W_1 = Y$ and
there are three key equations as described in equation \eqref{eq::Cox::EE}: $U(\beta), s^{(1)}(t;\beta), s^{(0)}(t;\beta)$.
\begin{enumerate}
\item $U(\beta) $. For this quantity, $U(\beta) =  \E[h(X,W,A)]$, so $f_0(X,W_0) = 0$ and $f_1(X,W_1) =  \left(X - \frac{s^{(1)}(Y; \beta)}{s^{(0)}(Y;\beta)}\right)$.

\item $s^{(1)}(t;\beta)$. In this case, by setting $s^{(1)}(t;\beta) = \E[h(X,W,A)]$, we obtain $f_0(X,W_0) = 1(Y\geq t)\exp(\beta^TX)X$
and $f_1(X,W_1) = 1(Y\geq t)\exp(\beta^TX)X$.
\item $s^{(0)}(t;\beta)$. This case is similar to $s^{(1)}(t;\beta)$, so that $f_0(X,W_0) = 1(Y\geq t)\exp(\beta^TX)$
and $f_1(X,W_1) = 1(Y\geq t)\exp(\beta^TX)$.
\end{enumerate}

\end{example}

\subsection{The IPW approach for missing covariates} \label{sect:weightedCox}

To handle missing covariates, we start with an IPW procedure based on using the \textit{complete odds} as defined below.  

\begin{definition}[Complete odds]
	For all missingness patterns $r$ and observed outcome patterns $a$, the complete odds is
	\begin{equation*}
		O_{r,a}(X,W_a) = \frac{P(R=r|X,W_a,A=a)}{P(R=1_d|X,W_a,A=a)}. \label{eq:complete-odds}
	\end{equation*}
\end{definition}

Thus, these complete odds $O_{r,a}(X,W_a)$ are a function of all of the outcome variables completely observed under outcome pattern $a$ and the covariates. With the complete odds, $\E[f_a(X,W_a)1(A=a)]$ can be decomposed via the following lemma.


\begin{lem}[Complete odds decomposition] \label{lemma:completeodds}
	The complete odds can be used to decompose the following two expectations as follows
	\begin{align*}
		\E[f_a(X,W_a)1(A=a)1(R=r)] &= \E\left[f_a(X,W_a)1(A=a)O_{r,a}(X,W_a)1(R=1_d)\right], 
	\end{align*}
	\begin{align}
		\E[f_a(X,W_a)1(A=a)] &= \sum_r\E\left[f_a(X,W_a)1(A=a)O_{r,a}(X,W_a)1(R=1_d)\right] \label{eq:lemma-completeodds-r}.
	\end{align}
\end{lem}

This lemma shows that the expectation can be estimated using the complete cases since we have the indicator $1(R=1_d)$ in the expression on the RHS.  Thus, we see that $\E[f_a(X,W_a)1(A=a)]$ is identifiable, provided that the complete odds $O_{r,a}(X,W_a)$ are identifiable for all $r$.  So, it is clear what we need to show before estimation.  Additionally, note that
\begin{equation}
	\sum_r O_{r,a}(X,W_a) = \sum_r \frac{P(R=r|X,W_a,A=a)}{P(R=1_d|X,W_a,A=a)} = \frac{1}{P(R=1_d|X,W_a,A=a)}
	\label{eq:PS}
\end{equation}
is the inverse propensity score, so we see that \eqref{eq:lemma-completeodds-r} recovers the form of the IPW estimator.  In light of this, one can view the complete odds as an alternate version of a propensity score.  In this situation, it is not required to use the complete odds approach since one may use the propensity score directly.  However, 
modeling via complete odds has several nice properties; we discuss this in Remark \ref{remark:completeodds}.

In the following proposition, we show that a set of identifiable complete odds is equivalent to a set of identifiable selection probabilities.  This connection offers one key insight: modeling the covariate missingness mechanism through the complete odds is equivalent to modeling the covariate missingness mechanism using the selection model \citep{LittleRubin02}.  Specific missing data assumptions may make the former approach easier than the latter, so this can be a useful result.

\begin{proposition}[Equivalence of complete odds and selection probabilities] \label{prop:odds-propensity}
	Fix an observed outcome pattern $a$.  The complete odds $O_{r,a}(X,W_a)$
	are identifiable for all $r$ if and only if the selection probabilities $P(R=r|X,W_a,a)$ are identifiable for all $r$.
\end{proposition}

Furthermore, another consequence of this proposition is that a set of identifiable complete odds implies that the distribution $p(x,r,w_a,a)$ is identifiable for all $a$.  Marginalizing out $r$ implies that the distribution $p(x,w_a,a)$ is nonparametrically identified \citep{robins2000sensitivity}, which is sufficient for showing the effect of interest is identifiable under the traditional assumptions that assume fully observed covariates.  This implies that with a set of identifiable complete odds, we can use existing problem-specific assumptions to identify the parameter of interest. 

\begin{remark}[Advantages to modeling the complete odds] \label{remark:completeodds}
	\emph{A modeling approach using the complete odds can be preferable to one using the selection probabilities directly.  Suppose $r$ and $a$ are fixed.  Then, $O_{r,a}(X,W_a)$ is variation independent of any $O_{r',a}(X,W_a)$ for $r'\neq r$.  Indeed, $O_{r,a}(X,W_a)$ can take any value in $[0,\infty)$ and is not constrained by any other odds function $O_{r',a}(X,W_a)$.
	On the other hand, the selection probabilities $P(R=r|x,w_a,a)$ have an explicit sum to 1 constraint, so the selection probabilities for all missing patterns $r$ need to be modeled at the same time.  In principle, one can use different methods to model each odds $O_{r,a}(X, W_a)$, provided it is identifiable under a missing data assumption for $X$.  These allows for some flexibility because one can choose to have different binary classifiers for each pattern $r$.  Due to the variation independence, they can all be compatible with each other.}
\end{remark}

Because of Proposition \ref{prop:odds-propensity},
we only need to identify the complete odds. 
Here we consider a (conditional)
complete case missing value (CCMV) assumption \citep{little1993pattern, tchetgen2018discrete}.

\begin{definition}[CCMV missing data assumption]
	Using a pattern mixture model factorization, the complete case missing value (CCMV) assumption equates the following two distributions
	\begin{equation}
		p(x_{\bar{r}} | x_r, r, w_a, a) \stackrel{\text{CCMV}}{=} p(x_{\bar{r}} | x_r, R=1_d, w_a, a) \label{eq:CCMV}
	\end{equation}
	for all covariate missingness patterns $r$ and observed outcome patterns $a$.
	\label{def:ccmv}
\end{definition}

Note that the above condition is a conditional CCMV in the sense that
we are using the CCMV for the missing $X$ 
conditioned on the observed outcomes. 
We call it CCMV for abbreviation.
Under Proposition \ref{prop:oddsCCMV}, this particular missing data assumption implies a simple closed form expression $Q_{r,a}(X_r,W_a)$ for the odds.  

\begin{proposition}[Odds under CCMV] \label{prop:oddsCCMV}
	Under the CCMV assumption for the covariates $X$, the odds $O_{r,a}(X,W_a)$ has the following simplified form in terms of observable quantities $X_r$ and $W_a$, namely
	\begin{equation}
		O_{r,a}(X,W_a) := \frac{P(R=r|X,W_a,a)}{P(R=1_d|X,W_a,a)} \stackrel{\text{CCMV}}{=} \frac{P(R=r|X_r,W_a,a)}{P(R=1_d|X_r,W_a,a)} =: Q_{r,a}(X_r,W_a). \label{eq:CCMV-odds}
	\end{equation}
\end{proposition}

Because each $Q_{r,a}(X_r,W_a)$ is identifiable from the data,
Proposition \ref{prop:oddsCCMV} shows that the CCMV assumption allows us to estimate the parameter
of interest under missing covariates.
The modeling $Q_{r,a}(X_r,W_a)$ is easy because it is comparing the proportions between
two missing patterns. 

We provide some comments on generalizing the CCMV assumption to other assumptions in Appendix \ref{appendix:beyondCCMV}.  Other missing data assumptions can yield other identifiable and estimable expressions for the odds, but one would simply obtain a different equality than the one in \eqref{eq:CCMV-odds}.  For example, \citet{chen2020pattern} exploits a pattern graph identifying restriction to present a recursive form for computing the odds.

\begin{remark}[Relation to density ratio estimation]
	\emph{\citet{wang2021propensity} have examined the problem of estimating an inverse propensity score using a density ratio representation under a missing at random assumption.  Interestingly, because they assume the variables can be partitioned into one set of always missing variables and one set of always observed variables, the missing at random assumption is equivalent to CCMV.  Additionally, our complete odds can be expressed as a density ratio as well, so these problems are related.  We elect to model our complete odds directly rather than reexpress it as a ratio of densities, but note that one can borrow methods from the density ratio literature to also tackle this problem.  See \citet{sugiyama_suzuki_kanamori_2012} for various approaches.}
\end{remark}

\begin{thm}[Nonparametric identification of CCMV]
The CCMV assumption in Definition \ref{def:ccmv} nonparametrically identifies the full-covariate distribution $p(x,r,w_a, a)$.
\label{thm:CCMV_ID}
\end{thm}

With Theorem~\ref{thm:CCMV_ID}, we are able to identify the full-covariate distribution $p(x,r,w_a, a)$. This implies
that the problem reduces to scenarios where the covariates are observed. 
Thus, we can estimate the parameter of interest as if all covariates are observed. 
In what follows, we give an example of Cox model, and in Appendix \ref{appendix:missingresp}, we provide examples
of missing response problems.

\subsubsection{Example: Cox model and survival problems}
Recall that in the survival analysis problem, $\Delta = A$ and $W_0 = W_1 = Y$.  In the case of $s^{(1)}(t;\beta)$, we have
\begin{equation}
	f_0(X,W_0) = f_1(X,W_1) = 1(Y\geq t) \exp(\beta^\top X) X. \label{eq:survival-f0-f1}
\end{equation}

For any fixed time point $t$, a direct application of Lemma \ref{lemma:completeodds} yields
\begin{align*}
	s^{(1)}(t;\beta) &= \E[1(Y\geq t) \exp(\beta^\top X) X] \\ 
	&= \E[1(Y\geq t) \exp(\beta^\top X) X 1(\Delta = 1)] + \E[1(Y\geq t) \exp(\beta^\top X) X 1(\Delta = 0)] \\
	&= \sum_r \E[1(Y\geq t) \exp(\beta^\top X) X 1(\Delta = 1)O_{r,1}(X,Y)1(R=1_d)] \\
	&\quad + \sum_r \E[1(Y\geq t) \exp(\beta^\top X) X 1(\Delta = 0)O_{r,0}(X,Y)1(R=1_d)] \\
	&= \sum_{r,\delta} \E[1(Y\geq t) X e^{\beta^\top X}O_{r,\delta}(X,Y)1(\Delta=\delta)1(R=1_d)],
\end{align*}
where $O_{r,\delta}(X,Y) = P(R=r|X,Y,\delta)/P(R=1_d|X,Y,\delta)$ are the complete odds.  A similar expression can be obtained for $s^{(0)}(t;\beta)$.  Next, we consider the estimating equation, which rewrites as
\begin{align}
	U(\beta) &= \E\left[\Delta \left(X - \frac{s^{(1)}(Y; \beta)}{s^{(0)}(Y;\beta)}\right)\right] \nonumber \\
	&= \sum_r \E\left[ \Delta \left(X - \frac{s^{(1)}(Y; \beta)}{s^{(0)}(Y;\beta)}\right) O_{r,1}(X,Y)1(R=1_d)\right] \label{eq:U-weighted}
\end{align}
using Lemma \ref{lemma:completeodds}.

With  Proposition \ref{prop:oddsCCMV}, $O_{r,\delta}(X,Y) = Q_{r,\delta}(X_r,Y)$, which provides an elegant way of computing the odds because the quantity $ Q_{r,\delta}(X_r,Y)$ is a ratio of two conditional probabilities that no longer depend on all of the covariates but rather observed covariates $X_r$.  We now discuss the construction of a multistage estimator under CCMV using the complete odds.  Henceforth, we will use $Q_{r,\delta}(X_r,Y)$ rather than $O_{r,\delta}(X,Y)$ because they are equal by Proposition \ref{prop:oddsCCMV}.  This choice of an identifying restriction allows for us to compute the inverse propensity score in a fairly straightforward manner.

There are three steps to our multistage estimator, which we outline below.  

\begin{enumerate}
	\item[1.] {\sc Stage 1 (Modeling the missingness of $X$ using the complete odds).}  We start by modeling $Q_{r,\delta}(X_r,Y)$ for every $r$ and $\delta$.  For example, we can impose a parametric model $Q_{r,\delta}(X_r,Y; \alpha_{r,\delta})$ such as a logistic regression.  Because this essentially becomes a binary classification problem (comparing the classes $R=r$ and $R=1_d$ based on the observed data $X_r$ and $Y$), one may also appeal to various machine learning methods such as decision trees and support vector machines.  Other parametric binary classification tools are also available.
	
	\item[2.] {\sc Stage 2 (Estimating intermediate quantities).} Next, we estimate the intermediate quantities of interest that are dependent on the potentially missing covariates $X$.  Because of the model described in Stage 1, this estimation becomes possible.  For example, assuming that the model for $Q_{r,\delta}(X_r,Y;\alpha_{r,\delta})$ is correctly specified and $\hat{\alpha}_{r,\delta}$ is a consistent estimator of $\alpha_{r,\delta}^*$, under the suitable regularity conditions, the following is a consistent estimator for $s^{(1)}(t;\beta)$
	\begin{align}
		\hat{s}^{(1)}(t;\beta) &= \sum_{r,\delta} \left(\frac{1}{n} \sum_{i=1}^n 1(Y_i\geq t) X_i e^{\beta^\top X_i} Q_{r,\delta}(X_{i,r}, Y_i; \hat{\alpha}_{r,\delta}) 1(\Delta_i=\delta) 1(R_i=1_d)\right) \nonumber \\
		&= \frac{1}{n}\sum_{i=1}^n\biggr(1(Y_i\geq t) X_i e^{\beta^\top X_i} \cdot \underbrace{\biggr[\sum_{r,\delta} Q_{r,\delta}(X_{i,r}, Y_i; \hat{\alpha}_{r,\delta}) 1(\Delta_i=\delta) 1(R_i=1_d)\biggr]}_{\lambda_i} \biggr) \label{eq:weighted-s1}\\
		&= \frac{1}{n}\sum_{i=1}^n \lambda_i \cdot 1(Y_i\geq t) X_i e^{\beta^\top X_i},  \nonumber
	\end{align}
	where $\lambda_i$ is a weight for the $i$th individual, as defined in \eqref{eq:weighted-s1}.  Similarly, we can mimic the same construction to estimate $s^{(0)}(t;\beta)$, yielding
	\begin{equation*}
		\hat{s}^{(0)}(t;\beta) = \frac{1}{n}\sum_{i=1}^n \lambda_i \cdot 1(Y_i\geq t) e^{\beta^\top X_i}.
	\end{equation*}
	
	\item[3.] {\sc Stage 3 (Constructing the final estimator).} Lastly, we assemble the intermediate quantities of interest together to create the final estimator.  Following the decomposition in \eqref{eq:U-weighted}, we have
	\begin{align*}
		U_n(\beta) &= \sum_{r,\delta} \left(\frac{1}{n}\sum_{i=1}^n \Delta_i \left(X_i - \frac{\hat{s}^{(1)}(Y_i; \beta)}{\hat{s}^{(0)}(Y_i;\beta)} \right) Q_{r,\delta}(X_{i,r}, Y_i; \hat{\alpha}_{r,1}) 1(R_i=1_d)\right) \\
		&= \frac{1}{n} \sum_{i=1}^n \biggr(\Delta_i \left(X_i - \frac{\hat{s}^{(1)}(Y_i; \beta)}{\hat{s}^{(0)}(Y_i;\beta)} \right)\cdot \underbrace{\sum_r Q_{r,1}(X_{i,r}, Y_i; \hat{\alpha}_{r,1}) 1(\Delta_i=1) 1(R_i=1_d)}_{\lambda_{i,1}} \biggr) \\
		&= \frac{1}{n} \sum_{i=1}^n \lambda_{i} \cdot \Delta_i \left(X_i - \frac{\hat{s}^{(1)}(Y_i; \beta)}{\hat{s}^{(0)}(Y_i;\beta)} \right) ,
	\end{align*}
	where the second to last equality follows because $\Delta_i \in \{0,1\}$.  The last equality follows because $\lambda_i \Delta_i = \lambda_{i,1} \Delta_i$.  Thus, $U_n(\beta)=0$ is a reweighted estimating equation and can be solved to provide a consistent estimator $\hat{\beta}$ for $\beta^*$.
	
\end{enumerate}

A nice property is that the weights $\lambda_i$ and $\lambda_{i,1}$ need only to be computed from the data once and can be reused thereafter.  We summarize the implementation of this reweighted Cox model in Algorithm \ref{alg:weighted-Cox}.

\begin{algorithm}[ht]
	\caption{Fitting a weighted Cox model under CCMV for $X$}
	\begin{enumerate}
		\setlength{\itemsep}{0pt}
		\item Input: $\{(X_{i,R_i}, R_i, Y_i, \Delta_i)\}_{i=1}^n$
		\item For each pair $(r,\delta)$, fit the parametric model $Q_{r,\delta}(X_r,Y; \alpha_{r,\delta}).$
		
		\item For $i=1,\ldots,n$, do
		
		\begin{enumerate}
			\item [3-1.] Form the weights
			$$\lambda_{i} = \sum_\delta \sum_r Q_{r,\delta}(X_{i,r}, Y_i; \hat{\alpha}_{r,\delta}) 1(\Delta_i=\delta) 1(R_i=1_d).$$
		\end{enumerate}
		
		\item Estimate the intermediate quantities
		$$\hat{s}^{(0)}(t;\beta) = \frac{1}{n}\sum_{i=1}^n \lambda_i \cdot 1(Y_i\geq t) e^{\beta^\top X_i},$$
		$$\hat{s}^{(1)}(t;\beta) = \frac{1}{n}\sum_{i=1}^n \lambda_i \cdot 1(Y_i\geq t) X_i e^{\beta^\top X_i}.$$
		
		\item Find $\hat{\beta}$ by finding a near-solution to the finite sample estimating equation
		$$U_n(\beta) = \frac{1}{n} \sum_{i=1}^n \lambda_{i} \cdot \Delta_i \left(X_i - \frac{\hat{s}^{(1)}(Y_i; \beta)}{\hat{s}^{(0)}(Y_i;\beta)} \right) = 0.$$
		\item Return: $\hat{\beta}$.
	\end{enumerate}
	\label{alg:weighted-Cox}
\end{algorithm}

\subsection{Regression adjustment and imputation for missing covariates} \label{sect:raCox}


In addition to the IPW approach for handling missing covariates,
we can use a regression adjustment approach as well.
Moreover, this regression adjustment naturally leads to an imputation procedure. 
The following lemma is a key step to the regression adjustment (RA).

\begin{lem}[RA decomposition] \label{lemma:ra}
	Regression adjustment can be used to decompose the following two expectations
	\begin{align*}
		\E[f_a(X,W_a)1(A=a)1(R=r)] &= \E[m_{r,a}(X_r, W_a) 1(R=r)1(A=a)],\\
		\E[f_a(X,W_a)1(A=a)] &= \sum_r \E[m_{r,a}(X_r, W_a) 1(R=r)1(A=a)],
	\end{align*}
	where $m_{r,a}(X_r, W_a) = \E[f_a(X,W_a)|X_r,R=r,W_a,A=a]$ is a regression function.
\end{lem}

Thus, in order to estimate the expectation, it suffices to model the regression function $m_{r,a}(X_r,W_a)$, which is a conditional expectation.

	In practice, placing a parametric model for each regression function via $m_{r,a}(x_r,w_a;\beta_{r,a})$ may be problematic because naively fitting models can lead to incompatible regression functions and uncongenial sources of input \citep{meng1994multiple}.  Instead, we recommend placing models on the extrapolation densities $p(x_{\bar{r}} | x_r, r, a, w_a)$ for all $r$.  This ensures that the modeled regression functions will be compatible with each other.  

Under the CCMV assumption (equation \ref{eq:CCMV}), we have 
$$
p(x_{\bar{r}} | x_r, r, w_a, a) = p(x_{\bar{r}} | x_r, R=1_d, w_a, a),
$$
which further implies
$$
m_{r,a}(X_r, W_a) = \E[f_a(X,W_a)|X_r,R=r,W_a,A=a] \stackrel{\text{CCMV}}{=} \E[f_a(X,W_a)|X_r,R=1_d,W_a,A=a].
$$
The RHS is identifiable using observations with complete covariates. 
In this case, there are two ways of estimating the regression function:

\begin{enumerate}
	\item {\bf Method 1: Closed-form computation.} 
	In some good scenarios, we may be able to derive the 
	the closed-form of the expectation $m_{r,a}(X_r, W_a) = \E[f_a(X,W_a)|X_r,R=r,W_a,A=a]$
	from a parametric model of the extrapolation densities $p(x_{\bar{r}} | x_r, r, a, w_a)$.	
	For instance, a multivariate Gaussian model implies that the conditional distributions will be conditional Gaussians.  
	
	\item {\bf Method 2: Imputation (Monte Carlo approximation).} 
	Alternatively, we can approximate the conditional expectation by
	a Monte Carlo method.  This is equivalent to the frequentist multiple imputation. 
	We generate from the extrapolation density $p(x_{\bar{r}} | x_r, r, a, w_a)$
	and then evaluate the conditional expectation $\E[f_a(X,W_a)|X_r,R=r,W_a,A=a]$.	
	 Under CCMV \eqref{eq:CCMV}, the extrapolation density $p(x_{\bar{r}} | x_r, r, y, \delta)$ is equated to the identifiable distribution $p(x_{\bar{r}} | x_r, R=1_d, y, \delta)$.  The latter distribution can be estimated using the complete cases.  
	
\end{enumerate}


\subsubsection{Example: Cox model and survival problems}

Using Lemma \ref{lemma:ra} with the functions defined in \eqref{eq:survival-f0-f1}, we have
\begin{align*}
	s^{(1)}(t;\beta) &= \E[1(Y\geq t) \exp(\beta^\top X) X] \\
	&= \sum_{r,\delta} \E[m^{(1)}_{r,\delta}(X_r,Y;t,\beta) 1(R=r) 1(\Delta= \delta)].
\end{align*}

Thus, to estimate $s^{(1)}(t;\beta)$, it suffices to model $m^{(1)}_{r,\delta}(X_r,Y;t,\beta) := \E[1(Y\geq t) \exp(\beta^\top X) X | X_r, R=r, Y, \Delta = \delta]$ for all $(r,\delta)$ pairs.  We obtain a similar expression for $s^{(0)}(t;\beta)$ via
\begin{equation*}
	s^{(0)}(t;\beta) = \sum_{r,\delta} \E[m^{(0)}_{r,\delta}(X_r,Y;t,\beta) 1(R=r) 1(\Delta= \delta)]
\end{equation*}
for $m^{(0)}_{r,\delta}(X_r,Y;t,\beta) := \E[1(Y\geq t) \exp(\beta^\top X) | X_r, R=r, Y, \Delta = \delta]$.  

Applying Lemma \ref{lemma:ra} on the overall estimating equation, we obtain
\begin{align*}
	U(\beta) = \sum_r \E\left[\E\left[X - \frac{s^{(1)}(Y;\beta)}{s^{(0)}(Y;\beta)} \biggr |X_r, R,Y, \Delta\right]\cdot 1(R=r) \cdot 1(\Delta=1) \right].
\end{align*}
Therefore, the regression adjustment method is equivalent to finding a near-solution to the following finite sample estimating equation
\begin{align}
	&U_n(\beta) = \sum_r \left[\frac{1}{n} \sum_{i=1}^n \left(\hat{\E}\left[X - \frac{\hat{s}^{(1)}(Y;\beta)}{\hat{s}^{(0)}(Y;\beta)} \biggr |X_{i,r}, R_i, Y_i, \Delta_i\right] \right) \cdot 1(R_i=r) \cdot 1(\Delta_i=1)\right] = o_p(1/\sqrt{n}), \nonumber \\
	& \hat{s}^{(1)}(t;\beta) = \sum_{r,\delta} \left[ \frac{1}{n} \sum_{i=1}^n \hat{m}^{(1)}_{r,\delta}(X_{i,r},Y_i;t,\beta) 1(R_i=r)1(\Delta_i=\delta) \right], \nonumber \\
	& \hat{s}^{(0)}(t;\beta) = \sum_{r,\delta} \left[ \frac{1}{n} \sum_{i=1}^n \hat{m}^{(0)}_{r,\delta}(X_{i,r},Y_i;t,\beta) 1(R_i=r)1(\Delta_i=\delta) \right].\label{eq:U-ra}
\end{align}

This can be achieved using a Monte Carlo procedure.  We describe the multistage process below, and we summarize our sampling method in Algorithm \ref{alg:mi}.  This algorithm is the same as performing a frequentist (type B) multiple imputation approach described in \citet{Wang1998} and \citet{LittleRubin02}.

\begin{enumerate}
	\item[1.] {\sc Stage 1 (Modeling the missingness of $X$ with extrapolation densities).} As previously discussed, instead of modeling the regression functions $m^{(1)}_{r,\delta}(X_r,Y;t,\beta)$ and $m^{(0)}_{r,\delta}(X_r,Y;t,\beta)$ for all $r$ and $\delta$ directly, we will instead focus on the extrapolation densities.  Under CCMV, we simply need to model the joint distribution $p(x,R=1_d,y,\delta)$.  This will induce a model $p(x_{\bar{r}}|x_r,R=1_d,y,\delta)$ for any covariate missing pattern $r$.  By CCMV, that conditional distribution equates exactly to the extrapolation distribution for the missing data.
	
	\item[2.] {\sc Stage 2 (Estimating intermediate quantities).} Next, we again outline how to estimate the nuisance functions $s^{(1)}(t;\beta)$ and $s^{(0)}(t;\beta)$ that are dependent on the potentially missing covariates $X$.  We estimate these using a Monte Carlo procedure through iterative imputation of the missing covariates.
	
	By estimating the regression functions using $M$ draws from the extrapolation distribution, the following is a consistent estimator for $s^{(1)}(t;\beta)$
	\begin{align*}
		\hat{s}^{(1)}(t;\beta) &= \sum_{r,\delta} \left[\frac{1}{n}\sum_{i=1}^n \hat{m}^{(1)}_{r,\delta}(X_{i,r},Y_i;t,\beta) 1(R_i=r) 1(\Delta_i= \delta)\right] \\
		&= \frac{1}{n}\sum_{i=1}^n \left(\sum_{r,\delta}  \hat{m}^{(1)}_{r,\delta}(X_{i,r},Y_i;t,\beta) 1(R_i=r) 1(\Delta_i= \delta)\right) \\
		&= \frac{1}{n} \sum_{i=1}^n \left ( \frac{1}{M}\sum_{j=1}^M  1(Y_i\geq t) \tilde{X}_i^{(j)} e^{\beta^\top \tilde{X}_i^{(j)}} \right), \\
		&= \frac{1}{Mn}\sum_{j=1}^M \sum_{i=1}^n  1(Y_i\geq t) \tilde{X}_i^{(j)} e^{\beta^\top \tilde{X}_i^{(j)}},
	\end{align*}
	where $\tilde{X}_i^{(j)} = \left(X_{i,R_i}, \tilde{X}_{i,\bar{R}_i}^{(j)}\right)$.  Here $\tilde{X}_{i,\bar{R}_i}^{(j)}$ denotes the $j$th imputed tuple of missing covariates for the $i$th observation, and $\tilde{X}_i^{(j)}$ is the result of attaching the imputed variables to the observed covariates $X_{i,R_i}$.  A similar construction applied to $s^{(0)}(t;\beta)$ yields
	\begin{equation*}
		\hat{s}^{(0)}(t;\beta) = \frac{1}{Mn}\sum_{j=1}^M \sum_{i=1}^n  1(Y_i\geq t) e^{\beta^\top \tilde{X}_i^{(j)}}.
	\end{equation*}
	
	\item[3.] {\sc Stage 3 (Constructing the final estimator).} Lastly, we again merge everything together in the final stage.  Following the decomposition in \eqref{eq:U-ra}, we have
	\begin{align*}
		U_n(\beta) &= \frac{1}{Mn} \sum_{j=1}^M \sum_{i=1}^n \Delta_i \left(\tilde{X}_i^{(j)} - \frac{\hat{s}^{(1)}(Y_i; \beta)}{\hat{s}^{(0)}(Y_i;\beta)} \right).
	\end{align*}
	We solve the above estimating equation to obtain the final estimate $\hat \beta$.

\end{enumerate}

The terms we have derived suggest a simple procedure for the estimation.  These estimators $\hat{s}^{(1)}(t;\beta), \hat{s}^{(0)}(t;\beta),$ and $U_n(\beta)$ are all formed by merging the $M$ imputed datasets together to form one large completed dataset with $Mn$ entries.  Thus, this regression adjustment method with a Monte Carlo approximation is equivalent to a stacked multiple imputation approach utilized by \citet{RobinsWang2000, BeesleyTaylor2020}.  Note that this approach differs from a traditional imputation approach where $M$ imputed datasets are generated separately, analysis is performed on each one, and each analysis is pooled to form a final estimator.  In contrast, in the stacked imputation approach, the pooling operation is done before the analysis such that $M$ imputed datasets are pooled first and analysis is performed on all $Mn$ observations to form a final point estimator.
	
	An advantage of the traditional imputation approach is that the separate $M$ analyses allows for parallelization.  On the other hand, the stacked imputation does not exhibit this nice property, but a single analysis yields a more straightforward implementation, which can be helpful in situations where we perform the bootstrap.

\begin{algorithm}
	\caption{Cox model using multiple imputation under CCMV for $X$}
	\label{alg:mi}
	\begin{enumerate}
		\setlength{\itemsep}{0pt}
		\item Input: $\{(X_{i,R_i}, R_i, Y_i, \Delta_i)\}_{i=1}^n$
		\item For $\delta=0,1$, fit the parametric model $p(x|R=1_d,y,\delta).$
		
		\item For $j=1,\ldots, M$, do
		
		\begin{enumerate}
			\item [3-1.] For $i=1,\ldots, n$, do
			\begin{enumerate}
				\item Impute the missing covariates $\tilde{X}_{i,\bar{R}_i}^{(j)}$ for observation $(X_{i,R_i}, R_i, Y_i, \Delta_i)$ using the estimated extrapolation distribution $p(x_{\bar{r}} | x_r, R=1_d, y, a)$.
				\item Set the $j$th imputation for the $i$th observation to be $(\tilde{X}_i^{(j)}, R_i, Y_i, \Delta_i)$, where $\tilde{X}_i^{(j)} = \left(X_{i,R_i}, \tilde{X}_{i,\bar{R}_i}^{(j)}\right)$.
			\end{enumerate}
			\item [3-2.] Let $\tilde{D}^{(j)} = \{(\tilde{X}_i^{(j)}, R_i, Y_i, \Delta_i)\}_{i=1}^n$ denote the $j$th imputed data set.
		\end{enumerate}
		
		\item Pool all $\tilde{D}^{(j)}, j=1,\ldots, M$ data sets together into a single large one $\check{D}$.
		
		\item Fit a Cox model using the completed data set $\check{D}$ to get an estimator $\hat{\beta}$.
		
		\item Return: $\hat{\beta}$.
	\end{enumerate}
	\label{alg:ra-Cox}
\end{algorithm}

	{In a frequentist multiple imputation approach, we first use the observed data to estimate the extrapolation densities and thereby, perform the imputation.  Therefore, all further analysis is based on the estimated extrapolation densities, and it is important to incorporate this uncertainty in the final variance calculation.  Rubin's estimator for the variance is commonly used in practice, but it will underestimate the variance in this situation.  Specifically, there will be an additional term in the asymptotic variance related to the estimator for the extrapolation distribution that will not be included.  See Sections 14.4 and 14.5 of \citet{tsiatis2007semiparametric} for more details.}
	
	{To avoid this issue in practice, we recommend using the nonparametric bootstrap to estimate the variance.  We bootstrap the data (including the missing values) many times and repeat the same analysis on each bootstrapped sample to obtain many point estimates.  Each analysis on a bootstrapped sample will utilize a different estimated extrapolation distribution, so this uncertainty will be taken into account.  \cite{chen2019nonparametric} provide additional related details in their work.}

\subsection{A multiply-robust estimator}

In the previous two sections, we have described two approaches based on an inverse probability weighting and a regression adjustment.  In the seminal paper by \citet{Robins1994}, the authors introduced a general framework of an augmented inverse probability weighted (AIPW) estimating function under a MAR assumption for the covariates, and thus, the selection probability did not depend on any missing covariates.  \citet{WangChen2001} discussed further details on constructing an AIPW estimator with augmented functions for the Cox model under MAR.  Additionally, in their paper, the covariates took the form of $(X_{\text{miss}}, X_{\text{obs}})$, where $X_{\text{miss}}$ was a covariate vector possibly missing for some individuals and $X_{\text{obs}}$ was a covariate vector that was always observed, thereby inducing only two possible missingness patterns for the covariates.  These AIPW methods rely on two parametric models for the selection probabilities and the missing covariates.  To relax this parametric assumption, \citet{Prentice2005} outlined a fully augmented weighted estimator using a kernel regression approach for a one-dimensional covariate.  
However, it is unclear how to handle the case with multiple covariates that may all be missing.


We extend the previous work by describing a multiply-robust estimator that allows missing not at random (MNAR) for a multidimensional covariate and arbitrary missing patterns.  The following lemma shows that there exists an estimator that is robust to the misspecification of at most one of $Q_{r,a}(X_r,W_a)$ or $m_{r,a}(X_r,W_a)$ for each $r$.

\begin{lem}[Doubly robust linear form under CCMV] \label{lemma:DR-form}
	Fix a missing pattern $r$ and an observed outcome pattern $a$.  Define
	\begin{align}
		g_r(X,W_a) &= f_a(X,W_a)1(A=a) 1(R=1_d) Q_{r,a}(X_r,W_a) \label{eq:odds-exp} \\
		&\quad+ m_{r,a}(X_r,W_a)1(R=r)1(A=a) \label{eq:ra-exp}\\
		&\quad- m_{r,a}(X_r,W_a)1(A=a) 1(R=1_d) Q_{r,a}(X_r,W_a). \label{eq:dr-exp}
	\end{align}
	If $Q_{r,a}(X_r,W_a)$ or $m_{r,a}(X_r,W_a)$ is correctly specified, then under a CCMV assumption of equation \eqref{eq:CCMV}, we have
	\begin{equation*}
		\E[f_a(X,W_a)1(A=a)] = \sum_r \E[g_r(X,W_a)].
	\end{equation*}
\end{lem}

%

The expression of Lemma~\ref{lemma:DR-form} shows an interesting connection to equation (10.95) of \citet{tsiatis2007semiparametric}.
In the case where the there is only two missing patterns of the covariates, i.e.,
the covariates can be expressed as $(X_{\text{miss}}, X_{\text{obs}})$, where $X_{\text{miss}}$ are possibly missing covariates and $X_{\text{obs}}$ are completely observed covariates, the MAR and CCMV assumptions on the covariates can be shown to be equivalent.  Our expression and equation (10.95) in \citet{tsiatis2007semiparametric} will be the same.

\begin{thm}[Multiply-robust Cox model]
	\label{theorem:multiplyrobust-cox}
	Under CCMV, the estimator $\hat{\beta}_\text{mr}$ that solves the augmented estimating function is consistent provided either $Q_{r,\delta}(X_r,Y)$ or $m_{r,\delta}(X_r,Y)$ is correctly specified for each $r$.
\end{thm}

Thus, this implies that we can construct a multistage estimator based on Lemma \ref{lemma:DR-form} that is multiply-robust since we have doubly robustness for each covariate pattern $r\neq1_d$.  However, this comes at a cost.  We see that the doubly robust expression above contains the terms from Lemmas \ref{lemma:completeodds} and \ref{lemma:ra} along with an additional term.  Term \eqref{eq:dr-exp} is of interest because it shows that in order to compute it, we will have to perform imputation even for cases with completely observed covariates.  Therefore, we can see explicitly that this doubly robust estimator is even more computationally intensive than the regression adjustment estimator in Lemma \ref{lemma:ra}.

\section{Simulation} \label{sect:simulation}

We examine two scenarios of the Cox model and the binary treatment problem to observe the consistency of our proposed estimators as well as behavior of the estimators under perturbations of the assumed identifying restriction.

\subsection{Cox model simulation}

We consider a Cox model with two binary covariates
$X=(X_1,X_2)\in\{0,1\}^2$ from the following generative model:
\begin{enumerate}
	\item Sample $X= (X_1,X_2)$ such that $P(X_1=1)=1/2$, $P(X_2=1)=3/10$, and $\text{Corr}(X_1,X_2)=0.3$.
	\item For $T$ and $C$ (and $Y$ and $\Delta$), we do the following:
	\begin{itemize}
		\item 
		Generate $T|X$ from an Exponential distribution with rate $\exp(x^\top \beta)$  and $\beta = \begin{bmatrix} -0.5 \\ 2\end{bmatrix}$. This is based on the baseline hazard $\lambda_0(t) = 1$.
		\item Generate $C|X$ from an distribution with rate $2$.
		\item Compute $Y = \min\{T,C\}$ and $\Delta = 1(T\leq C)$.
	\end{itemize}
	\item Generate the missing pattern of covariates based on CCMV:
	\begin{align*}
		\dfrac{P(R=01|x,y,\delta)}{P(R=11|x,y,\delta)} = \dfrac{P(R=01|x_2,y,\delta)}{P(R=11|x_2,y,\delta)} = \exp(-0.5x_2 + 0.75y + 0.5\delta), \\
		\dfrac{P(R=10|x,y,\delta)}{P(R=11|x,y,\delta)} = \dfrac{P(R=10|x_1,y,\delta)}{P(R=11|x_1,y,\delta)} = \exp(-1 + 0.5x_1 + y + \delta),
	\end{align*}
	and remove the covariates accordingly.
\end{enumerate}

Roughly, we expect approximately $40\%$, $30\%$, and $30\%$ of the observations to fall in the classes $R=01$, $R=10$, and $R=11$, respectively.  This data generating mechanism completely specifies the joint model, and we  consider three different estimators.  The first estimator is the IPW estimator, which is based off the weighted Cox model from Section \ref{sect:weightedCox}, and the complete odds are estimated using logistic regression under CCMV.  The second estimator is multiple imputation (MI) / regression adjustment estimator as discussed in Section \ref{sect:raCox} with $M=50$ imputations.  Notably, the complete odds and the imputation distribution are not variation independent in this problem (and will generally not be in most problems of interest), so we have to fit the imputation distribution using knowledge of the complete odds.
The third estimator is called the transformed MLE estimator.  The specific parametric model assumed for $p_{Y,\Delta|X}(y,\delta|x)$ implies that we can recover the terms
$$\nu_1, \quad \nu_1\exp(\beta_1), \quad \nu_1\exp(\beta_2), \quad \text{and} \quad \nu_1\exp(\beta_1+\beta_2)$$
via maximum likelihood through the EM algorithm (details provided in Appendix \ref{appendix:sim:survival}).  Let $\hat{\gamma}_i$ denote the estimator for the $i$th term above ($i=1,2,3,4$).  Two naive estimators, which we term the transformed MLEs, are as follows:
\begin{align*}
	\hat{\beta}_{1,\text{Transformed MLE}} = \frac{1}{2} \log(\hat{\gamma}_2 / \hat{\gamma}_1) + \frac{1}{2} [\log(\hat{\gamma}_4 / \hat{\gamma}_1) - \log(\hat{\gamma}_3 / \hat{\gamma}_1)], \\
	\hat{\beta}_{2,\text{Transformed MLE}} = \frac{1}{2} \log(\hat{\gamma}_3 / \hat{\gamma}_1) + \frac{1}{2} [\log(\hat{\gamma}_4 / \hat{\gamma}_1) - \log(\hat{\gamma}_2 / \hat{\gamma}_1)].
\end{align*}
We did not include the DR estimator in comparison because it has a very complex form (see Lemma~\ref{lemma:DR-form}).

\begin{figure}[!h]
	\centering
	\includegraphics[scale=0.5]{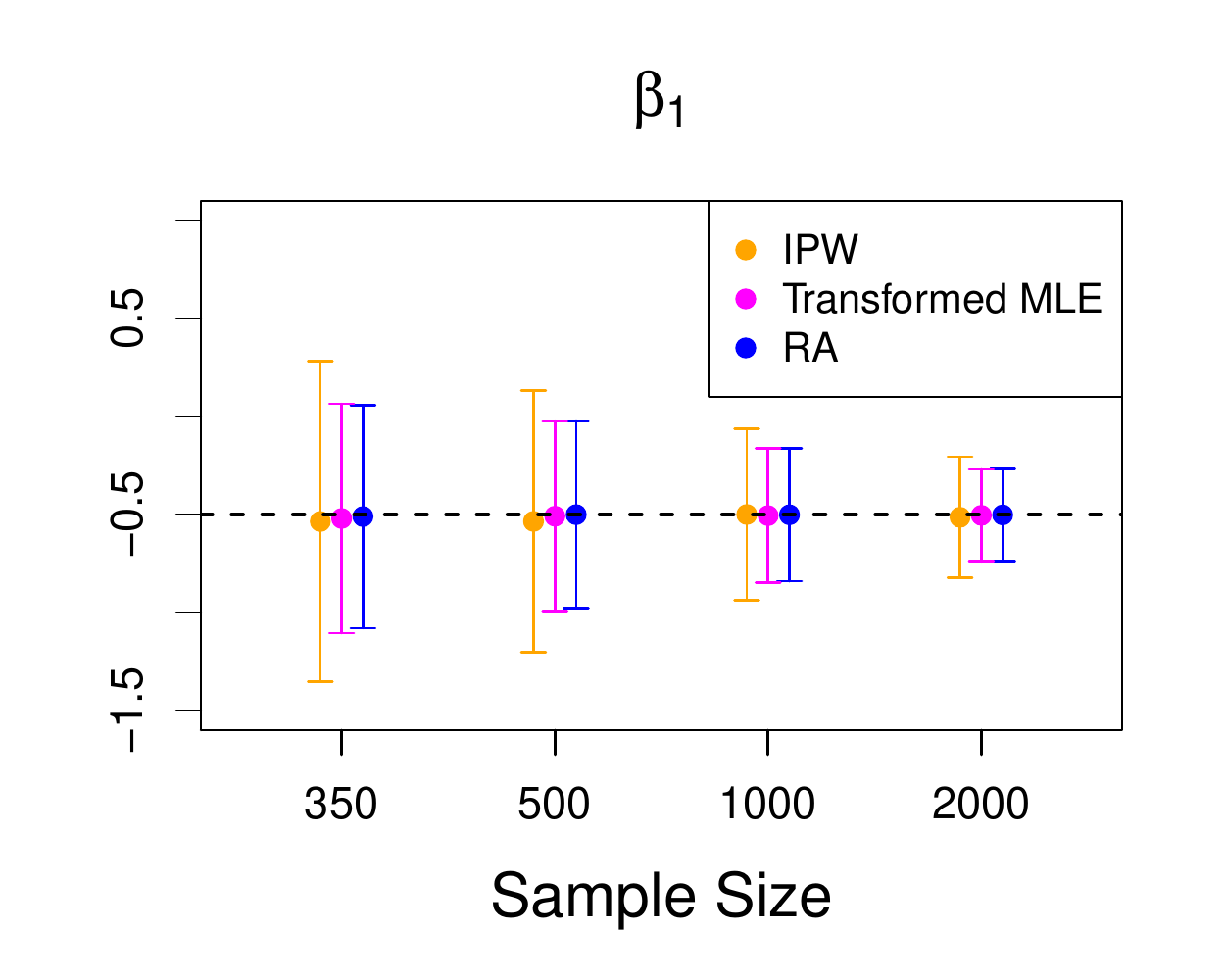}
	\includegraphics[scale=0.5]{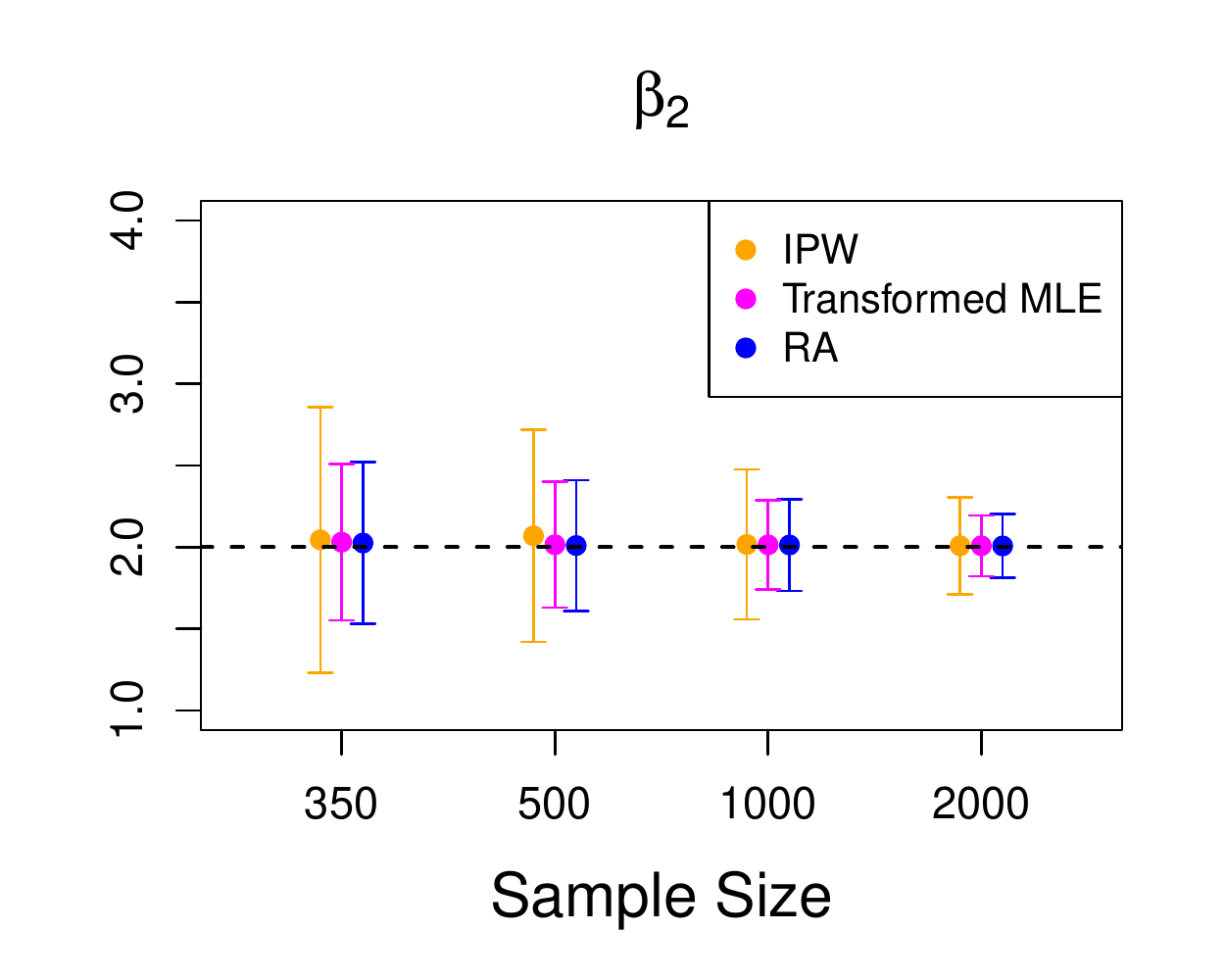}
	\caption{Estimation of $\beta$ in a simulated Cox model for varying sample sizes.}
	\label{fig:betas}
\end{figure}

For each of the sample sizes (350, 500, 1000, 2000), we generate 1000 data sets and construct the three estimators.  The means and the standard errors are calculated across the 1000 replicates, and we show the 95\% confidence interval.  The results are summarized in Figure \ref{fig:betas}.  We first observe that all three estimators are consistent and have standard errors that appear to roughly decline at the appropriate $\sqrt{n}$-rate.  The IPW estimator (equivalent to the reweighted Cox model) has the highest variance, which is to be expected because it uses the least information and inverse probabilities are known to introduce instability.  The multiple imputation / regression adjustment (MI/RA) estimator uses slightly more information than the IPW estimator by assuming a parametric model for $p_{Y,\Delta|X}(y,\delta|x)$.  As the imputation distribution and the complete odds are typically not variation independent, we are forced to fit them together.  We see that the RA estimator appears to have similar variance to the transformed MLE, which is to be expected for large enough $M$ (number of imputations).

\subsection{Binary treatment problem}

\begin{figure}[!h]
	\centering
	\includegraphics[scale=0.5]{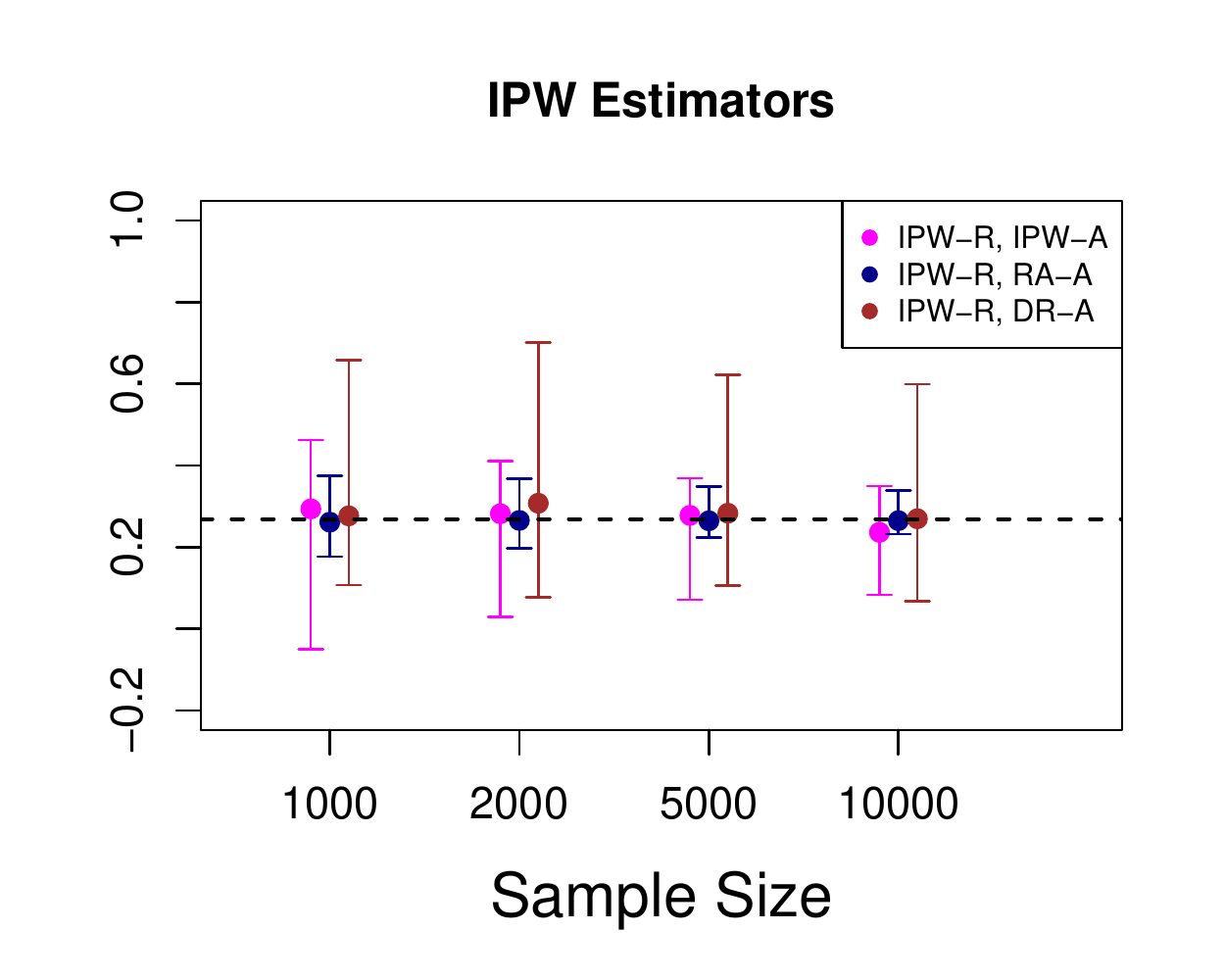}
	\includegraphics[scale=0.5]{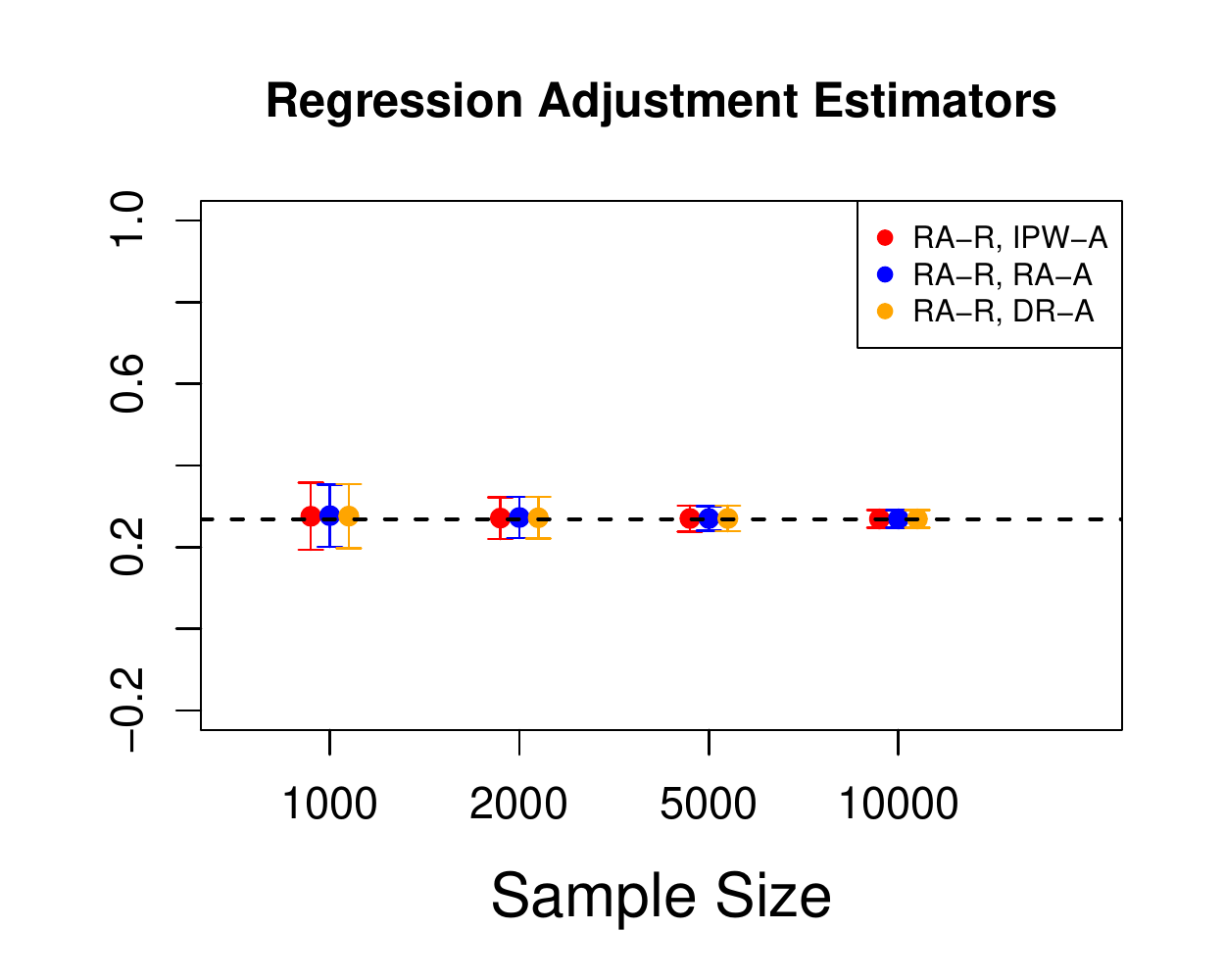}
	\caption{Estimation of the ATE in a simulated causal inference problem for varying sample sizes.}
	\label{fig:binary}
\end{figure}

We perform another simulation under the binary treatment setting from causal inference to show consistency under a new setting.  The generating process is described in Appendix \ref{appendix:sim:binary}.  For each of the samples sizes (1000, 2000, 5000, 10000), we generate 1000 data sets and construct $6$ estimators. We consider two methods for accounting for the missing covariates (IPW-$R$ and RA-$R$) and three methods for estimating the ATE (IPW-$A$, RA-$A$ and DR-$A$). So, there are 6 combinations of one method for missing covariates and one method for the ATE.
The left panel of Figure \ref{fig:binary} displays the results of
three estimators  formed by using an IPW approach for the missing covariates: (IPW-$R$, IPW-$A$), (IPW-$R$, RA-$A$), and (IPW-$R$, DR-$A$).  
In this case, we use the bootstrap percentile method for computing the confidence intervals.
The right panel of Figure \ref{fig:binary} provides the results of
three estimators using a RA (imputation) approach for the missing covariates: (RA-$R$, IPW-$A$), (RA-$R$, RA-$A$), and (RA-$R$, DR-$A$).  We use $M=50$ imputations and report the 95\% confidence intervals.
We simply construct the confidence intervals by computing the standard errors across all 1000 generated data sets.  Again, we also report the means across the 1000 replicates.

In the left panel of Figure \ref{fig:binary}, we observe some surprising results.  First, in finite samples, the (IPW-$R$, IPW-$A$) estimator can lead to unstable estimates for some data sets due to the fact that the complete odds are estimated using a density ratio.  Although we expect the doubly-robust estimator (IPW-$R$, DR-$A$) to have smaller variance than the (IPW-$R$, IPW-$A$) estimator, we see that it has larger variance. This is because the unstable part of IPW-$R$ appears in each of the 3 terms of the DR-$A$ estimator.


In the right panel of Figure \ref{fig:binary},
we see very similar performance across all 3 estimators.  These 3 estimators dominate the other 3 estimators in the right panel, which  suggests that a multiple imputation approach can provide more stable results in practice.  
We discuss further sensitivity analysis for this simulation in Appendix \ref{appendix:sens-binary-imp}.

\section{Data analysis}  \label{sect:data}

We apply our methods on the data from the National Alzheimer's Coordinating Center's (NACC) Uniform Data Set\footnote{\url{https://naccdata.org/}}.  This data set contains longitudinal data collected at Alzheimer's Disease Research Centers, funded by the National Institutes of Health and starting from the year 2005.  Participating individuals have varying cognitive statuses from no cognitive impairment to mild impairment to severe impairment, and data is collected on a yearly basis with certain covariates measured at entry into the study.

We consider individuals who display mild cognitive impairment (MCI) and follow their trajectory until they either display evidence of dementia (mild, moderate, severe) or are censored.  These individuals are binned into 5 age groups based on when they first display MCI: 60-64, 65-69, 70-74, 75-79, and 80-84.  In each age group, we fit a Cox model, regressing on the binary variables: sex (female or not), race (white or not), education (have graduate degree or not), depression within last 2 years (yes or no), and anxiety (yes or no).   
	Tracking the change in the $\beta$ across the different age groups allows us to observe the age effect.  The first three covariates (sex, race, and education) are generally measured for all individuals at the baseline.  There are a few individuals for which the education status is unknown, but because these represent less than $1\%$ of the observations, we exclude this individuals from the analysis for simplicity.  The last two covariates (depression and anxiety) are subject to missingness, yielding a total of $2^2=4$ covariate missing patterns.
	There are about $16\%$ participants in our data with complete observations.

\begin{figure}[!h]
	\centering
	\includegraphics[scale=0.35]{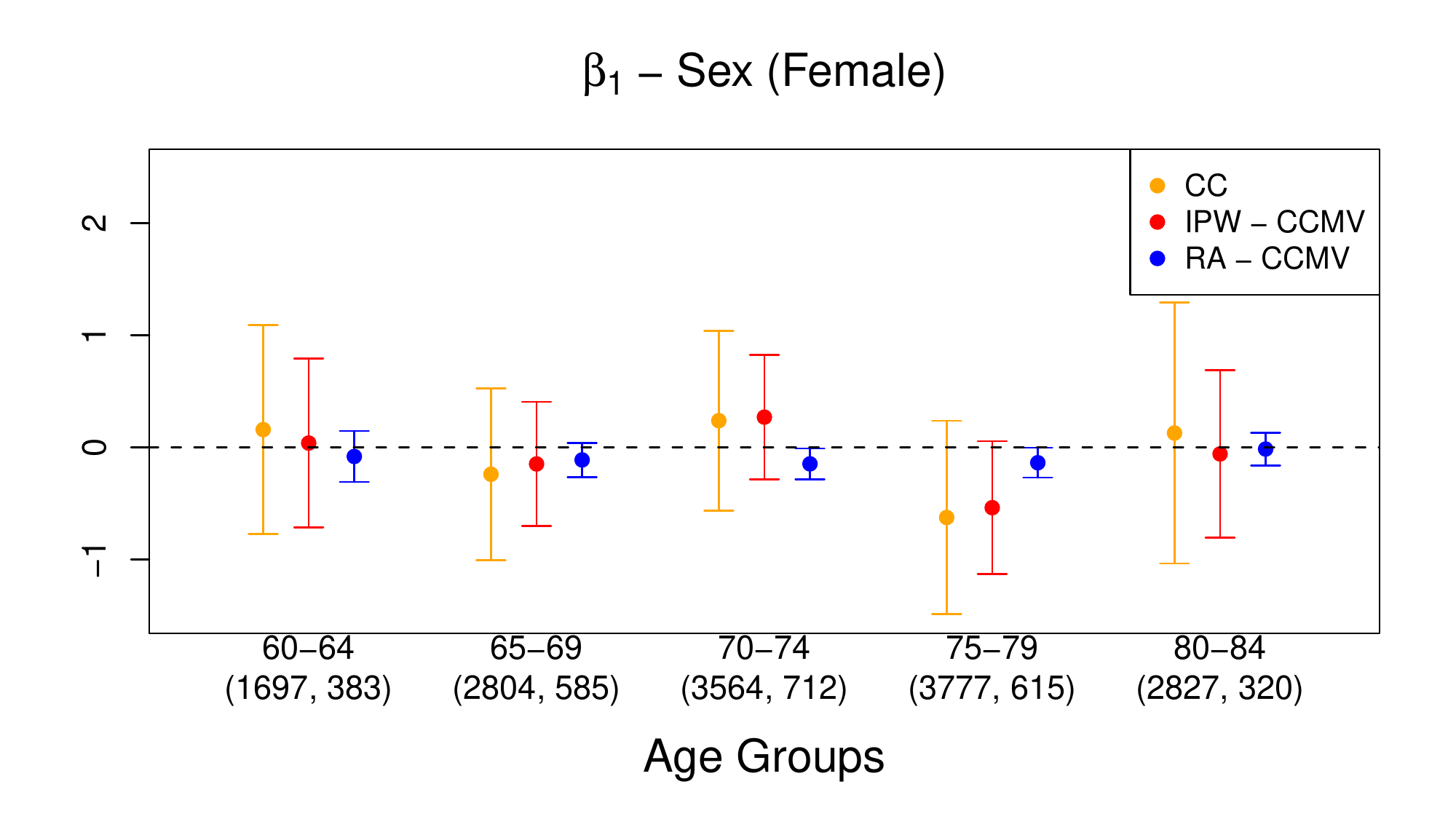}
	\includegraphics[scale=0.35]{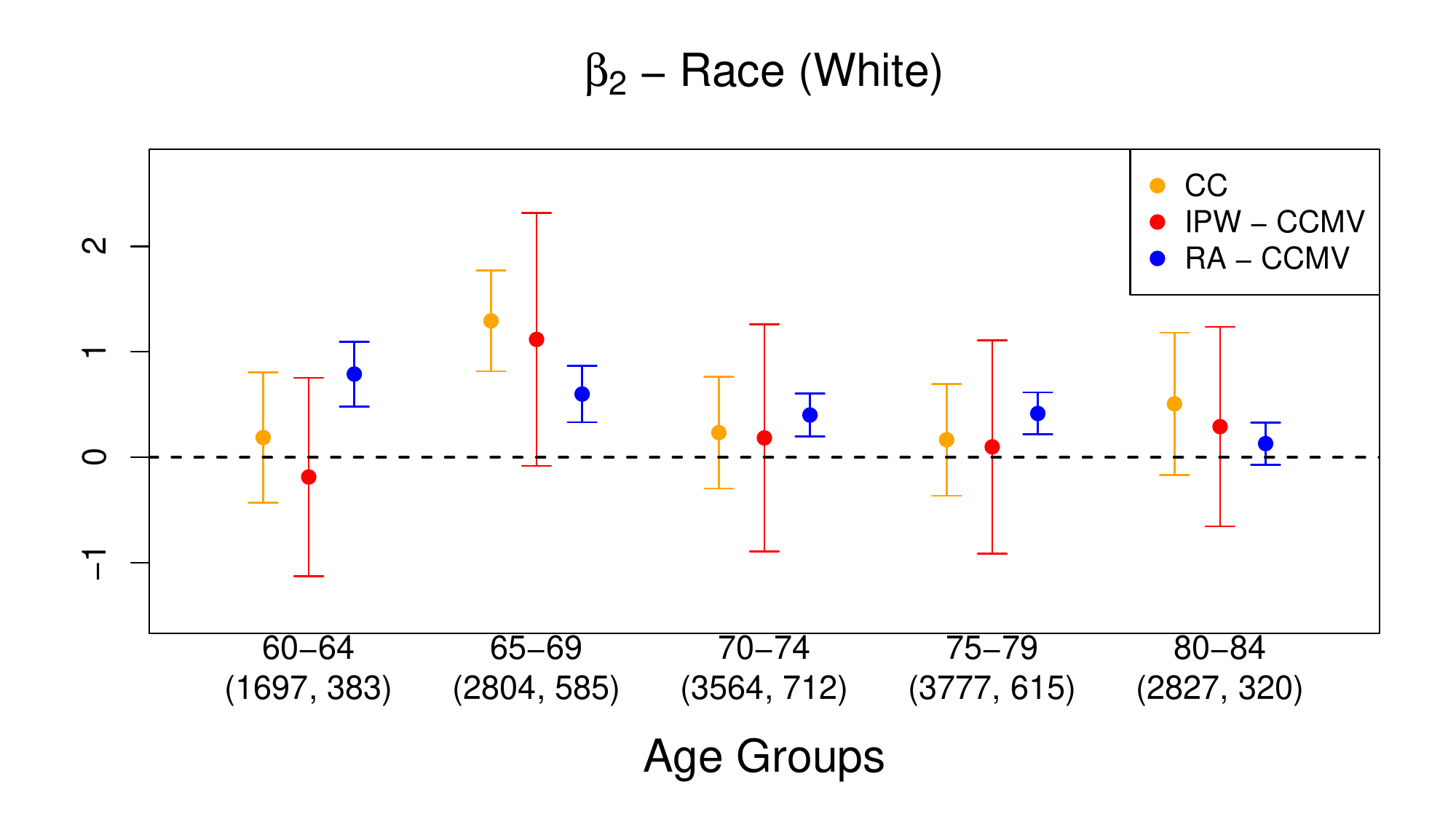}
	\includegraphics[scale=0.35]{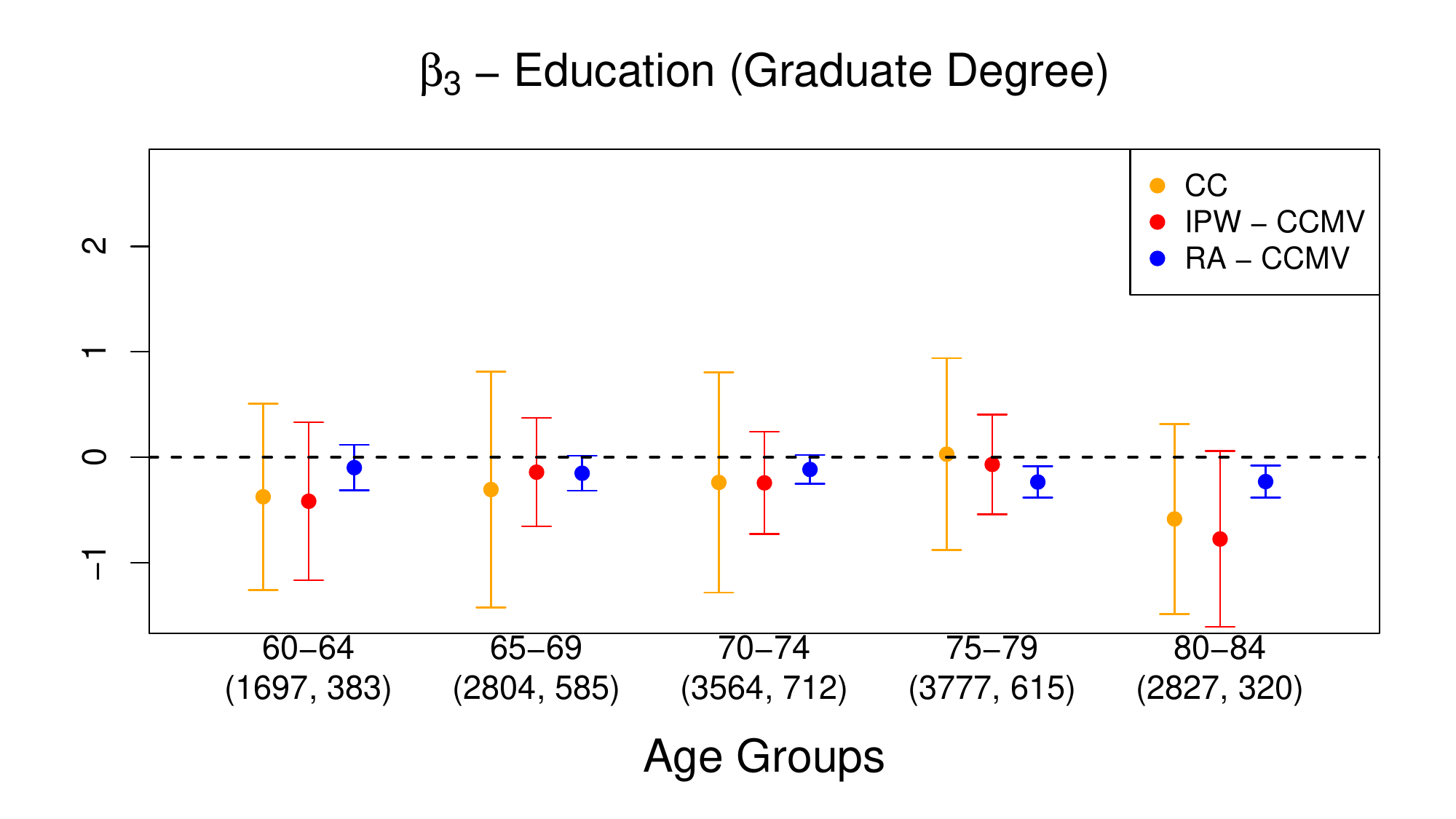}
	\includegraphics[scale=0.35]{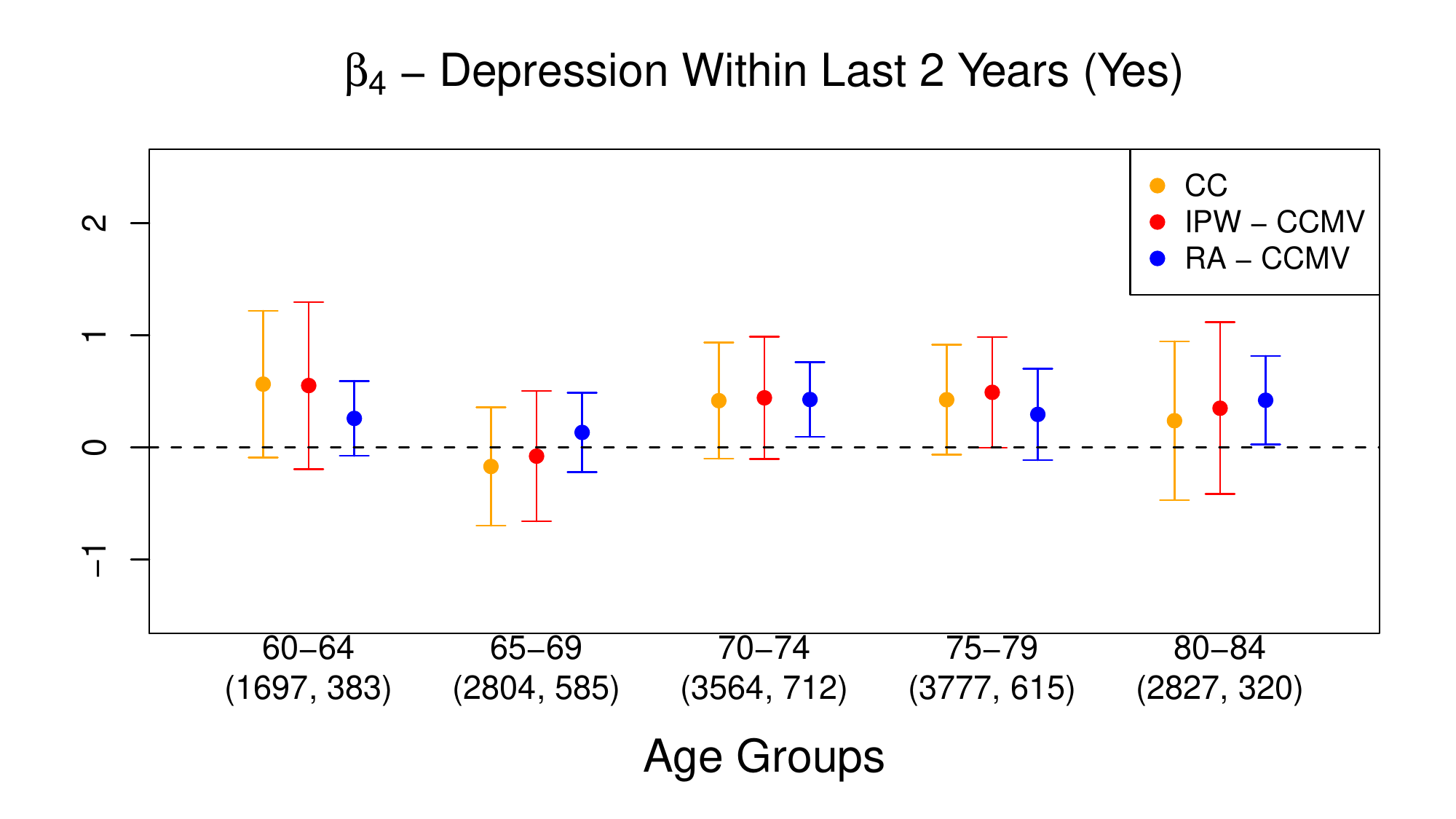}
	\includegraphics[scale=0.35]{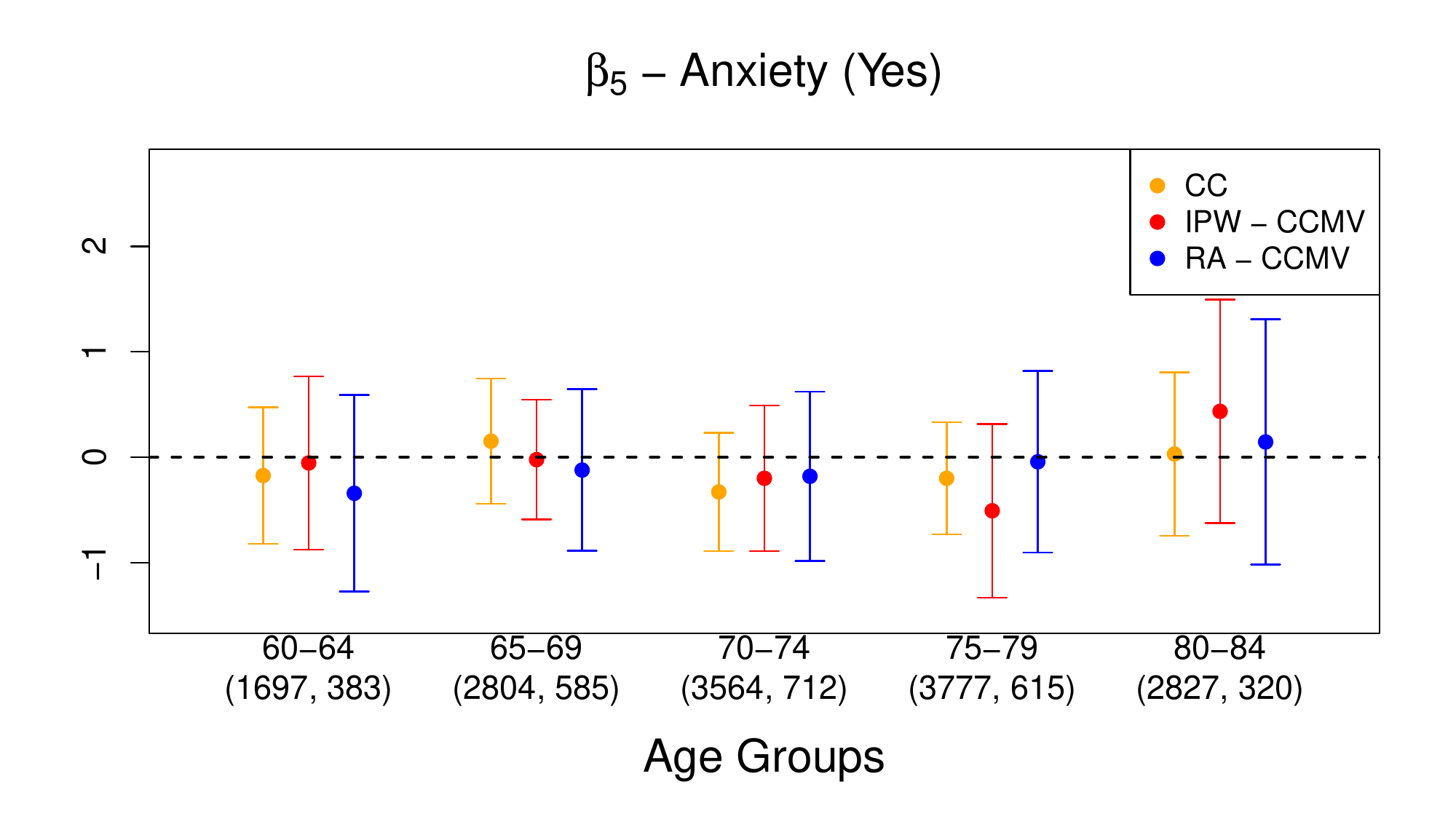}
	\caption{Point estimates and the $95\%$ confidence intervals for each $\beta$ parameter.  The total sample size and the number of complete cases is included on the $x$-axis for each age group.}
	\label{fig:realdata}
\end{figure}

We consider three different estimators: complete case (CC) analysis, IPW and RA estimators. 
	The CC analysis is a common approach used by practitioners. In the CC analysis, we remove any individuals with any missing  covariates and fit a standard Cox model. 
	For the IPW estimator, we use the CCMV assumption using logistic models for the complete odds. Since the variables are binary, the imputation distribution in the RA estimator is the usual nonparametric model. Point estimates and the 95\% confidence intervals (constructed using the nonparametric bootstrap) are provided in Figure \ref{fig:realdata}.

First, we observe that the results from a CC analysis are generally not significant.  For example, in the plot for $\beta_3$ (education), the CC confidence intervals all strongly cover $0$.  On the other hand, across the plots, the RA approach generally yields narrower confidence intervals while the IPW approach sometimes yields narrower confidence intervals; this is due to the instability introduced by the inverse probability weights.  Specifically, for $\beta_3$ (education), the RA provides clear significant results for several of the age groups.  These plots suggest that often, when applying a CC analysis, we suffer from such a loss in sample size that it becomes difficult to draw a significant conclusion.

Moreover, it is known that higher education provides a protection against cognitive impairment (known as the cognitive reserve;
\citealt{Meng2012EducationAD, ThowMeganE2018Feic}).
The point estimates for $\beta_3$ from all three estimation procedures are negative, which is what we would expect. As the RA approach allows one to attain significant negative point estimates, it highlights the potential benefits of accounting for the missing data.
We discuss further details for the analysis in Appendix \ref{appendix:data}.

\subsection{Sensitivity analysis}

\begin{figure}
	\centering
	\includegraphics[scale=0.35]{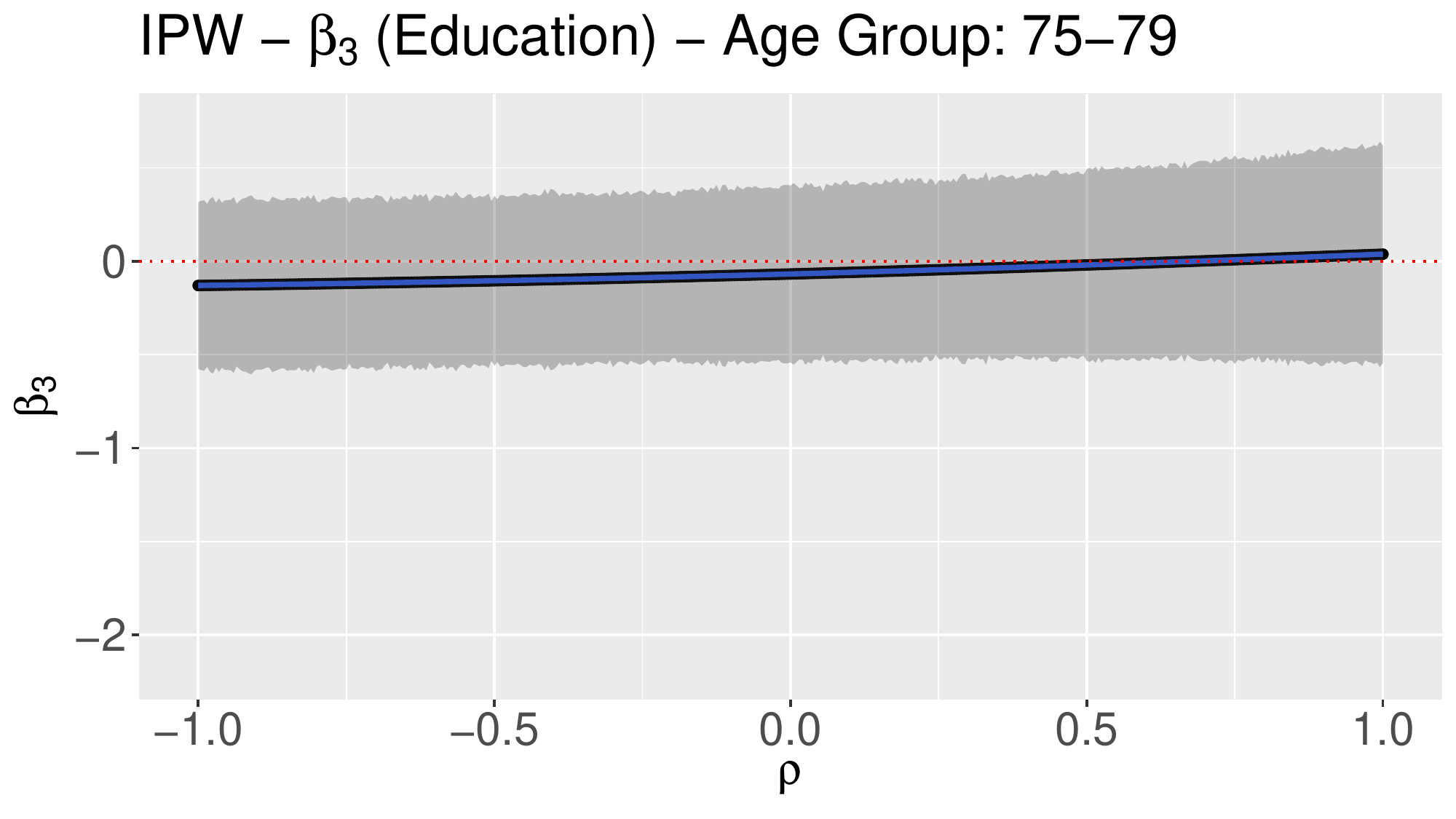}
	\includegraphics[scale=0.35]{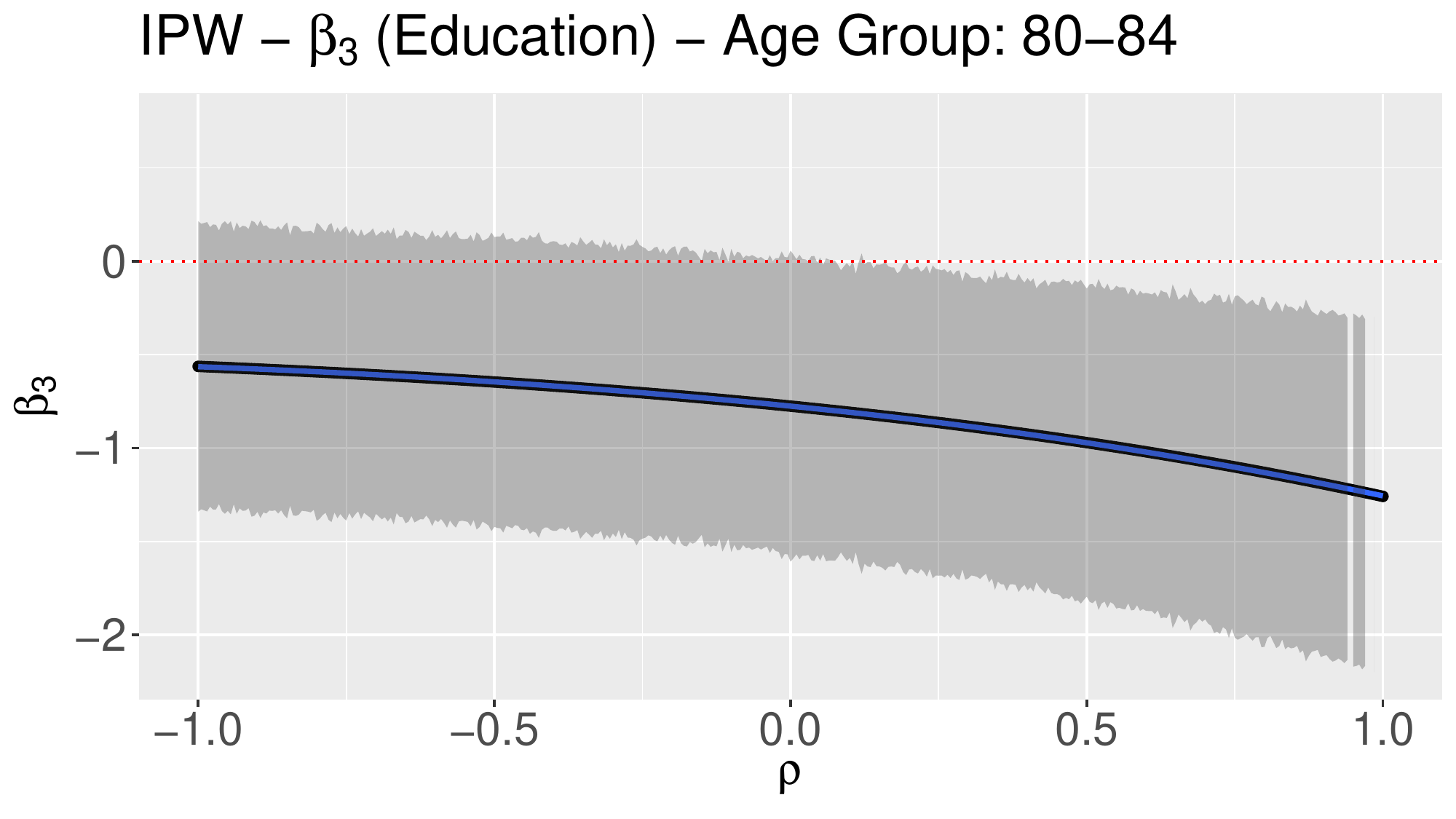}
	\includegraphics[scale=0.35]{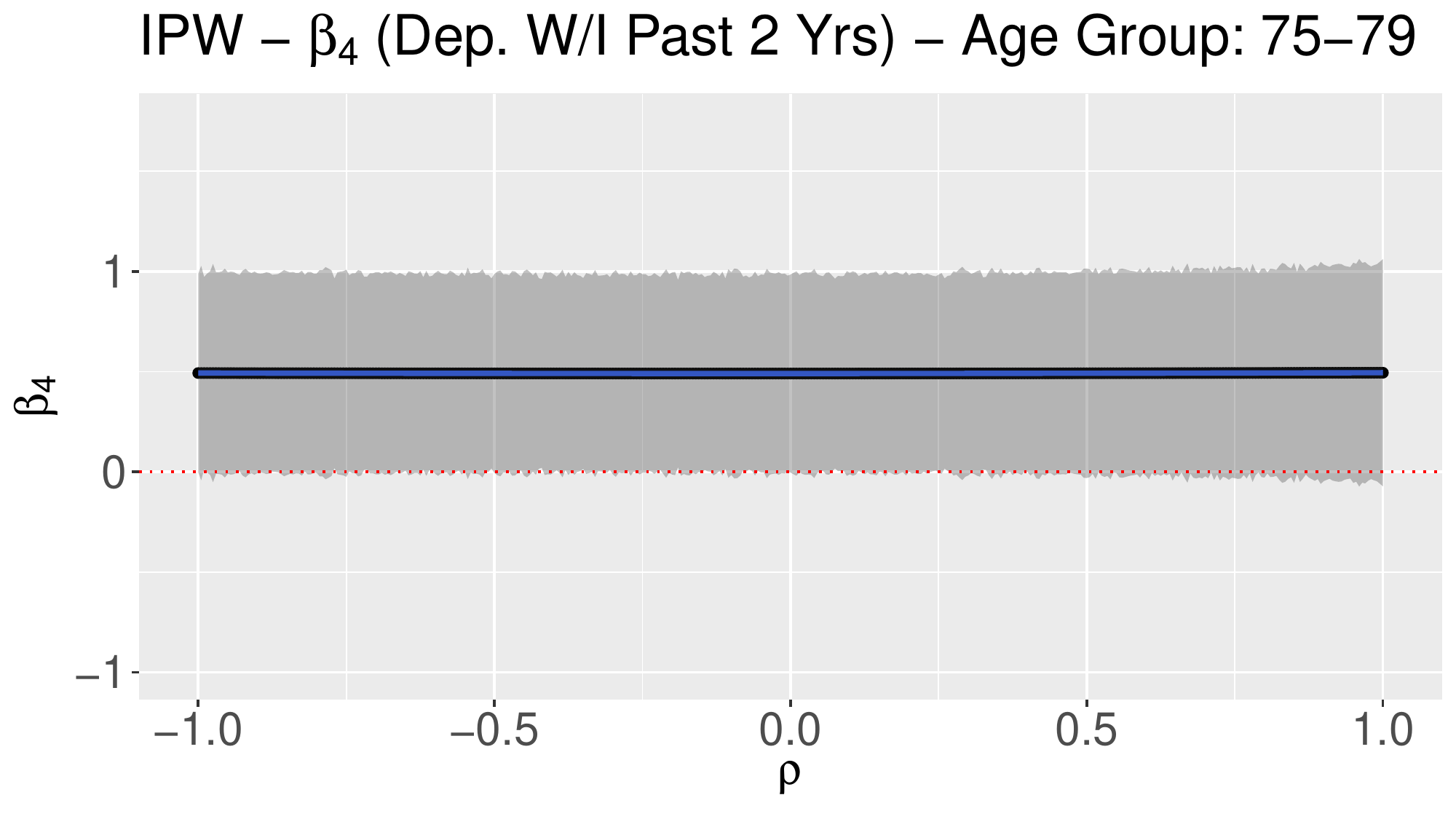}
	\includegraphics[scale=0.35]{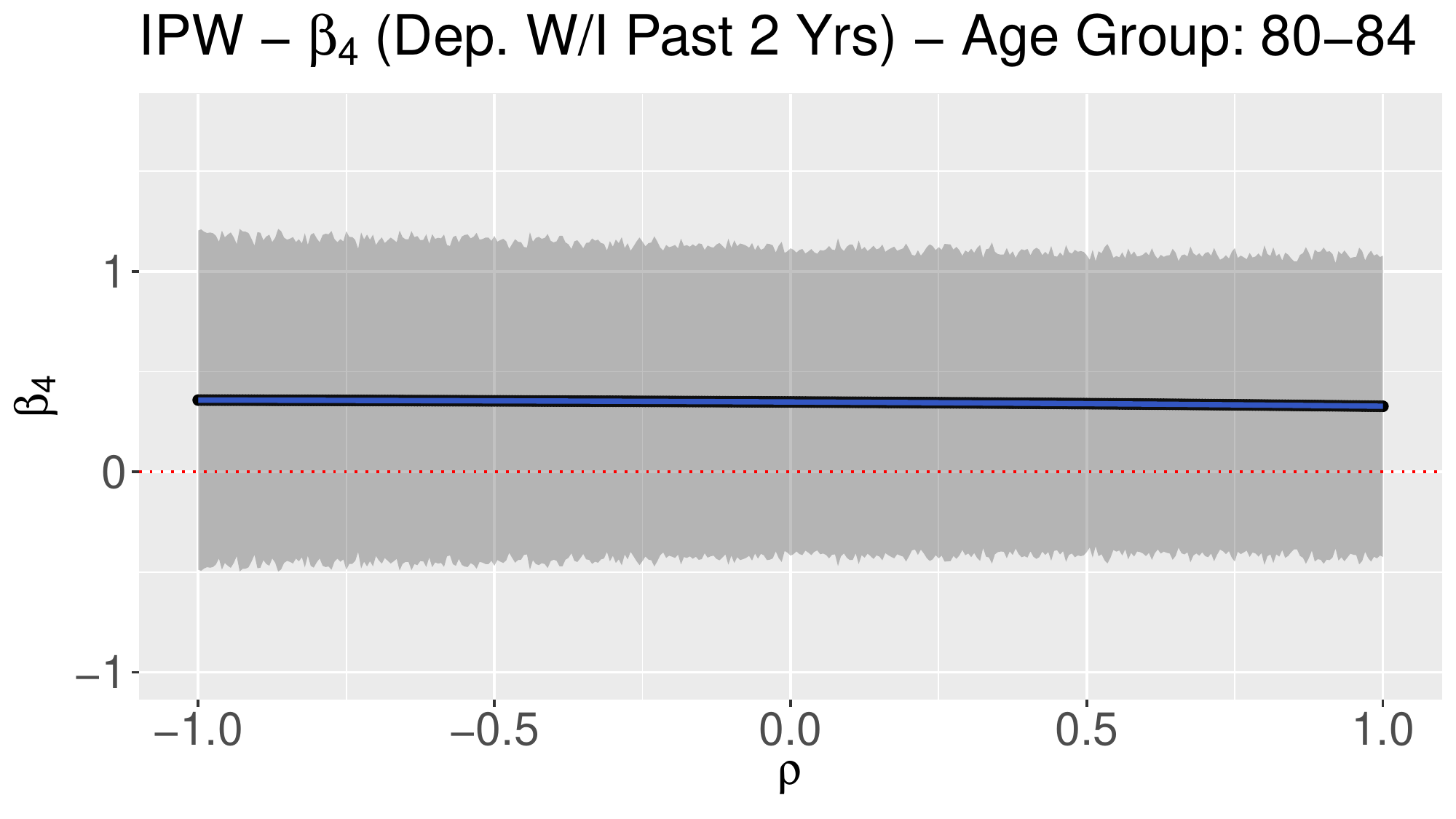}
	\caption{Point estimates and the $95\%$ confidence intervals for the IPW estimator ($\beta_3$ and $\beta_4$) as we vary the sensitivity parameter $\rho$ for the age groups $75-79$ and $80-84$.}
	\label{fig:realdata-ipw-sens}
\end{figure}

\begin{figure}
	\centering
	\includegraphics[scale=0.35]{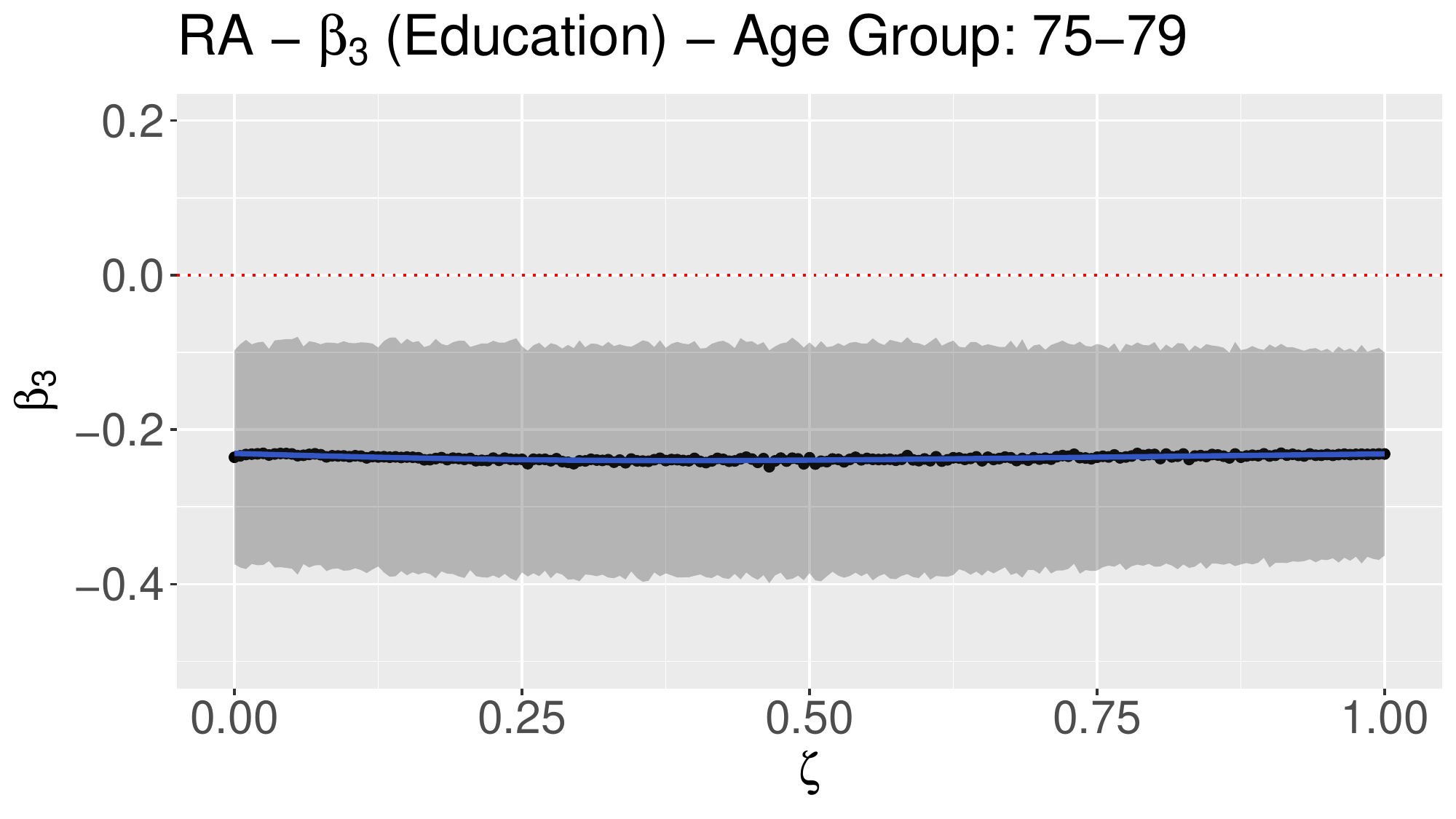}
	\includegraphics[scale=0.35]{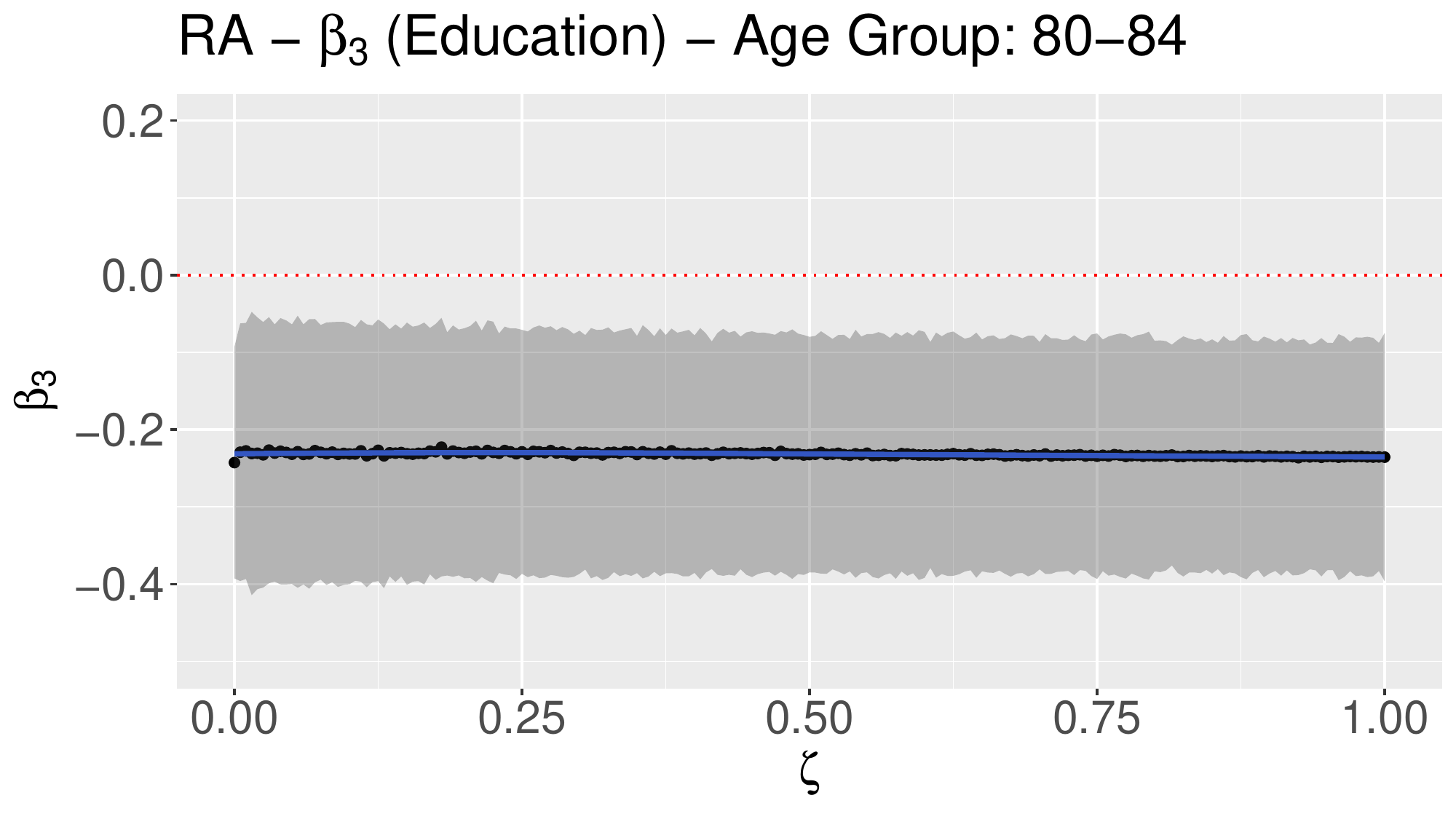}
	\includegraphics[scale=0.35]{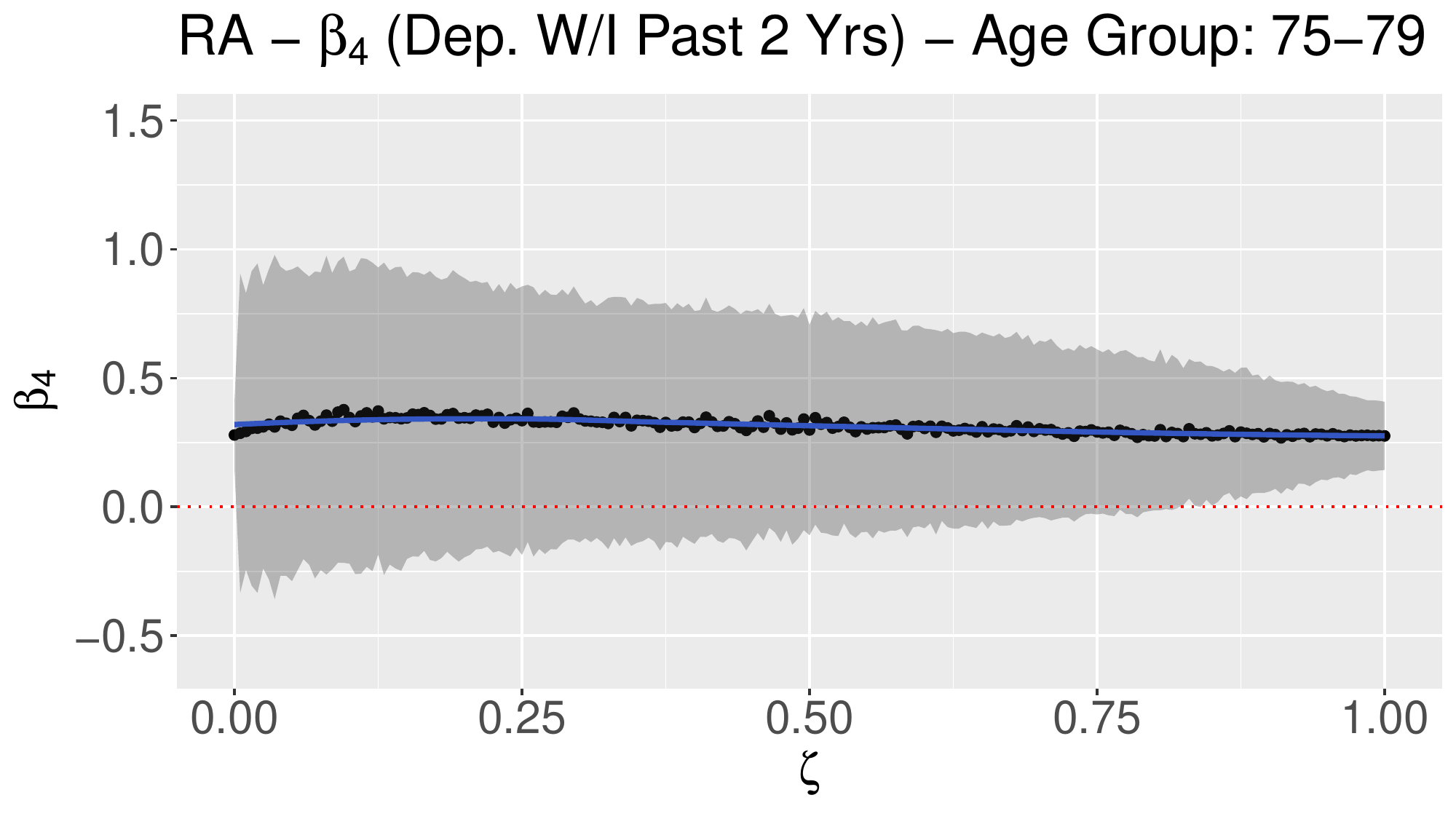}
	\includegraphics[scale=0.35]{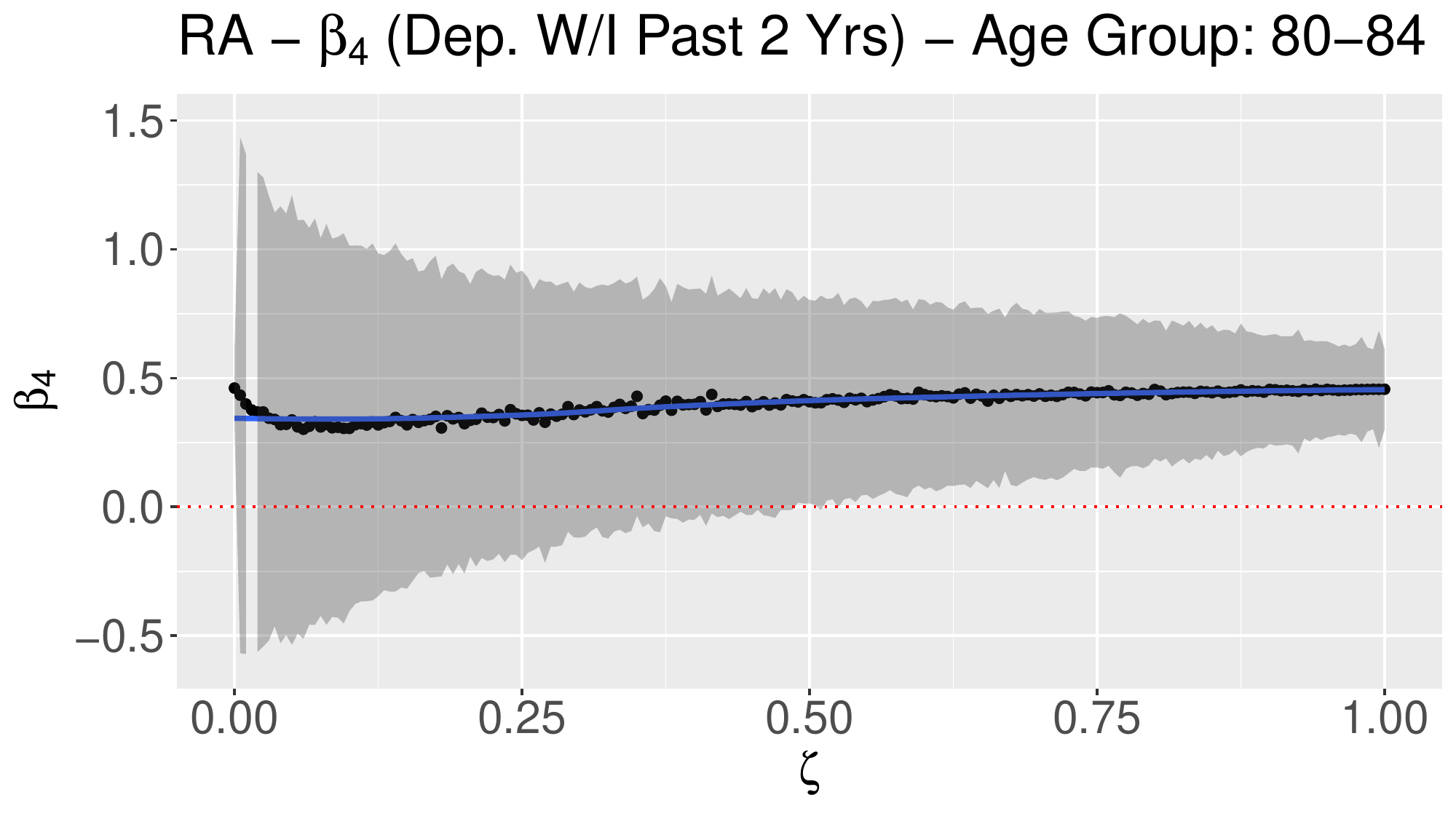}
	\caption{Point estimates and the $95\%$ confidence intervals for the RA estimator ($\beta_3$ and $\beta_4$) as we vary the sensitivity parameter $\zeta$ for the age groups $75-79$ and $80-84$.}
	\label{fig:realdata-ra-sens}
\end{figure}

We conduct a sensitivity analysis to investigate the robustness of our conclusion under CCMV.
We perturb the complete odds using an exponential tilting parameter $\rho$ such that
$$
	O_{r,a}(X,W_a) = Q_{r,a}(X_r,W_a) e^{\rho_{\bar{r}} X_{\bar{r}}}
$$
and every element in the vector $\rho_{\bar{r}}$ takes the same value $\rho$.
When $\rho=0$, there is no perturbation, but as $|\rho|\to\infty$, the scale of the perturbation increases.  For the imputation model, we use a sensitivity analysis parameter $\zeta \in [0,1]$.  When $\zeta \to 1$, the imputation probability moves towards $0$, and when $\zeta\to0$, the imputation probability moves towards $1$.  
If $\zeta=1/2$, then the imputation probability remains unchanged (i.e., the case of CCMV). 
We provide additional discussion of these sensitivity parameters in Appendix \ref{appendix:sens}.
The results are displayed in Figures \ref{fig:realdata-ipw-sens} and \ref{fig:realdata-ra-sens}.
The confidence intervals are shaded in gray. There is some slight wiggles in the upper and lower bands because 
of the Monte Carlo errors from the bootstrap (we construct each confidence interval using the nonparametric bootstrap).


In Figure~\ref{fig:realdata-ipw-sens}, we display the result of the IPW estimator.
The IPW estimator is robust when we perturb the missing data assumption in the sense that only the education effect
has 
a stronger dependency with the sensitivity parameter. 
Across all $\rho \in [-1,1]$, the education effect is still negative, which suggests that the cognitive reserve we 
obtain from the IPW estimator is resilient. 

In Figure~\ref{fig:realdata-ra-sens}, we observe that the RA estimator for $\beta_3$ (education effect) is robust to changes in the sensitivity parameter, which again reinforces our findings on the cognitive reserve.

On the other hand, we see that the confidence interval width of $\beta_4$ (depression) depends on the sensitivity parameter $\zeta$. While the sign of the effect remains the same and the effect (value of $\beta_4$) is also stable, the width of confidence intervals do change drastically. So, we have to be careful when making claims regarding $\beta_4$.


\section{Discussion}

In this paper, we introduced a process for constructing a multistage estimator for problems with missing not at random covariates and incomplete outcomes.  We discussed our method in the context of Cox regression from survival analysis and a missing response problem (in Appendix \ref{appendix:missingresp}).  Although we did not show the specific details, the missing response problem provides a bridge to the binary treatment problem from causal inference and a more general longitudinal study problem with dropout.  So, the arguments we provide for the missing response problem naturally extend to these two problems.  

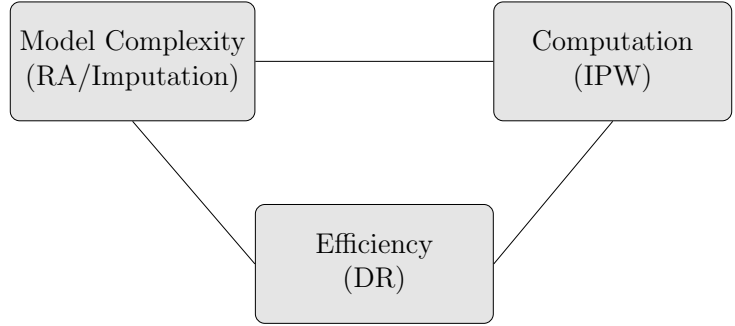
\begin{figure}
	\centering
	\scalebox{0.9}{
	\begin{tikzpicture}[>=latex,rect/.style={draw=black, 
			rectangle, 
			fill=gray,
			fill opacity = 0.2,
			text opacity=1,
			minimum width=100pt, 
			minimum height = 50pt, 
			align=center}]
		\node[rect,rounded corners] (a1) {Model Complexity \\ (RA/Imputation)};
		\node[rect,rounded corners,right=100pt of a1] (a2) {Computation \\ (IPW)};
		\path (a1) -- (a2) node[midway,below=60pt,rect,rounded corners] (a3) {Efficiency \\ (DR)};
		\draw[-] (a1)--(a2)node[midway,sloped, above]{};
		\draw[-] (a1.south)--(a3.west)node[midway,sloped,below]{};
		\draw[-] (a2.south)--(a3.east)node[midway,sloped,below]{};
	\end{tikzpicture}
	}
	\caption{The tradeoff between model complexity, computation, and efficiency in the CCMV setting.}
	\label{fig::tradeoff}
\end{figure}

Figure~\ref{fig::tradeoff} shows an interesting tradeoff between model complexity, computation, and efficiency in the special case of the CCMV assumption for the covariates. 
	The RA estimator is the easiest to model since we only need to place a model on $p(x|1_d,w_a,a)$ for all $a$.  But this comes at a cost  of computation since we generally need to perform imputation to fill in the missing data. Also, RA does not achieve 
	the (semiparametric) efficiency bound.
	The IPW estimator is easy to compute because it is a simple reweighting of the cases with completely observed covariates.  But we need to model every $Q_{r,a}(X_r,W_a)$, which would require many models.
	Also, IPW does not achieve the (semiparametric) efficiency bound and may be unstable if the weights are too large.
	The DR estimator achieves the semiparametric efficiency bound by utilizing components from both the regression adjustment and inverse probability weighted estimator.  
	Due to the augmenting term, imputation must be performed even for the complete cases,
	so the computational cost is high and we have to place models for every $Q_{r,a}(X_r,W_a)$.  
	 Thus, efficiency comes at the cost of complexity in modeling and high computation.  

There are several possible areas for future work.  
One direction is to study the problem of model compatibility issues between complete odds and regression functions.
When the assumption on missing covariates is not CCMV, the complete odds and regression function may again not be variationally
independent. 
Thus, extra caution has to be taken 
when placing models on them. 
Another direction is to study the use of nonparametric models in complete odds and regression functions.
This would lead to a more flexible modeling strategy and avoid the problem of model compatibility, but
overcoming the slow convergence rate will be a challenge.


\clearpage

\bibliographystyle{abbrvnat}
\bibliography{references}

\bigskip

\section*{Acknowledgements.} DS is supported by NSF DGE-2140004.  YC is supported by NSF DMS-1952781, NSF DMS-2112907, and NIH U24-AG072122.

\clearpage

	\begin{center}
		{\LARGE\bf 
			\vspace{1 in}
			Supplementary Materials: Multistage Estimators for Missing Covariates and Incomplete Outcomes}
	\end{center}
	\medskip

\bigskip

\appendix

{\Large \textbf{Appendices}}

\vskip 10pt

The appendix has several sections to provide additional theoretical results, proofs, details on simulations and data analyses, and further generalizations of work in the main body.  In Appendix \ref{appendix:missingresp}, we construct a multistage estimator for the missing response problem.  
The assumptions for the asymptotic theory are provided in Appendix \ref{appendix:asymptotic}.  Since we utilize CCMV as the main missing data assumption in the main body, we generalize this assumption in Appendix \ref{appendix:beyondCCMV} and show how to attain identifiable complete odds and imputation distributions.  Further details on the sensitivity analyses and simulations are included in Appendices \ref{appendix:sens} and \ref{appendix:sim}, respectively.  We include more data analyses in Appendix \ref{appendix:data}.
The proofs for any theoretical results are included in Appendix \ref{appendix:proofs}.

The table of contents is provided below.

\begin{enumerate}
	\item Appendix \ref{appendix:missingresp}: Missing response problem.
	\item Appendix \ref{appendix:asymptotic}: Technical assumptions for the asymptotic theory.
	\item Appendix \ref{appendix:beyondCCMV}: Beyond CCMV.
	\item Appendix \ref{appendix:sens}: Procedure of sensitivity analysis.
	\item Appendix \ref{appendix:sim}: Details of simulations.
	\item Appendix \ref{appendix:data}: Additional analysis of the NACC data.
	\item Appendix \ref{appendix:proofs}: Proofs of 1. identification theories (Appendix \ref{app:ID}), 2. efficiency theories (Appendix \ref{app:eff}), 3. asymptotic theories (Appendix \ref{app:asymp})
	
\end{enumerate}

\section{A multistage estimator for the missing response problem} \label{appendix:missingresp}

We now turn our attention to the missing response problem and show how to construct a multistage estimator in this setting. 
Note that this procedure can be easily generalized to the binary treatment effect in causal inference.
The parameter of interest is the marginal mean of the outcome variable $\E[Y]$.  Let $\pi_A(X) = P(A=1|X)$ and $m_1(X)$ denote the propensity score and regression function for $Y$, respectively.  We start with the classic DR estimator assuming fully-observed covariates for the missing response problem.  The efficient influence function takes the form
\begin{equation*}
	\frac{1(A=1)Y}{\pi_A(X)} + m_1(X) - \frac{1(A=1)m_1(X)}{\pi_A(X)}.
\end{equation*}
The expectation of the above quantity equals $\E[Y]$ as long as at least one of $\pi_A(X)$ and $m_1(X)$ is correctly specified.  
Using empirical average of the above quantity, we obtain the DR estimator under the missing response problem.

\subsection{Multiply-robust property}

In this section, we highlight the multiply-robust property of the multistage estimator by demonstrating the construction of one such multiply-robust estimator in the missing response problem.

For this problem, we also define the following regression functions
\begin{align}
	& M_{m,r,0}(X_r) = \E[m_1(X)|X_r,R=r,A=0], \label{regf1}\\
	& M_{m,r,1}(X_r,Y) = \E[m_1(X)|X_r,R=r,Y,A=1], \label{regf2} \\
	& M_{1/\pi_A,r,1}(X_r,Y) = \E\left[\frac{1}{\pi_A(X)} \biggr| X_r,R=r,Y,A=1\right], \label{regf3} \\
	& M_{m/\pi_A,r,1}(X_r,Y) = \E\left[\frac{m_1(X)}{\pi_A(X)} \biggr| X_r, R=r, Y, A=1\right]. \label{regf4} 
\end{align}

\begin{definition}[Semiparametric submodels]
	Define the following semiparametric submodels.
	\begin{itemize}
		\item $\M_R(r)$: The semiparametric model such that the parametric model $p(x_{\bar{r}}|x_r,r,w_a,a;\eta_{r,a})$ is correctly specified.
		\item $\M_O(r)$: The semiparametric model such that the parametric model $O_{r,a}(x,w_a;\alpha_{r,a})$ is correctly specified.
	\end{itemize}
\end{definition}

It follows from the previous definition that under model $\M_R(r)$, the regression functions \eqref{regf1}-\eqref{regf4} are correctly specified.  
We specify the semiparametric model in this manner to avoid the model congeniality issue described in Section \ref{sect:raCox}.  Note that one could instead define $\M_R(r)$ to be the semiparametric model under which these regression functions are correctly specified.

\begin{thm}[Double robustness for missing response under CCMV for given $r$] \label{theorem:multiplyrobust-missingresp}
	Let $\theta_r = \int y \ P(dx, r, dy, da)$ be the parameter of interest.
	Consider the uncentered influence function
	\begin{align*}
		&q_r(X,R,Y,A) = m_1(X)1(A=0)1(R=1_d) Q_{r,0}(X_r) \\
		&\quad + m_1(X)1(A=1)1(R=1_d) Q_{r,1}(X_r,Y) \\
		&\quad + M_{m,r,0}(X_r) 1(A=0) [1(R=r) - 1(R=1_d) Q_{r,0}(X_r)] \\
		&\quad+ M_{m,r,1}(X_r,Y) 1(A=1) [1(R=r) - 1(R=1_d) Q_{r,1}(X_r,Y)] \\
		&\quad+ \frac{1(R=1_d)1(A=1)Y}{\pi_A(X)}Q_{r,1}(X_r,Y) \\
		&\quad + M_{1/\pi_A,r,1}(X_r,Y) Y 1(A=1) [1(R=r) - 1(R=1_d)Q_{r,1}(X_r,Y)] \\
		&\quad -\frac{1(R=1_d)1(A=1)m_1(X)}{\pi_A(X)} Q_{r,1}(X_r,Y) \\
		&\quad - M_{m/\pi_A,r,1}(X_r,Y) 1(A=1) [1(R=r)-1(R=1_d)Q_{r,1}(X_r,Y)].
	\end{align*}
	Under the model $\M_R(r) \cup \M_O(r)$ (i.e., either the complete odds or the regression functions are correct), $\E[q_r(X,R,Y,A)] = \theta_r$.
\end{thm}

Theorem \ref{theorem:multiplyrobust-missingresp} shows that the estimator for $\theta_r$ based on the provided influence function is doubly robust.  
One key takeaway is that for a fairly simple problem, the estimation procedure for the doubly robust estimator is rather complex.  In particular, there are 12 terms in total.  With this in mind, we expect that this doubly robust estimator to be more of a theoretical construct than one that will generally be implemented in practice.

We further develop the theory by considering a doubly robust estimator for $\theta_r$ for each $r$.  Merging this together, we are able to estimate
\begin{equation*}
	\theta = \int y \ P(dx, dr, dy, da) = \sum_r \theta_r.
\end{equation*}
We see that in the following corollary, this yields a multiply-robust estimator for $\theta = \E[Y]$.

\begin{cor}[Multiply-robustness for missing response under CCMV] \label{cor:mr-missresp}
	Define the parameter of interest
	\begin{equation*}
		\theta = \sum_r \theta_r = \sum_r \int y \ P(dx,r,dy,da) = \int y \ P(dx,dr,dy,da). \label{eq:marginalmean}
	\end{equation*}
	The uncentered influence function $\sum_r q_r(X,R,Y,A) $ yields a consistent estimator of $\theta = \sum_r \theta_r$ under the model $\bigcap_{r} (\M_R(r) \cup \M_O(r))$.
\end{cor}

Corollary \ref{cor:mr-missresp} provides a strong result in that the estimator is $2^{|\mathcal{R}|-1}$-robust, where $|\mathcal{R}|$ is the number of missing covariate patterns.  As shown in Theorem \ref{theorem:multiplyrobust-missingresp}, for an estimator of $\theta_r$ to be consistent, it is sufficient for at least one of $O_{r,a}(X,W_a)$ or the distribution $p(x_{\bar{r}}|x_r,w_a,a)$ to be correctly specified.  In other words, we have two chances to obtain a correct estimator for each missing covariate pattern $r\neq 1_d$.

Since the missing response and binary treatment problems are related and well-studied problems, we omit the construction of our multistage estimator for the binary treatment problem.  The included arguments can easily be extended to the binary treatment problem and the problem of longitudinal data with monotone missingness in the response.

\subsection{Semiparametric efficiency}

We now show that the multiply-robust estimator developed in Theorem \ref{theorem:multiplyrobust-missingresp} and Corollary \ref{cor:mr-missresp} is the semiparametric efficient estimator that attains the minimum possible variance in the model $\bigcap_{r} (\M_R(r) \cap \M_O(r))$, where all the regression functions and complete odds are correctly specified.

\begin{thm}[EIF for missing response under CCMV for $X$] \label{theorem:eif}
	For the mean statistical functional
	\begin{equation*}
		\psi(P) = \int y \ P(dx,dr,dy,da),
	\end{equation*}
	the multiply-robust influence function in Corollary \ref{cor:mr-missresp} is the efficient influence function (EIF),
	and it achieves the semiparametric efficiency bound in the model $\bigcap_{r} (\M_R(r) \cap \M_O(r))$.
\end{thm}

This theorem further shows that the multiply-robust estimator is also full semiparametric efficient.

\section{Asymptotic theory} \label{appendix:asymptotic}


In this section, we discuss the asymptotic theory for our multistage estimators.  We direct our attention to estimating the population quantity
\begin{equation*}
	\theta = \E[h(X,W,A)] = \sum_{a\in \mathcal{A}}\E[f_a(X,W_a)1(A=a)], \label{eq:theta}
\end{equation*}

where is an ODF as described in Definition \ref{def:ODF}.  We will show that the estimation methods that we have previously outlined lead to consistent, asymptotically normal estimators under appropriate conditions.  This theory will be useful for constructing confidence intervals for data analysis and demonstrating consistency.  If $h$ is an estimating equation, then the asymptotic normality of the solution to the finite-sample version of the estimating equation follows under standard regularity conditions for $Z$-estimators.

\subsection{Assumptions for estimating $f_a$}

Fix an observed outcome pattern $a$.  These assumptions hold for estimating the function $f_a$.

\begin{enumerate}[label=(A\arabic*)]
	\item \label{A1} There exists $\tau_{a}^*$ in the interior of the parameter space $T_a$ such that
	\begin{equation*}
		f_a(x,w_a; \tau_a^*) = f_a(x,w_a).
	\end{equation*}
	We assume that there exists $\hat{\tau}_{a}$, which is a regular asymptotically linear estimator such that
	\begin{equation*}
		\hat{\tau}_{a} - \tau_{a}^* = \frac{1}{n}\sum_{i=1}^n \phi_{r,a}(X_i, R_i, W_{i,A_i}, A_i) + o_p(1/\sqrt{n}),
	\end{equation*}
	and the following convergence in distribution holds
	\begin{equation*}
		\hat{\tau}_{a} - \tau_{a}^* \stackrel{d}{\to} N(0, \sigma_{f_a}^2)
	\end{equation*}
	for some $\sigma_{f_a}^2 > 0$.
	
	
	\item \label{A2} Assume that $f_a$ is uniformly bounded for all values of its parameter such that
	\begin{equation*}
		\sup_{x,w_a,\tau_a} |f_a(x,w_a; \tau_a)| < M
	\end{equation*}
	for some constant $M>0$.
	
	\item \label{A3} Let $\int (f_{a}(x, w_a; \hat{\tau}_{a}) - f_{a}(x, w_a; \tau_{a}^*))^2 \ P(dx,dr,dw_a,da) = o_p(1)$.
	
	\item \label{A4} Let $f_a(x,r,w_a;\tau_a)$ have a nonzero first derivative with respect to $\tau_a$, and assume that
	\begin{equation*}
		\int |\nabla_{\tau_{a}} f_{a}(x,w_a;\tau_a)| \ P(dx,dr,dw_a,da)
	\end{equation*}
	is finite in all of its entries.
\end{enumerate}

The assumption \ref{A1} says that the $f_a$ functions belong to a parametric class and can be estimated at $\sqrt{n}$-consistency.  We elect to use a parametric assumption to make the analysis easier.  Assumption \ref{A2} is a bounded assumption, which is reasonable if the input variables have a compact support.  Assumption \ref{A3} describes the $L_2(P)$ convergence of the estimated $f_a$ functions.  Finally, Assumption \ref{A4} is a technical assumption to ensure that asymptotic normality can be achieved and allows for application of the Dominated Convergence Theorem.

\subsection{The IPW estimator}

\subsubsection{Assumptions}  We now introduce the following 4 assumptions for the IPW theorem below.  Each assumption holds for each pair of fixed covariate pattern $r$ and observed-outcome pattern $a$.

\begin{enumerate}[label=(B\arabic*)]
	\item \label{B1} For all missing covariate patterns $r$ and observed outcome patterns $a$, there exist constants $C_1$ and $C_2$ such that
	\begin{equation*}
		0 < C_1 \leq O_{r,a}(x,w_a; \alpha_{r,a}) \leq C_2 < 1.
	\end{equation*}
	
	\item \label{B2} There exists $\alpha_{r,a}^*$ in the interior of the parameter space $A_{r,a}$ such that
	\begin{equation*}
		O_{r,a}(x,w_a; \alpha_{r,a}^*) = \frac{P(R=r|x,a,w_a)}{P(R=1_d|x,a,w_a)}.
	\end{equation*}
	
	We assume that there exists $\hat{\alpha}_{r,a}$, which is a regular asymptotically linear estimator
	$$\hat{\alpha}_{r,a} - \alpha_{r,a}^* \stackrel{d}{\to} N(0, \sigma_{\text{odds},r,a}^2)$$
	for some $\sigma_{\text{odds},r,a}^2 > 0$.
	
	\item \label{B3} The following class of functions is Donsker:
	\begin{equation*}
		\{O_{r,a}(x,w_a; \alpha_{r,a} : \alpha_{r,a} \in A_{r,a}\}.
	\end{equation*}
	
	\item \label{B4} Assume that $O_{r,a}(x, w_a; \alpha_{r,a})$ have a nonzero first derivative with respect to $\alpha_{r,a}$.  Also, let
	\begin{equation*}
		\int |\nabla_{\alpha_{r,a}} O_{r,a}(x, w_a; \alpha_{r,a})| \ P(dx,dr,dw_a,da)
	\end{equation*}
	be finite in all of its entries, and assume that
	\begin{equation*}
		\int (O_{r,a}(x, w_a; \hat{\alpha}_{r,a}) - O_{r,a}(x, w_a; \alpha_{r,a}^*))^2 \ P(dx,dr,dw_a,da) = o_p(1).
	\end{equation*}
\end{enumerate}

Assumption \ref{B1} is to ensure that the complete odds are sufficiently bounded away from 0, which is true if the variables belong to a compact support.  Assumption \ref{B2} says that the complete odds belong to a parametric model, and there exists an estimator that is $\sqrt{n}$-consistent.  This is reasonable for most common parametric models under consideration.  Assumption \ref{B3} is a Donsker condition, which is also reasonable if the parametric model is sufficiently smooth.  Assumption \ref{B4} ensures that the estimated regression function $O_{r,a}(x,w_a;\hat{\alpha}_{r,a})$ converges in the $L_2(P)$ norm to the true regression function $O_{r,a}(x,w_a;\alpha_{r,a}^*)$.  The finite first derivative ensures that we can apply the Dominated Convergence Theorem to simplify the expression in the proof.	

\subsubsection{Asymptotic theory}

We continue our goal of estimating $\theta$ by using an inverse probability weighting approach.  
First, using Lemma \ref{lemma:completeodds}, the parameter of interest writes as
\begin{align*}
	\theta &= \sum_{a}\sum_r \E[f_a(X,W_a) O_{r,a}(X,W_a) 1(R=1_d) 1(A=a)], \label{eq:theta-ipw}
\end{align*}
where $O_{r,a}(X,W_a) = P(R=r|X,W_a,A=a)/P(R=1_d|X,W_a,A=a)$ is a complete odds.  

\begin{thm}[Inverse probability weighting]\label{theorem:ipw}
	Suppose the assumptions \ref{A1}-\ref{A4} and \ref{B1}-\ref{B4} hold.  Then, the estimator
	\begin{align*}
		\hat{\theta}_{\text{IPW}} &= \sum_r \sum_a \hat{\theta}_{IPW,r,a} \\
		&= \sum_r \sum_a \left( \frac{1}{n}\sum_{i=1}^n f_a(X_i,W_{i,a}) O_{r,a}(X_i, W_{i,a}) 1(R_i=1_d) 1(A_i=a) \right)
	\end{align*}
	is consistent, and for some $\sigma^2_{\text{IPW}}>0$, we have the following convergence in distribution
	$$\sqrt{n}(\hat{\theta}_{\text{IPW}}-\theta) \stackrel{d}{\to} N(0, \sigma^2_{\text{IPW}}).$$
\end{thm}

\subsection{The RA estimator}

\subsubsection{Assumptions}  We introduce the following 3 assumptions for the RA estimator.

\begin{enumerate}[label=(C\arabic*)]
	\item \label{C1} For all covariate patterns $r$ and observed-outcome patterns $a$, there exists $\eta_{r,a}^*$ in the interior of the parameter space $\mathcal{H}_{r,a}$ such that
	\begin{equation*}
		P_{ex}(x_{\bar{r}} | x_r, r, w_a, a) = P_{ex}(x_{\bar{r}} | x_r, r, w_a, a; \eta_{r,a}^*).
	\end{equation*}
	Let $\hat{P}_{ex}(x_{\bar{r}}|x_r,r,w_a,a) \equiv P_{ex}(x_{\bar{r}}|x_r,r,w_a,a;\hat{\eta}_{r,a})$.  We assume that $\hat{\eta}_{r,a}$ is a regular asymptotically linear (RAL) estimator satisfying
	\begin{equation*}
		\hat{\eta}_{r,a} - \eta_{r,a}^* = \frac{1}{n}\sum_{i=1}^n \psi_{r,a}(X_i, R_i, W_{i,A_i}, A_i) + o_p(1/\sqrt{n}).
	\end{equation*}
	Moreover, $\sqrt{n}(\hat{\eta}_{r,a}-\eta_{r,a}^*) \stackrel{d}{\to} N(0, \sigma_{\text{RA},r,a}^2)$ for some $\sigma_{\text{RA},r,a}^2>0$.\\
	
	\item \label{C2} For a fixed covariate pattern $r$ and observed-outcome pattern $a$, define
	\begin{align*}
		m_{r,a}(x_r,w_a;\eta_{r,a}) &= \E[f_a(X,W_a)|X_r=x_r,R=r,W_a=w_a,A=a] \\
		&= \int f_a(x,w_a) \ P_{ex}(dx_{\bar{r}}|x_r,r,w_a,a; \eta_{r,a}).
	\end{align*}
	
	The following class of regression functions is Donsker:
	\begin{equation*}
		\{m_{r,a}(x_r,w_a; \eta_{r,a}) : \eta_{r,a}\in \mathcal{H}_{r,a}\}.
	\end{equation*}
	
	\item \label{C3} For all covariate patterns $r$ and observed-outcome patterns $a$, assume that $m_{r,a}(x_r,w_a;\eta_{r,a})$ has a nonzero first derivative with respect to $\eta_{r,a}$.  Also, let
	\begin{equation*}
		\int |\nabla_{\eta_{r,a}} m_{r,a}(x_r,w_a;\eta_{r,a})| \ P_{obs}(dx_r,r,dw_a,a)
	\end{equation*}
	be finite in all of its entries, and assume that
	\begin{equation*}
		\int (m_{r,a}(x_r,w_a;\hat{\eta}_{r,a}) - m_{r,a}(x_r,w_a; \eta_{r,a}^*))^2 \ P_{obs}(dx_r,r,dw_a,a)= o_p(1).
	\end{equation*}
\end{enumerate}

These are fairly mild assumptions.  Assumption \ref{C1} says that the extrapolation distribution belongs to a parametric family and that the estimator $\hat{\eta}_{r,a}$ admits an asymptotically linear expansion.  Provided the parametric model holds, the $\sqrt{n}$-consistency of $\hat{\beta}_{r,a}$ is a reasonable assumption since $\eta^*_{r,a}$ is often estimated using $M$-estimators.  Assumptions \ref{C2} and \ref{C3} are similar to assumptions \ref{B3} and \ref{B4} made in the IPW estimation theorem.


%
%

\subsection{Asymptotic theory}

We can estimate $\theta$ using a regression adjustment approach.  Using Lemma \ref{lemma:ra}, the parameter writes as 
\begin{align*}
	\theta &= \E[h(X,W,A)] \\ 
	&= \sum_{a}\E[f_a(X,W_a)1(A=a)] \\
	&= \sum_a \sum_r \E[ m_{r,a}(X_r,W_a) 1(R=r) 1(A=a)] \\
	&= \sum_a \sum_r \theta_{\text{RA},r,a}.
\end{align*}
Thus, the main strategy will be to show asymptotic results for every fixed $r$ and $a$ pair.

\begin{thm}[Regression adjustment]\label{theorem:ra}
	Suppose the assumptions \ref{A1}-\ref{A4} and \ref{C1}-\ref{C3} hold.  Then, the estimator
	\begin{align*}
		\hat{\theta}_{\text{RA}} &= \sum_r \sum_a \hat{\theta}_{\text{RA},r,a} \\
		&= \sum_r \sum_a \left(\frac{1}{n}\sum_{i=1}^n m_{r,a}(X_r,W_a; \hat{\eta}_{r,a}) 1(R_i=r) 1(A_i=a)\right)
	\end{align*}
	is consistent, and for some $\sigma^2_{\text{RA}}$, we have the following convergence in distribution
	$$\sqrt{n}(\hat{\theta}_{\text{RA}}-\theta) \stackrel{d}{\to} N(0, \sigma^2_{\text{RA}}).$$
\end{thm}

Theorem \ref{theorem:ra} establishes that a population quantity expressible in the form \eqref{eq:theta} can be estimated via a regression adjustment approach.  The asymptotic normality of the provided estimator can be used to construct a confidence interval for inference.  Although it is possible to construct a sandwich estimator, we again recommend using the bootstrap to estimate the variance because writing a closed form expression may not be so elegant.




\section{Beyond CCMV}

\label{appendix:beyondCCMV}

We extend the work in the main body of the paper to see the analysis under an identifying restriction that is not CCMV.
We consider the case with 2 covariate variables $(X_1,X_2)$ subject to missingness for a total of $2^2=4$ possible patterns.  
For simplicity, let $v = (w_a,a).$
The CCMV assumption (Definition \ref{def:ccmv}) is equivalent to:
\begin{equation}
	\begin{aligned}
		& p(x_1 | x_2, R=01, v) = p(x_1 | x_2, R=11,v), \\
		& p(x_2 | x_1, R=10, v) = p(x_2 | x_1, R=11,v), \\
		& p(x_1, x_2 | R=00, v) = p(x_1, x_2 | R=11,v).
	\end{aligned}
	\label{eq:ccmv:PMM}
\end{equation}
The last equality can be relaxed by other assumptions.
Here we consider the following identifying restriction:
\begin{equation}
	\begin{aligned}
		& p(x_1 | x_2, R=01,v) = p(x_1 | x_2, R=11,v), \\
		& p(x_2 | x_1, R=10,v) = p(x_2 | x_1, R=11,v), \\
		& p(x_1, x_2 | R=00,v) = \kappa_1 \cdot p(x_1, x_2 | R=01,v) + \kappa_2 \cdot p(x_1, x_2 | R=10,v) + \kappa_3 \cdot p(x_1, x_2 | R=11,v)
	\end{aligned}
	\label{eq:general}
\end{equation}
for nonnegative constants $\kappa_1, \kappa_2$, and $\kappa_3$ satisfying $\kappa_1+\kappa_2+\kappa_3 = 1$.  
When $\kappa_3 =1$, equation \eqref{eq:general} reduces to the CCMV in equation \eqref{eq:ccmv:PMM}.
If there exist variables that are always observed, then the above identifying restrictions can be generalized to allow $\kappa_1$, $\kappa_2$, and $\kappa_3$ to be real-valued functions of the observed variables.
Under suitable choice of $\kappa_1,\kappa_2, \kappa_3$ (as a function of observed variables), 
the last restriction in equation \eqref{eq:general}
becomes
\begin{align*}
	& p(x_1, x_2 | R=00,v) = p(x_1, x_2 | R\in\{01,10,11\},v),
\end{align*}
which shows connections to the available-case missing value assumption in \citep{ACMV}.

\subsection{IPW estimators}

We now show that the (complete) odds admits a nice identifiable representation.  
Let $V  = (W_A, A)$.
The two odds
\begin{equation*}
	\frac{P(R=10|X,V)}{P(R=11|X,V)} \quad \text{and} \quad \frac{P(R=01|X,V)}{P(R=11|X,V)}
\end{equation*}
are identifiable via the same formulas under CCMV.  We now decompose the odds $P(R=00|X,V)/P(R\in\{01,10,11\}|X,V)$ via
\begin{align*}
	&\kappa_1  \cdot \frac{p(x_1,x_2|R=01,v)P(R=00|v)}{p(x_1,x_2,R\in\{01,10,11\}|v)} + \kappa_2 \cdot \frac{p(x_1,x_2|R=10,v) P(R=00|v)}{p(x_1,x_2,R\in\{01,10,11\}|v)} \\
	&\quad + \kappa_3 \cdot \frac{p(x_1,x_2|R=11,v)P(R=00|v)}{p(x_1,x_2,R\in\{01,10,11\}|v)}.
\end{align*}
The first term can be decomposed by
\begin{align*}
	&\kappa_1  \cdot \frac{p(x_1,x_2|R=01,v)P(R=00|v)}{p(x_1,x_2,R\in\{01,10,11\}|v)} \\
	&= \kappa_1 \cdot \frac{p(x_1|x_2,R=01,v)p(x_2|R=01,v)P(R=00|v)}{p(x_2|x_1,R=11,v)p(x_1,R=10|v) + p(x_1|x_2,R=11|v)p(x_2,R=01|v) + p(x_1,x_2,R=11|v)} \\
	&= \kappa_1 \cdot \frac{p(x_2|R=01,v) P(R=00|v)}{\dfrac{p(x_2|x_1,R=11,v)p(x_1,R=10|v)}{p(x_1|x_2,R=01,v)} + p(x_2,R=01|v) + p(x_2,R=11|v)} \\
	&= \kappa_1 \cdot \frac{p(x_2|R=01,v) P(R=00|v)}{\dfrac{p(x_1,x_2,R=11|v)p(x_1,R=10|v)p(x_2,R=01|v)}{p(x_1,x_2,R=01|v)p(x_1,R=11|v)} + p(x_2,R=01|v) + p(x_2,R=11|v)} \\
	&= \kappa_1 \cdot \frac{p(x_2|R=01,v) P(R=00|v)}{\dfrac{P(R=11|x_2,v)}{P(R=01|x_2,v)} \cdot \dfrac{P(R=10|x_1,v)}{P(R=11|x_1,v)} \cdot p(x_2,R=01|v) + p(x_2,R=01|v) + p(x_2,R=11|v)} \\
	&= \kappa_1 \cdot \frac{P(R=00|v)/P(R=01|v)}{\dfrac{P(R=11|x_2,v)}{P(R=01|x_2,v)} \cdot \dfrac{P(R=10|x_1,v)}{P(R=11|x_1,v)} + \dfrac{P(R=11|x_2,v)}{P(R=01|x_2,v)} + 1}, \\
\end{align*}
which is identifiable.

A similar argument for the second term yields
\begin{equation*}
	\kappa_2 \cdot \frac{P(R=00|v)/P(R=10|v)}{\dfrac{P(R=11|x_1,v)}{P(R=10|x_1,v)} \cdot \dfrac{P(R=01|x_2,v)}{P(R=11|x_2,v)} + \dfrac{P(R=11|x_1,v)}{P(R=10|x_1,v)} + 1}.
\end{equation*}

The third term yields
\begin{align*}
	&\kappa_3  \cdot \frac{p(x_1,x_2|R=11,v)P(R=00|v)}{p(x_1,x_2,R\in\{01,10,11\}|v)} \\
	&= \kappa_3 \cdot \frac{P(R=00|v)/P(R=11|v)}{\dfrac{P(R=10|x_1,v)}{P(R=11|x_1,v)} + \dfrac{P(R=01|x_2,v)}{P(R=11|x_2,v)} + 1}, \\
\end{align*}
which is identifiable.

Thus, this implies that $P(R=00|x,v)/P(R=11|x,v)$ is identifiable.  We can model the odds $P(R=r|x_r,v)/P(R=11|x_r,v)$, and then, plug them into the formulas above to obtain an identifiable expression for $P(R=00|x,v)/P(R=11|x,v)$.

\subsection{RA estimator (imputation)}

First, as the identifying restriction is expressed in terms of the extrapolation distribution, the regression adjustment estimator is fairly straightforward.  Since the extrapolation distribution $p(x_1,x_2|R=00)$ is a mixture of three identifiable distributions, we can again utilize a Monte Carlo approximation to construct the regression adjustment estimator.  This approximation is essentially the same as under CCMV except we sample from this mixture distribution for pattern $R=00$.

\section{Sensitivity analysis}

\label{appendix:sens}

We present details on the sensitivity analyses utilized in the paper.

\subsection{Sensitivity analysis for the odds under logistic regression}

\label{appendix:sens-odds}

We have showed with Proposition \ref{prop:oddsCCMV} that under CCMV, the odds satisfies
\begin{equation*}
	O_{r,a}(X,W_a) := \frac{P(R=r|X,W_a,a)}{P(R=1_d|X,W_a,a)} \stackrel{\text{CCMV}}{=} \frac{P(R=r|X_r,W_a,a)}{P(R=1_d|X_r,W_a,a)} =: Q_{r,a}(X_r,W_a).
\end{equation*}

Thus, the complete odds depend only on the variables observed under pattern $R=r$.  We can introduce a sensitivity parameter to perturb this CCMV assumption using an exponential tilting method with logistic regression \citep{kim2011semiparametric, shao2016semiparametric, zhao2017semiparametric}.  This results in
\begin{equation}
	O_{r,a}(X,W_a) = Q_{r,a}(X_r,W_a) e^{\rho_{\bar{r}} X_{\bar{r}}},
	\label{eq::SA1}
\end{equation}
where $\rho_{\bar{r}}$ is a sensitivity parameter that consists of a vector of real numbers (one for each missing covariate under pattern $r$).
In practice, we often let every element of $\rho_{\bar{r}}$ to be the same for simplicity.

We now consider the special case of logistic regression.  Under CCMV with an assumed logistic regression model, the complete odds satisfies
\begin{equation*}
	\log O_{r,a}(X,W_a;\alpha_{r,a}) = \log Q_{r,a}(X_r,W_a;\alpha_{r,a}) = \alpha_{r,a}^\top X_r.
\end{equation*}
Under the sensitivity model (equation \eqref{eq::SA1}), this becomes
\begin{equation*}
	\log O_{r,a}(X,W_a;\alpha_{r,a},\rho_{\bar{r}}) = \alpha_{r,a}^\top X_r + \rho_{\bar{r}}^\top X_{\bar{r}}.
\end{equation*}
Note that the first term on the RHS is the traditional term obtained by fitting a logistic regression model, and the second term is the tilting factor to push the missing data assumption away from CCMV.

Varying the elements of $\rho_{\bar{r}}$ will modify the missing data assumption and change the weights.  This  allows one to determine the robustness of the estimator to changes in this sensitivity parameter.  This sensitivity parameter has a nice interpretation because it is the coefficient in a linear model, and the range can be specified based on one's belief about the relation of impact of the missing variables to the impact of the observed variables.  For example, if the coefficient of the $X_1$ variable is $1$ and if $X_2$ is subject to missingness, then the sensitivity parameter corresponding to $X_2$ can be chosen to be within the interval $[-1,1]$ by assuming that $X_2$ has at most the same impact in magnitude as $X_1$.

\subsection{Sensitivity analysis for binary imputation via rejection sampling}

\label{appendix:sens-binary-imp}

We also present an approach for sensitivity analysis under imputation model.
We first consider the case where the variable to be imputed $X$ is binary.  
Suppose that under an identifying restriction (such as the CCMV), we will impute the variable $X$
by drawing $X \sim \text{Binomial}(p)$, where $p$
is the probability parameter from a particular imputation model. 



To perturb the imputation model, we consider the following procedure:
\begin{itemize}
	\item Draw $X \sim \text{Bernoulli}(p)$.
	\item Consider a sensitivity parameter $\zeta \in [0,1]$.  We iterate the following steps until a draw is accepted:
	\begin{itemize}
		\item If $X=1$, with probability $\zeta$, redraw $X \sim \text{Bernoulli}(p)$.  Otherwise, accept the draw.
		\item If $X=0$, with probability $1-\zeta$, redraw $X \sim \text{Bernoulli}(p)$.  Otherwise, accept the draw.
	\end{itemize}
\end{itemize}

We denote this new random variable by $Y$.  We can see when $\zeta=1$, we force $Y$ to be deterministically $0$, and when $\zeta=0$, we force $Y$ to be deterministically $1$.  When $\zeta=1/2$, we resample exactly half of the time, so we obtain $Y=X$. Namely, when $\zeta = 1/2$, we are under the unperturbed model.
Every other value of $\zeta$ interpolates between these extremes.
Thus, the parameter $\zeta$ can be viewed as a preference toward $X=0$.

It turns out that the above procedure admits a simpler expression.
Given that the final output is a binary number, the accepted value $Y$ follows from a Bernoulli random variable.
Since $Y$ is binary, $\E[Y]$ is the probability of success.  On the initial iteration, we draw $1$ and accept it with probability $p(1-\zeta)$.  Otherwise, we redraw with probability $p\zeta + (1-p)(1-\zeta)$ and return to the same state.  This yields the following equation
\begin{equation*}
	\E[Y] = p(1-\zeta) + p\zeta \E[Y] + (1-p)(1-\zeta) \E[Y],
\end{equation*}
so we have $\E[Y] = \dfrac{p(1-\zeta)}{1-p\zeta-(1-p)(1-\zeta)} = \dfrac{p(1-\zeta)}{p+\zeta-2p\zeta}$.\\
Thus, we actually do not need to perform the rejection sampling procedure. 
Instead, we can simply draw
$$
Y\sim \text{Bernoulli}\left(\dfrac{p(1-\zeta)}{p+\zeta-2p\zeta}\right).
$$


\subsection{Sensitivity analysis for continuous imputation via Gaussian tilting}

\label{appendix:sens:cont}

Now, we discuss the case of sensitivity analysis on imputing a continuous random variable. 
Suppose that without perturbations, we impute a continuous random variable $X$ from a PDF $p(x)$. 
We propose the following modified approach:

\begin{itemize}
	\item Draw $X \sim p(x)$.
	\item Consider a sensitivity parameter $\xi \in [0,\infty)$.  First, we sample $X\sim p(x)$.  We iterate the following steps until a draw is accepted:
	\begin{itemize}
		\item With probability $\exp(-\xi X^2)$, accept $X \sim p(x)$.
		\item Otherwise, redraw $X$.
	\end{itemize}
\end{itemize}

First, note that as $\xi \to \infty$, we almost always resample, and as $\xi \to 0$, we almost never resample.  We denote this new random variable by $Y$.  The distribution of $Y$ is given by
\begin{equation*}
	P(Y\leq y) = P(X \leq y \ | \ \text{accept } X) = P(X \leq y \ | \  U \leq \exp(-\xi X^2))
\end{equation*}
for a random variable $U \sim \text{Uniform}(0,1)$ independent of $X$.  We see that
\begin{align*}
	P(Y\leq y) &= P(X \leq y \ | \  U \leq \exp(-\xi X^2)) \\
	&\propto P(X \leq y, U \leq \exp(-\xi X^2)) \\
	&= P(U \leq \exp(-\xi X^2) \ | \ X \leq y) \cdot P(X\leq y) \\
	&= \int_{-\infty}^y \frac{P(U \leq \exp(-\xi x^2) \ | \  x)}{P(X\leq y)} \cdot p(x) \ dx \cdot P(X\leq y) \\
	&= \int_{-\infty}^y P(U \leq \exp(-\xi x^2) \ | \  x) \cdot p(x) \ dx \\
	&= \int_{-\infty}^y \exp(-\xi x^2) p(x) \ dx.
\end{align*}

Therefore, we see that this resampling method is equivalent to multiplying the original PDF $p(x)$ by $\exp(-\xi x^2)$ and renormalizing.  Since the multiplication factor $\exp(-\xi x^2)$ is an exponential, this action can be considered a form of \textit{Gaussian tilting}.

\textbf{$p(x)$ is Gaussian.}  Suppose that the original imputation distribution is Gaussian such that $X \sim N(\mu, \sigma^2)$.  A simple completing the square argument shows that 
\begin{align*}
	-\frac{1}{2\sigma^2} (x-\mu)^2 - \xi x^2 = -\left(\frac{1}{2\sigma^2} + \xi\right) \left(x- \frac{\mu}{2\sigma^2\xi + 1}\right)^2 + \left(\frac{\mu^2}{2\sigma^2 + 4\sigma^4 \xi} - \frac{\mu^2}{2\sigma^2}\right).
\end{align*}
Therefore, Gaussian tilting by $\exp(-\xi x^2)$ results in a random variable
\begin{equation*}
	Y \sim N\left(\frac{\mu}{2\sigma^2 \xi + 1}, \frac{\sigma^2}{2\sigma^2 \xi + 1}\right).
\end{equation*}
As $\xi \to \infty$, the distribution for $Y$ degenerates to a point mass of $1$ at $0$.  As $\xi \to 0$, we approach the original distribution $p(x)$, which is what we expect as we almost never resample.


\textbf{$p(x)$ is Exponential.}  Suppose $p(x)$ is Exponential$(\lambda)$ such that $p(x) = \lambda \exp(-\lambda x)$.  Then, Gaussian tilting by $\exp(-\xi x^2)$ results in a truncated Gaussian that is formed by taking the PDF over the positive support of a Gaussian with mean $-\lambda/(2\xi)$ and variance $1/\xi$.
As $\xi \to \infty$, again, the distribution for $Y$ degenerates to a point mass of $1$ at $0$.

\section{Details on survival analysis simulation and additional simulation studies}

\label{appendix:sim}

\subsection{Details on Cox model simulation}

\label{appendix:sim:survival}
Here we provide details on the Cox model simulation.
Note that the data generating process is described in the main paper.

\subsubsection{Estimation of the imputation model}

In our simulated Cox model problem,
there are two covariates $X = (X_1,X_2)\in\{0,1\}^2$.
In the complete case, the observed-data distribution can be decomposed into the following terms:
\begin{align*}
	p(x,R=11,y,\delta) &= \underbrace{p(x,R=11)}_{\text{Empirical}} \cdot \frac{\overbrace{P(R=11|x,y,\delta)}^{\text{Propensity score}}}{\underbrace{P(R=11|x)}_{\text{
				Integration}}} \cdot \underbrace{p(y,\delta|x)}_{\text{EM}}.
\end{align*}


Each of the above four components in the RHS is estimated as follows:
\begin{itemize}
	
	\item {\bf Empirical.} This component can be directly estimated from the data using empirical proportion.
	
	\item {\bf Propensity score.} This is estimated by combining estimators of the complete odds
	and use equation \eqref{eq:PS}.

	\item {\bf Integration.} The term $P(R=11|x)$ can be rewritten as follows
	\begin{align*}
		P(R=11|x) &= \sum_{\delta=0}^1 \int_0^\infty P(R=11|x,y,\delta)p(y,\delta|x) \ dy.
	\end{align*}
	In practice, this integral does not have an analytic form, so we use numerical integration in R to circumvent the problem.
	
	\item {\bf EM.} 
	The estimation of this component is more involved. We use an EM-like method to estimate it.
	We assume a parametric model $p_{Y,\Delta|X}(y,\delta|x) = [\nu_1 \exp(x^\top \beta)]^\delta \nu_2^{1-\delta} e^{-y(\nu_1 \exp(x^\top \beta) + \nu_2)}$; this model is derived from our design of the simulation.  Fitting $p(y,\delta|x)$ is a simple maximum likelihood procedure when the covariates are fully observed.  Consider a specific $\tilde{x}$.  The model becomes
	\begin{align*}
		p_{Y,\Delta|X}(y,\delta|\tilde x) &= [\nu_1 \exp(\tilde{x}^\top\beta)]^\delta \nu_2^{1-\delta} e^{-y(\nu_1 \exp(\tilde{x}^\top \beta) + \nu_2)} \\
		&= \gamma_{\tilde{x}}^\delta \nu_2^{1-\delta} e^{-y(\gamma_{\tilde{x}} + \nu_2)}
	\end{align*}
	for $\gamma_x = \nu_1 \exp(x^\top \beta)$.  The ML estimators are
	$$\hat{\gamma}_{\tilde{x}} = \frac{\sum_{i=1}^n \delta_i 1(x_i = \tilde{x})}{\sum_{i=1}^n y_i 1(x_i=\tilde{x})} \quad \text{and} \quad \hat{\nu}_2 = \frac{\sum_{i=1}^n (1(x_i=\tilde{x})-\delta_i 1(x_i = \tilde{x}))}{\sum_{i=1}^n y_i 1(x_i=\tilde{x})}.$$
	
	In practice, we do not know the covariates for all of the individuals.  So, we fit the models using an EM algorithm.  We replace the indicators $1(x_i=\tilde{x})$ with weights $w_i$.
	
	Suppose $\tilde{x}=(1,1)$.  The weights are as follows:
	$$w_i = \begin{cases} 1 & \text{ if $x_1=x_2=1$} \\ 0 & \text{ if $x_1=0$ or $x_2=0$} \\ \hat{P}(X_2=1|X_1=1,R=11,y,\delta) & \text{ if $x_1=1$ and $x_2$ missing} \\ \hat{P}(X_1=1|X_2=1,R=11,y,\delta) & \text{ if $x_2=1$ and $x_1$ missing} \end{cases}.$$
	
	The weights depend on the fitted parameters.  For the EM algorithm, we iterate between updating $w_i$ (E-step) and computing the MLEs (M-step).
\end{itemize}

With the above procedure, we are able to estimate the distribution $p(x,R=11,y,\delta)$,
which describes all imputation models under the CCMV assumption.

\subsubsection{Sensitivity analysis}
We include the following sensitivity analysis as part of our simulations.  We fix a sample size of $n=1000$ with $M=50$ imputations.  For the IPW approach, we vary the exponential tilting sensitivity parameter $\rho$ in the interval $[-3,3]$.
For the imputation/RA approach, we change the binary sensitivity analysis parameter $\zeta$ in the interval $[0,1]$.  For each corresponding value of the sensitivity parameter, we compute the IPW and RA estimators, respectively.  The true value of the parameter is indicated with the red line.

\begin{figure}
	\centering
	\includegraphics[scale=0.3]{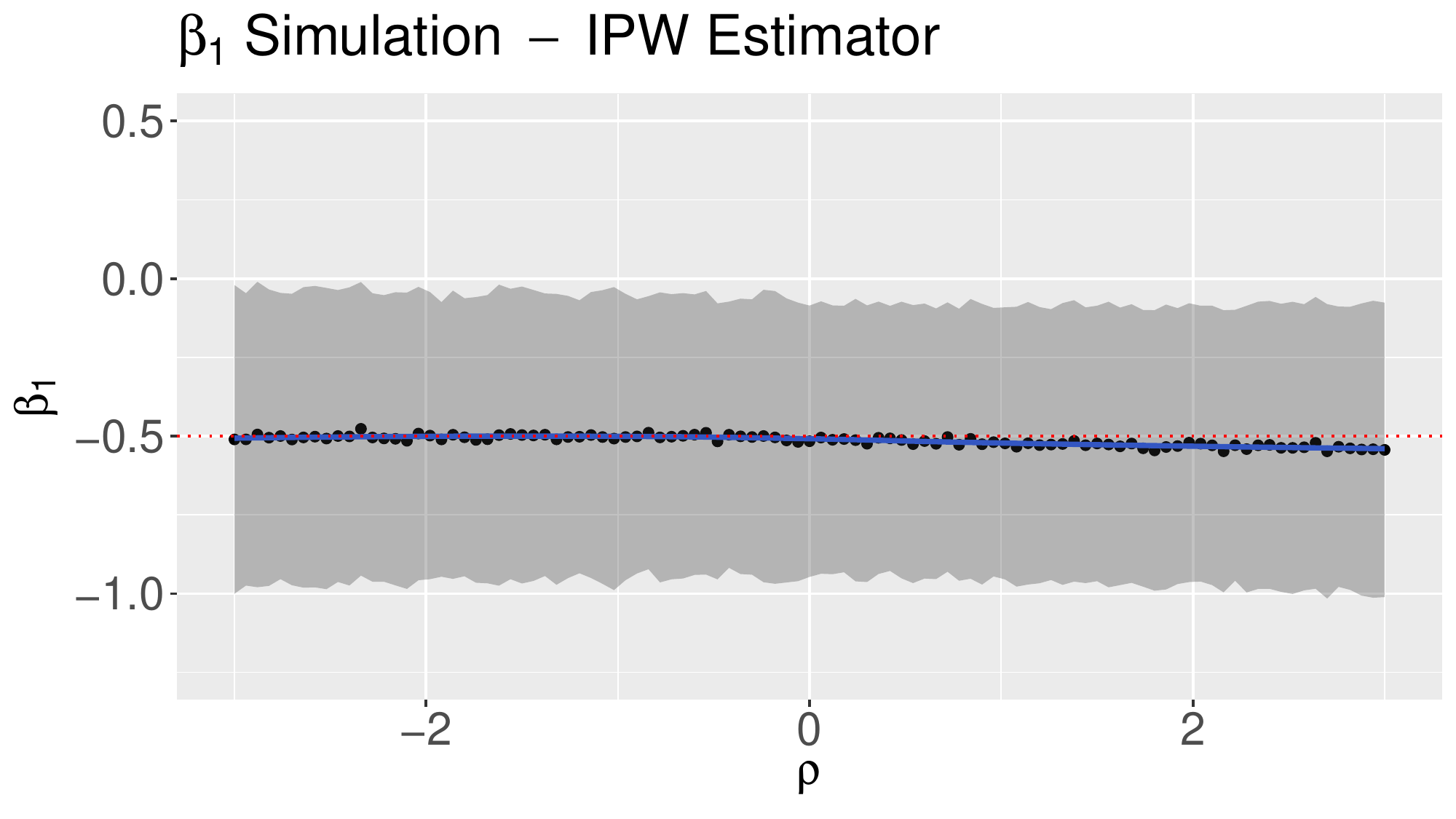}
	\includegraphics[scale=0.3]{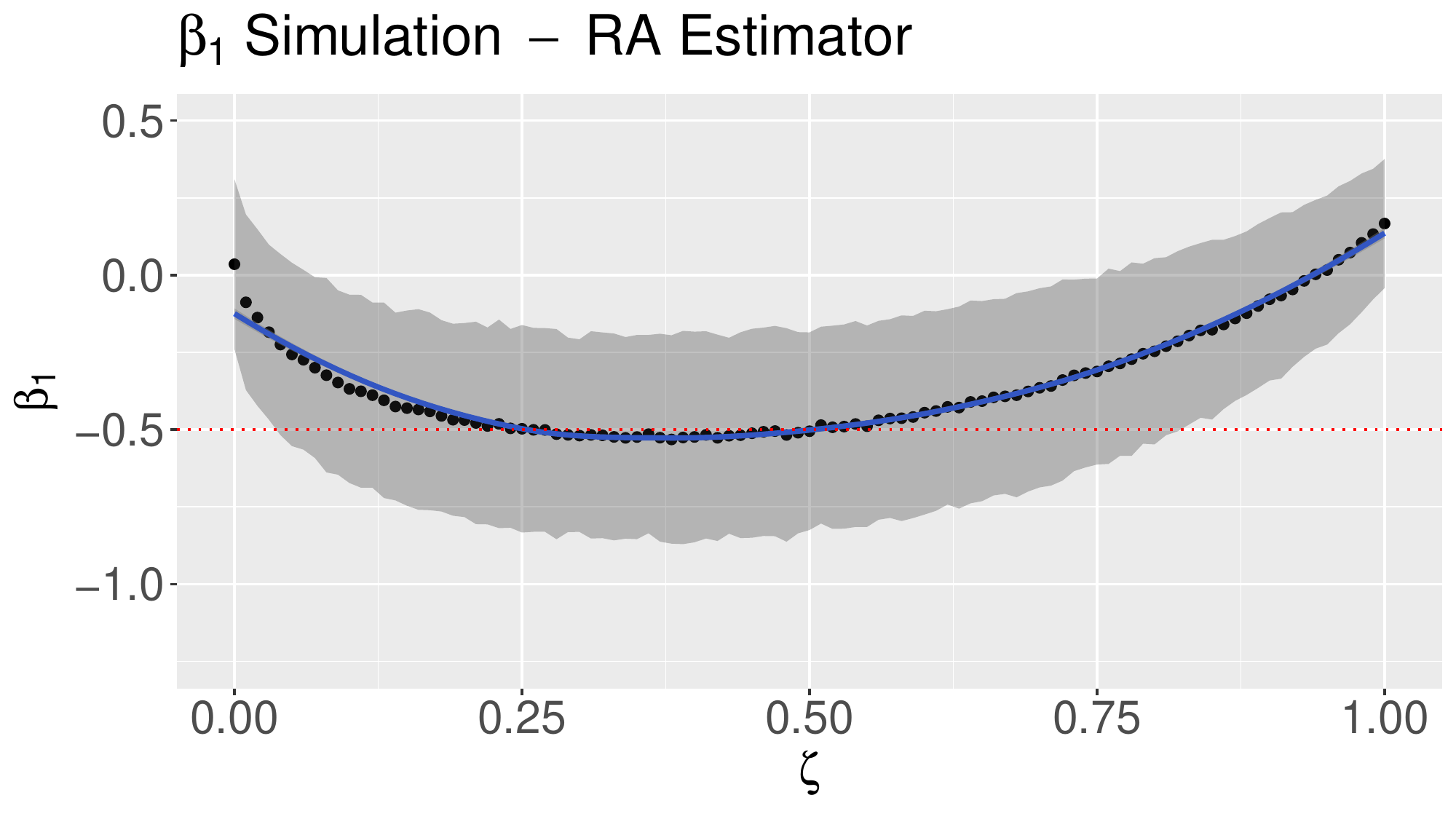}
	\includegraphics[scale=0.3]{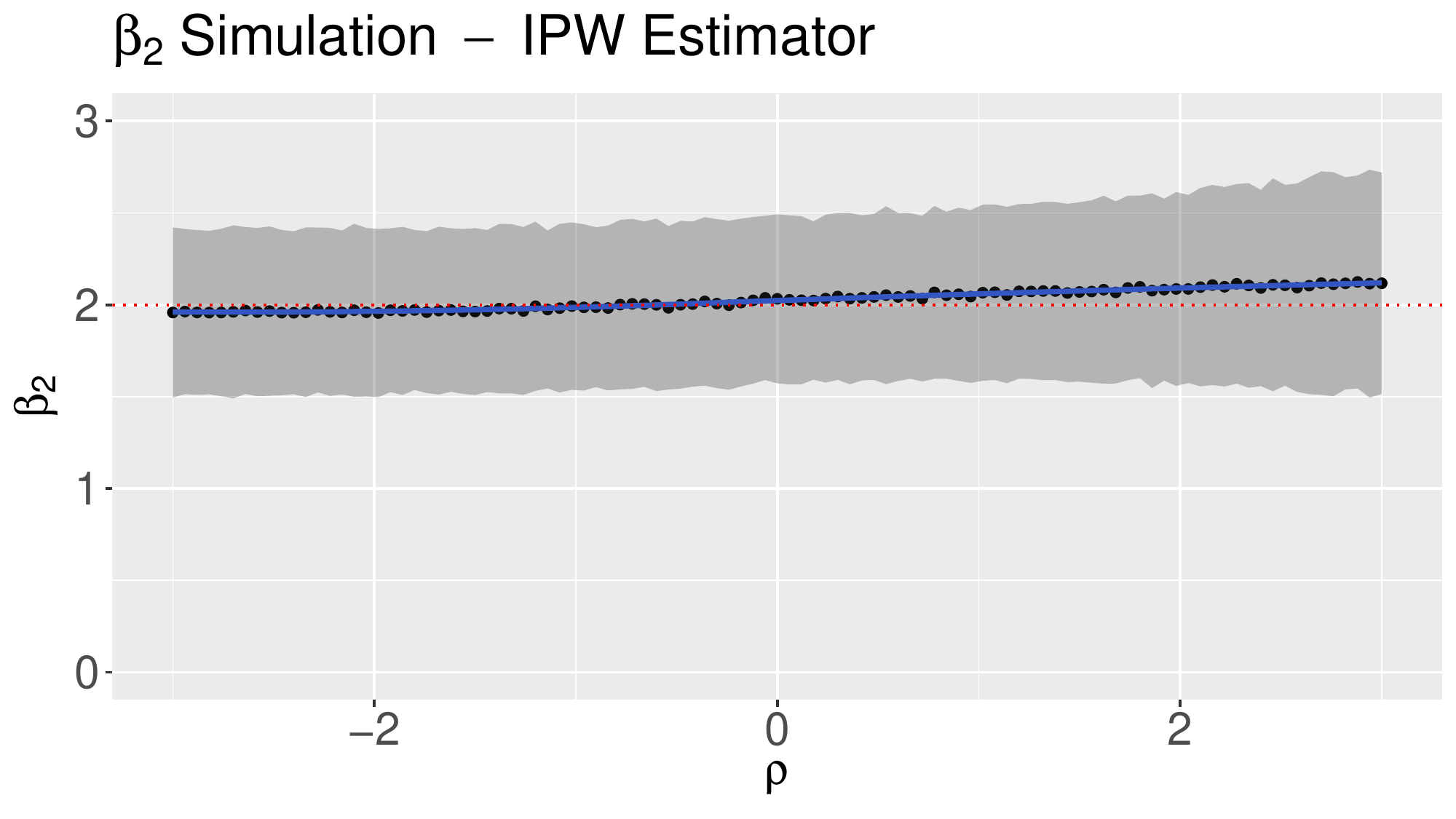}
	\includegraphics[scale=0.3]{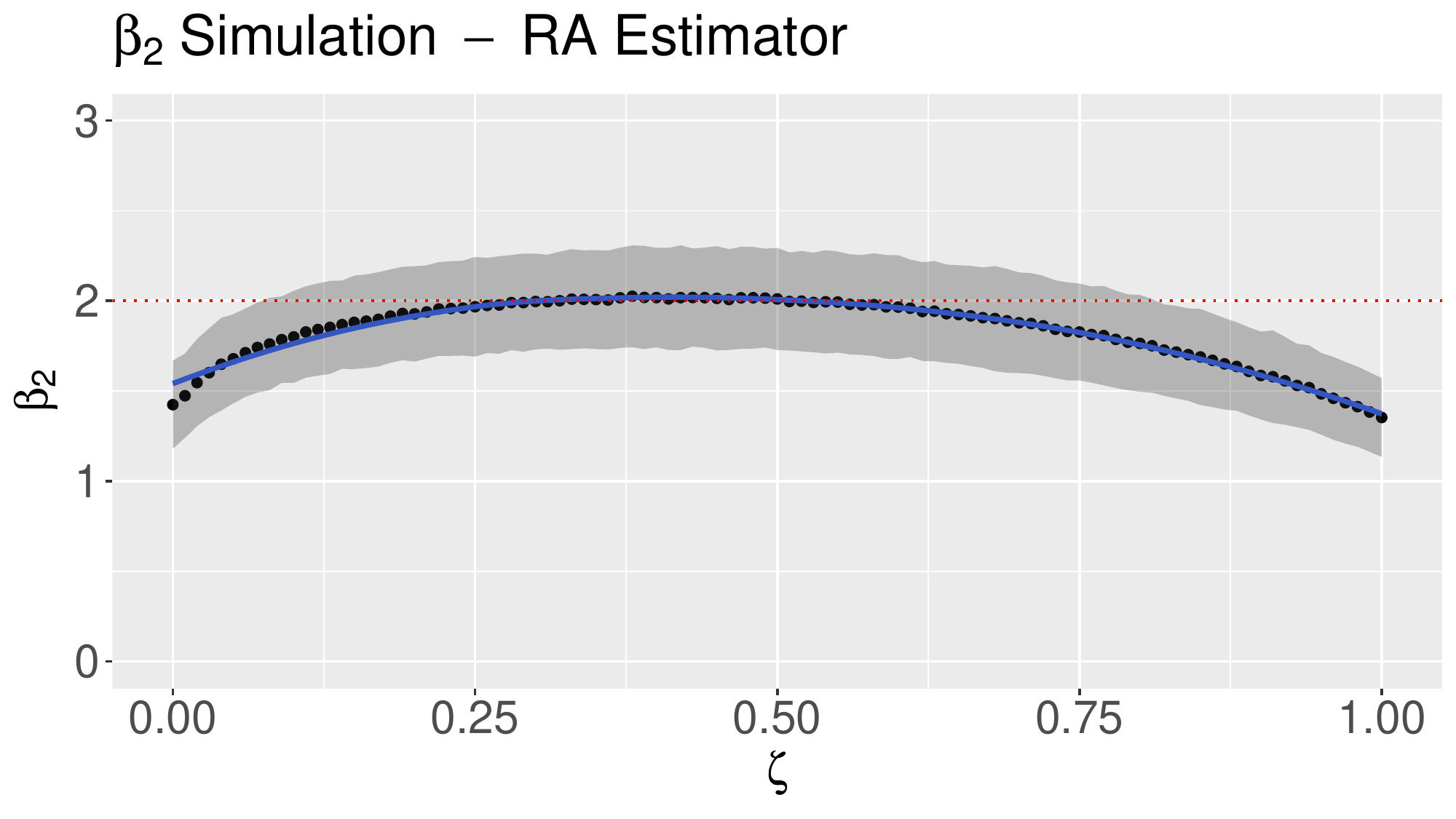}
	\caption{Sensitivity analysis for the Cox model simulation with both the IPW and RA estimators.}
\end{figure}

First, we observe that the IPW estimator appears to be fairly stable as we vary the sensitivity parameter, and the size of the confidence interval does not change drastically.  Therefore, we conclude that the $\beta_1$ parameter is significant even as we perturb the estimator.  On the other hand, the point estimates of the RA estimator do appear to change slightly as we vary the imputation model.  The $\beta_2$ parameter is large enough that any of these perturbations do not result in a change in conclusion.  However, the $\beta_1$ parameter is relatively close in magnitude to $0$, so for extreme values of the $\zeta$ parameter, the results are not significant.  This suggests that the choice of identifying restriction is critical, and a sensitivity analysis is crucial for a sound data analysis.  We note that the extreme values of the $\zeta$ parameter generally imply near almost sure deterministic imputation for both covariates, which generally will be unlikely to be true.

\subsection{Details on binary treatment simulation}

\label{appendix:sim:binary}

\subsubsection{Data generating process}

We generate the data as follows:
\begin{enumerate}
	\item We sample $(R,Y,A)$ as discrete random variables as follows:
	\begin{multicols}{2}
		\begin{itemize}[leftmargin=*]
			\small
			\item $P(R=11,Y=0,A=0) = 1/16$
			\item $P(R=11,Y=0,A=1) = 1/12$
			\item $P(R=11,Y=1,A=0) = 1/16$
			\item $P(R=11,Y=1,A=1) = 1/24$
			\vskip 8pt
			\item $P(R=10,Y=0,A=0) = 1/12$
			\item $P(R=10,Y=0,A=1) = 1/8$
			\item $P(R=10,Y=1,A=0) = 1/24$
			\item $P(R=10,Y=1,A=1) = 1/12$
			\vskip 8pt
			\item $P(R=01,Y=0,A=0) = 1/16$
			\item $P(R=01,Y=0,A=1) = 1/24$
			\item $P(R=01,Y=1,A=0) = 1/16$
			\item $P(R=01,Y=1,A=1) = 1/12$
			\vskip 8pt
			\item $P(R=00,Y=0,A=0) = 1/36$
			\item $P(R=00,Y=0,A=1) = 1/24$
			\item $P(R=00,Y=1,A=0) = 1/24$
			\item $P(R=00,Y=1,A=1) = 1/18$
		\end{itemize}
	\end{multicols}
	
	\item We sample $(X_1, X_2) | R, Y, A$ via the following distributions:
	
	\begin{itemize}[leftmargin=*]
		\item $X_1, X_2 | R=11, Y=0, A=0 \sim N\left(\begin{bmatrix} 3 \\ 4 \end{bmatrix}, \begin{bmatrix} 0.5 & 0.1 \\ 0.1 & 0.5 \end{bmatrix} \right)$
		
		\item $X_1, X_2 | R=11, Y=0, A=1 \sim N\left(\begin{bmatrix} 4 \\ 4 \end{bmatrix}, \begin{bmatrix} 0.5 & 0.2 \\ 0.2 & 0.5 \end{bmatrix} \right)$
		
		\item $X_1, X_2 | R=11, Y=1, A=0 \sim N\left(\begin{bmatrix} 3 \\ 2 \end{bmatrix}, \begin{bmatrix} 0.4 & 0.1 \\ 0.1 & 0.4 \end{bmatrix} \right)$
		
		\item $X_1, X_2 | R=11, Y=1, A=1 \sim N\left(\begin{bmatrix} 2 \\ 2 \end{bmatrix}, \begin{bmatrix} 0.5 & 0.1 \\ 0.1 & 0.5 \end{bmatrix} \right)$
		
		\item $X_1, X_2 | R=10, Y=0, A=0 \sim N\left(\begin{bmatrix} 2 \\ 19/5 \end{bmatrix}, \begin{bmatrix} 0.5 & 0.1 \\ 0.1 & 0.5 \end{bmatrix} \right)$
		
		\item $X_1, X_2 | R=10, Y=0, A=1 \sim N\left(\begin{bmatrix} 2 \\ 16/5 \end{bmatrix}, \begin{bmatrix} 0.5 & 0.2 \\ 0.2 & 0.5 \end{bmatrix} \right)$
		
		\item $X_1, X_2 | R=10, Y=1, A=0 \sim N\left(\begin{bmatrix} 1 \\ 3/2 \end{bmatrix}, \begin{bmatrix} 0.4 & 0.1 \\ 0.1 & 0.4 \end{bmatrix} \right)$
		
		\item $X_1, X_2 | R=10, Y=1, A=1 \sim N\left(\begin{bmatrix} 3 \\ 11/5 \end{bmatrix}, \begin{bmatrix} 0.5 & 0.1 \\ 0.1 & 0.5 \end{bmatrix} \right)$
		
		\item $X_1, X_2 | R=01, Y=0, A=0 \sim N\left(\begin{bmatrix} 13/5 \\ 2 \end{bmatrix}, \begin{bmatrix} 0.5 & 0.1 \\ 0.1 & 0.5 \end{bmatrix} \right)$
		
		\item $X_1, X_2 | R=01, Y=0, A=1\sim N\left(\begin{bmatrix} 14/5 \\ 1 \end{bmatrix}, \begin{bmatrix} 0.5 & 0.2 \\ 0.2 & 0.5 \end{bmatrix} \right)$
		
		\item $X_1, X_2 | R=01, Y=1, A=0 \sim N\left(\begin{bmatrix} 3.125 \\ 2.5 \end{bmatrix}, \begin{bmatrix} 0.4 & 0.1 \\ 0.1 & 0.4 \end{bmatrix} \right)$
		
		\item $X_1, X_2 | R=01, Y=1, A=1 \sim N\left(\begin{bmatrix} 1.9 \\ 1.5 \end{bmatrix}, \begin{bmatrix} 0.5 & 0.1 \\ 0.1 & 0.5 \end{bmatrix} \right)$
		
		\item $X_1, X_2 | R=00, Y=0, A=0 \sim N\left(\begin{bmatrix} 3 \\ 4 \end{bmatrix}, \begin{bmatrix} 0.5 & 0.1 \\ 0.1 & 0.5 \end{bmatrix} \right)$
		
		\item $X_1, X_2 | R=00, Y=0, A=1 \sim N\left(\begin{bmatrix} 4 \\ 4 \end{bmatrix}, \begin{bmatrix} 0.5 & 0.2 \\ 0.2 & 0.5 \end{bmatrix} \right)$
		
		\item $X_1, X_2 | R=00, Y=1, A=0 \sim N\left(\begin{bmatrix} 3 \\ 2 \end{bmatrix}, \begin{bmatrix} 0.4 & 0.1 \\ 0.1 & 0.4 \end{bmatrix} \right)$
		
		\item $X_1, X_2 | R=00, Y=1, A=1 \sim N\left(\begin{bmatrix} 2 \\ 2 \end{bmatrix}, \begin{bmatrix} 0.5 & 0.1 \\ 0.1 & 0.5 \end{bmatrix} \right)$

	\end{itemize}
\end{enumerate}

The generative model for $(X_1,X_2)|R,Y,A$ is constructed in such a way to ensure that the CCMV assumption holds.

\subsubsection{Details on the estimation}

Note that in the binary treatment problem, we have two estimators; 
one estimator to account for the missing covariates (denoted as IPW-R or RA-R)
and the other estimator for the ATE (denoted as IPW-A, RA-A, DR-A).

{\bf Final estimators.}
For ease of simulation, we can consider the following 6 estimators for $\E[Y(1)]$.
Note that here we describe the linear functional form of the estimator.
We take the empirical average of them to form the final estimate.
\begin{enumerate}
	\item \emph{IPW-R, IPW-A.}
	\begin{equation*}
		\frac{1(R=11)1(A=1)Y}{P(A=1|X)}\sum_r Q_{r,1}(X_r,Y)
	\end{equation*}
	
	\item \emph{IPW-R, RA-A.}
	\begin{equation*}
		\frac{1(R=11)m_1(X)}{P(A=1|X)}\sum_r\sum_a 1(A=a)Q_{r,a}(X_r,Y)
	\end{equation*}
	
	\item \emph{IPW-R, DR-A.}
	\begin{align*}
		&\frac{1(R=11)1(A=1)(Y-m_1(X))}{P(A=1|X)}\sum_r Q_{r,1}(X_r,Y) \\
		&\quad + \frac{1(R=11)m_1(X)}{P(A=1|X)}\sum_r\sum_a 1(A=a)Q_{r,a}(X_r,Y)
	\end{align*}
	
	\item \emph{RA-R, IPW-A.}
	\begin{equation*}
		1(A=1)Y \sum_r 1(R=r) M_{1/\pi_A,r,1}(X_r,Y)
	\end{equation*}
	
	\item \emph{RA-R, RA-A.}
	\begin{equation*}
		\sum_r \sum_a 1(R=r) 1(A=a) M_{m,r,a}(X_r, Y)
	\end{equation*}
	
	\item \emph{RA-R, DR-A.}
	\begin{align*}
		& 1(A=1)Y \sum_r 1(R=r) M_{m,r,1}(X_r,Y) \\
		&\quad + \sum_r \sum_a 1(R=r) 1(A=a) M_{m,r,a}(X_r, Y) \\
		&\quad - 1(A=1) \sum_r 1(R=r) M_{m/\pi_A,r,1}(X_r,Y)
	\end{align*}
\end{enumerate}

As a reminder, the notation $m_1$, $M_{m,r,0}$, $M_{m,r,1}$, $M_{1/\pi_A,r,1}$, and $M_{m/\pi_A,r,1}$ refer to the following regression functions:
\begin{align*}
	& m_1(X) = \E[Y|A=1,X], \\
	& M_{m,r,0}(X_r) = \E[m_1(X)|X_r,R=r,A=0], \\
	& M_{m,r,1}(X_r,Y) = \E[m_1(X)|X_r,R=r,Y,A=1], \\
	& M_{1/\pi_A,r,1}(X_r,Y) = \E\left[\frac{1}{\pi_A(X)} \biggr| X_r,R=r,Y,A=1\right], \\
	& M_{m/\pi_A,r,1}(X_r,Y) = \E\left[\frac{m_1(X)}{\pi_A(X)} \biggr| X_r, R=r, Y, A=1\right].
\end{align*}

Similar estimators can be constructed for $\E[Y(0)]$.  Merging estimators for $\E[Y(1)]$ and $\E[Y(0)]$ together, we can obtain $6$ estimators similar to the ones above for the ATE.  We choose not to construct the multiply-robust estimator for simplicity.  As seen in Theorem \ref{theorem:multiplyrobust-missingresp}, there will be numerous terms, so the multiply-robust estimator will generally be considered a theoretical construct.

Here we describe how to estimate the components in the above estimators.

{\bf IPW-R component.}
The above construction implies that the complete odds has the following form:
\begin{align*}
	Q_{r,a}(X_r, Y) &= \frac{P(R=r|X_r, Y, A)}{P(R=11| X_r, Y, A)} \\
	&= \frac{p(x_r | r, y, a) p(r,y,a)}{p(x_r | 11, y, a) p(11, y, a)}.
\end{align*}

In the special case where covariances are assumed to be the same for a fixed $y$ and $a$, one may simplify the complete odds accordingly
\begin{align*}
	Q_{10,a}(x_1, y)	&= \frac{p(x_1 | 10, y, a) p(r,y,a)}{p(x_1 | 11, y, a) p(11, y, a)} \\
	&= \frac{\exp(-(x_1-\mu_{1,10})^2/(2\sigma^2)) \cdot p(10,y,a)}{\exp(-(x_1-\mu_{1,11})^2/(2\sigma^2)) \cdot p(11,y,a)} \\
	&= \exp\left(\frac{\mu_{1,10}-\mu_{1,11}}{\sigma^2} x_1 + \frac{\mu_{1,11}^2 - \mu_{1,10}^2}{2\sigma^2}\right) \cdot \frac{p(10,y,a)}{p(11,y,a)},
\end{align*}
\begin{align*}
	Q_{01,a}(x_2, y)	&= \frac{p(x_2 | 01, y, a) p(r,y,a)}{p(x_2 | 11, y, a) p(11, y, a)} \\
	&= \frac{\exp(-(x_2-\mu_{2,01})^2/(2\sigma^2)) \cdot p(01,y,a)}{\exp(-(x_2-\mu_{2,11})^2/(2\sigma^2)) \cdot p(11,y,a)} \\
	&= \exp\left(\frac{\mu_{2,01}-\mu_{2,11}}{\sigma^2} x_1 + \frac{\mu_{2,11}^2 - \mu_{2,01}^2}{2\sigma^2}\right) \cdot \frac{p(01,y,a)}{p(11,y,a)},\\
	Q_{00,a}(y)	&= \frac{p(00,y,a)}{p(11, y, a)}.
\end{align*}
The probabilities $p(r,y,a)$ can be estimated by the empirical frequencies, and the means and variances are estimated using the sample versions.
Finally, the propensity score of $R=11$ is 
$$
\pi_R(x,y,a) = P(R=11|x,y,a) = \frac{1}{1 + Q_{10,a}(x_1, y) + Q_{01,a}(x_1, y) +Q_{00,a}(y)}.
$$

{\bf  RA-R component.}
We estimate the regression functions $M_{m,r,0}$, $M_{m,r,1}$, $M_{1/\pi_A,r,1}$, and $M_{m/\pi_A,r,1}$ using a Monte Carlo approach, described in Section \ref{sect:raCox}.  Specifically, in the simulation procedure, we make the assumption that the distribution $p(x_1,x_2|r,y,a)$ is a multivariate Gaussian and that the missing mechanism for the covariates satisfies the CCMV assumption.  This implies that the conditional distributions $p(x_1|x_2,r,y,a)$ and $p(x_2|x_1,r,y,a)$ are conditional Gaussians.  Therefore, we first can estimate the distributions $p(x_1|R=10,r,y,a)$, $p(x_2|R=01,r,y,a)$, and $p(x_1,x_2|R=11,r,y,a)$ with a maximum likelihood procedure directly (plug in the sample means and variances/covariances as parameters).  Then, under the CCMV assumption, we can construct the estimates of the distribution $p(x_1,x_2|r,y,a)$ in the following way:
\begin{align*}
	& \hat{p}(x_1,x_2|R=00,y,a) = \hat{p}(x_1,x_2|R=11,y,a), \\
	& \hat{p}(x_1,x_2|R=01,y,a) = \hat{p}(x_1|x_2,R=11,y,a)\hat{p}(x_2|R=01,y,a), \\
	& \hat{p}(x_1,x_2|R=10,y,a) = \hat{p}(x_2|x_1,R=11,y,a)\hat{p}(x_1|R=10,y,a).
\end{align*}

This further implies that we can use the estimated distributions
\begin{align*}
	& \hat{p}(x_1,x_2|R=00,y,a) = \hat{p}(x_1,x_2|R=11,y,a), \\
	& \hat{p}(x_1|x_2,R=01,y,a) = \hat{p}(x_1|x_2,R=11,y,a), \\
	& \hat{p}(x_2|x_1,R=10,y,a) = \hat{p}(x_2|x_1,R=11,y,a)
\end{align*}
to complete our data.  We use the above extrapolation densities as an estimated imputation model.  We then sample the missing covariates $M = 50$ times to generate multiple complete data sets $\tilde{D}^{(j)} = \{(\tilde{X}_i^{(j)}, R_i, Y_i, A_i)\}_{i=1}^n$, where $\tilde{X}_i^{(j)} = \left(X_{i,R_i}, \tilde{X}_{i,\bar{R}_i}^{(j)}\right)$ refers to the $j$th imputed missing covariates appended to the observed covariates corresponding to the $i$th observation.

We pool all $\tilde{D}^{(j)}$ data sets together to form one large data set $\check{D}$.  This enables us to take the empirical average of the RA-$R$ estimators directly using all of the completed data $\check{D}$.  For example, for the (RA-$R$, IPW-$A$) estimator, we have
\begin{align*}
	&\frac{1}{n}\sum_{i=1}^n \left[ 1(A_i=1)Y_i \sum_r 1(R_i=r) \hat{M}_{1/\pi_A,r,1}(X_{R_i,i}, Y_i) \right]\\
	&\quad = \frac{1}{n}\sum_{i=1}^n \left[1(A_i=1)Y_i  \cdot \frac{1}{M} \sum_{j=1}^M \frac{1}{\hat{\pi}_A(\tilde{X}_{i}^{(j)})} \right] \\
	&\quad = \frac{1}{Mn}\sum_{i=1}^n \sum_{j=1}^M \left[ 1(A_i=1)Y_i  \cdot \frac{1}{\hat{\pi}_A(\tilde{X}_{i}^{(j)})}\right].
\end{align*}

Similar logic implies that

\begin{align*}
	&\frac{1}{n}\sum_{i=1}^n \left[ \sum_r \sum_a 1(R_i=r)1(A_i=a) \hat{M}_{m,r,a}(X_{R_i,i}, Y_i) \right] \\
	&\quad = \frac{1}{n}\sum_{i=1}^n \left[ \sum_a 1(A_i=a) \frac{1}{M}\sum_{j=1}^M \hat{m}_1(\tilde{X}_{i}^{(j)}) \right] \\
	&\quad = \frac{1}{Mn}\sum_{i=1}^n \sum_{j=1}^M \hat{m}_1(\tilde{X}_{i}^{(j)}), \\
	&\frac{1}{n}\sum_{i=1}^n \left[1(A_i=1) \sum_r  1(R_i=r) \hat{M}_{m/\pi_A,r,a}(X_{R_i,i}, Y_i) \right] \\
	&\quad = \frac{1}{n}\sum_{i=1}^n \left[1(A_i=1) \frac{1}{M}\sum_{j=1}^M \frac{\hat{m}_1(\tilde{X}_{i}^{(j)})}{\hat{\pi}_A(\tilde{X}_{i}^{(j)})} \right] \\
	&\quad = \frac{1}{Mn}\sum_{i=1}^n \sum_{j=1}^M 1(A_i=1) \frac{\hat{m}_1(\tilde{X}_{i}^{(j)})}{\hat{\pi}_A(\tilde{X}_{i}^{(j)})}.
\end{align*}

With these equations, we can construct the (RA-$R$, RA-$A$) and (RA-$R$, DR-$A$) estimators.

{\bf IPW-A component.}
The propensity score can be calculated via
\begin{align*}
	\pi_A(x) = P(A=1|x_1,x_2) &= \frac{\sum_r \sum_{y=0}^1 p(x_1,x_2|r,y,A=1)p(r,y,A=1)}{\sum_r \sum_{y=0}^1 \sum_{a=0}^1 p(x_1,x_2|r,y,a)p(r,y,a)}.
\end{align*}

We describe how to estimate the distribution $\hat{p}(x_1,x_2|r,y,a)$ in the previous discussion on the \textbf{RA-R component}.

{\bf RA-A component.}
Since $Y$ is binary, the regression function can be calculated via
\begin{align*}
	m_a(x) = \E[Y|A=a,x_1,x_2] &= P(Y=1|A=a,x_1,x_2) \\
	&= \frac{\sum_r p(x_1,x_2|r,Y=1,A=a)p(r,Y=1,A=a)}{\sum_r \sum_{y=0}^1 p(x_1,x_2|r,y,a)p(r,y,a)}.
\end{align*}

We describe how to estimate the distribution $\hat{p}(x_1,x_2|r,y,a)$ in the previous discussion on the \textbf{RA-R component}.

\subsubsection{Sensitivity analysis}

We perform a sensitivity analysis for this simulation by varying the sensitivity parameters $\rho\in[-2,0.5]$ and $\xi\in[0,4]$ based on Appendices \ref{appendix:sens-odds} and \ref{appendix:sens:cont}.  For each sensitivity parameter, we generate 1000 data sets with $n=1000$ and perturb the 6 estimators.  We use $M=50$ imputations for the RA estimator.  In keeping with what was done in the primary analysis, the confidence intervals for the IPW-$R$ estimators are constructed by pooling the results across all 1000 iterations and then taking the 0.025\% and 97.5\% quantiles.  This was done to reduce the effect of outliers (created by unstable inverse probability weights) on the standard error estimate.  The confidence intervals for the RA-$R$ estimators are generated by calculating the standard error across all 1000 iterations.  The true ATE is plotted in red.

\begin{figure}[!h]
	\centering
	\includegraphics[scale=0.3]{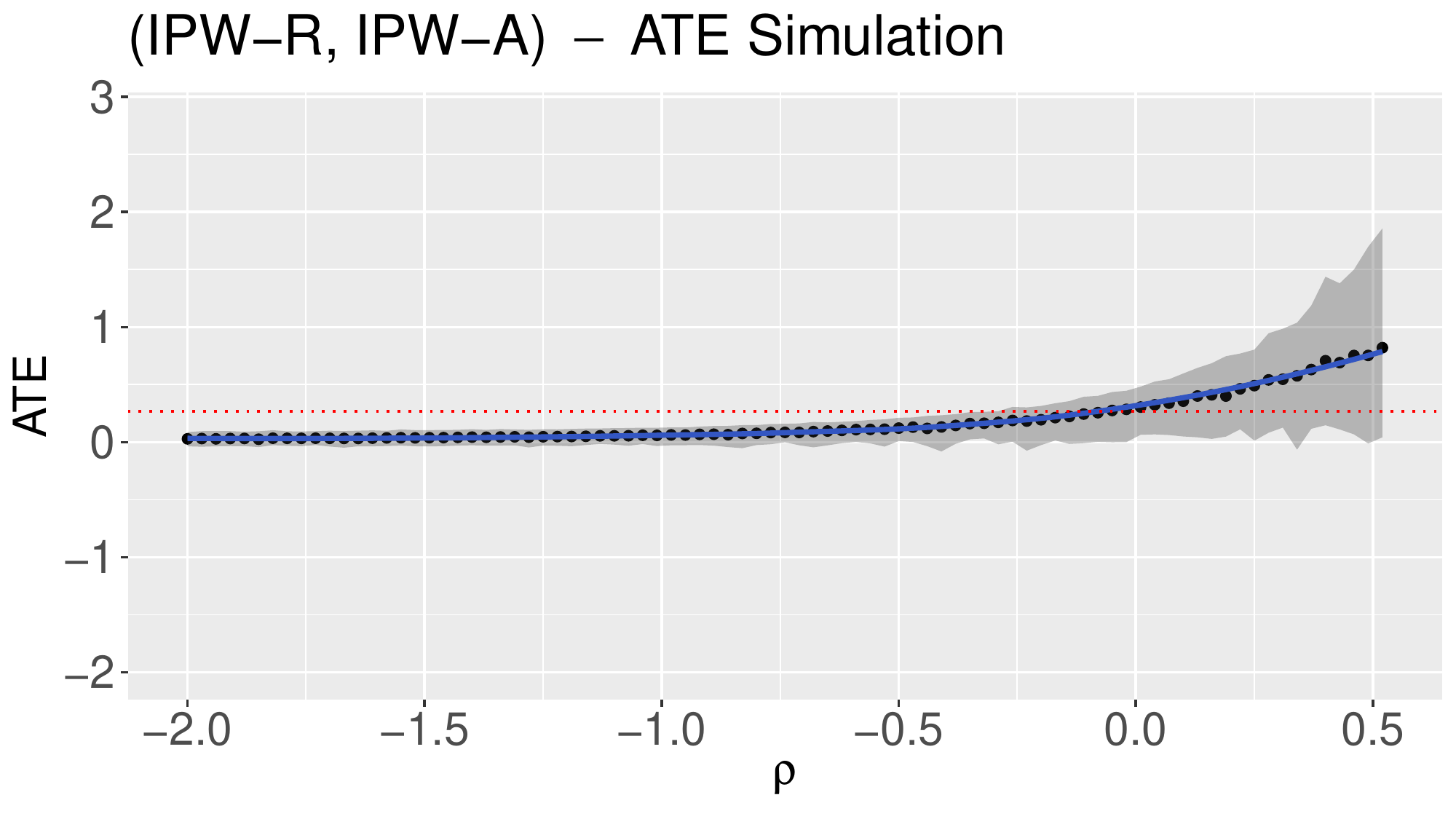}
	\includegraphics[scale=0.3]{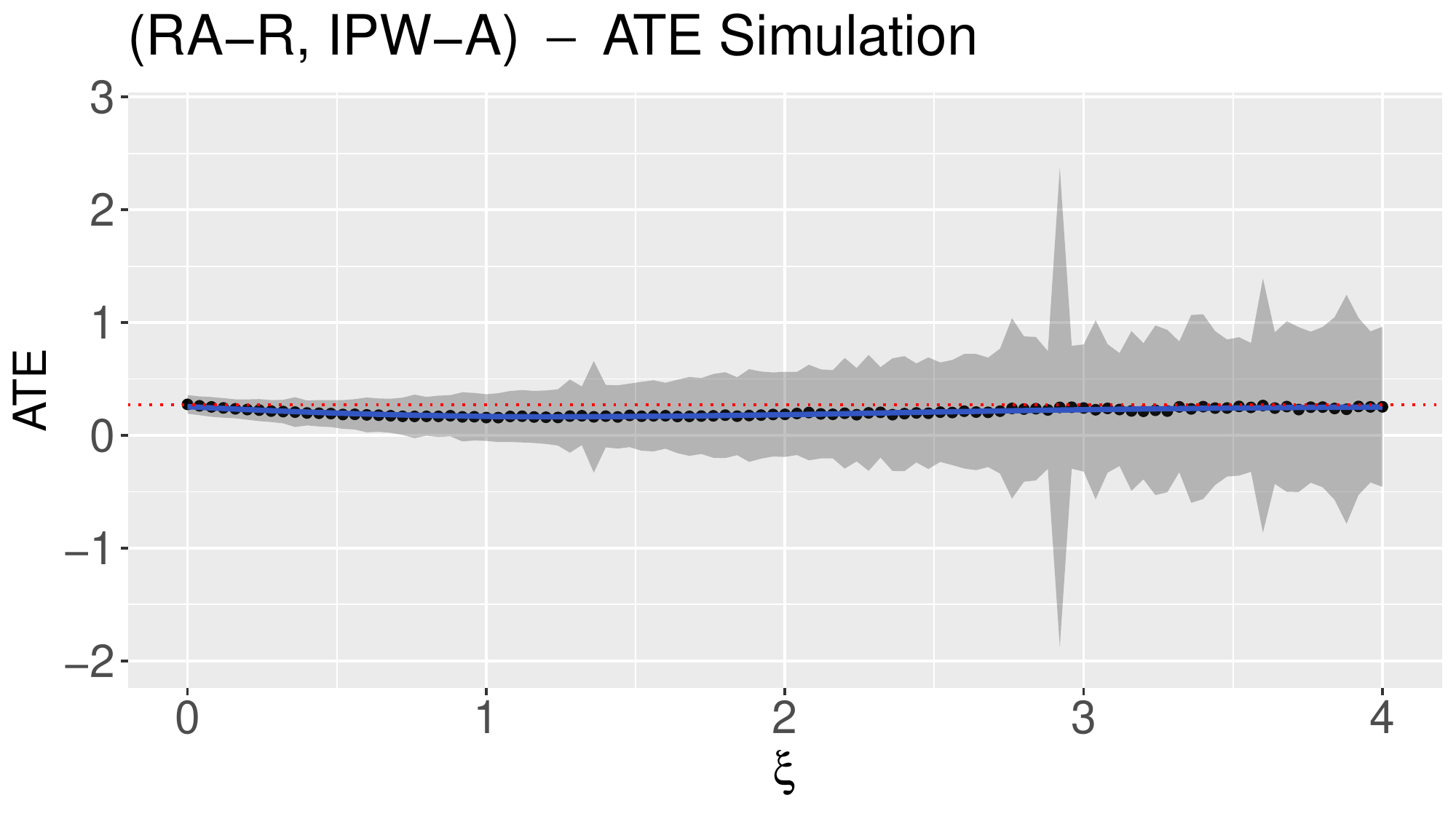}
	\includegraphics[scale=0.3]{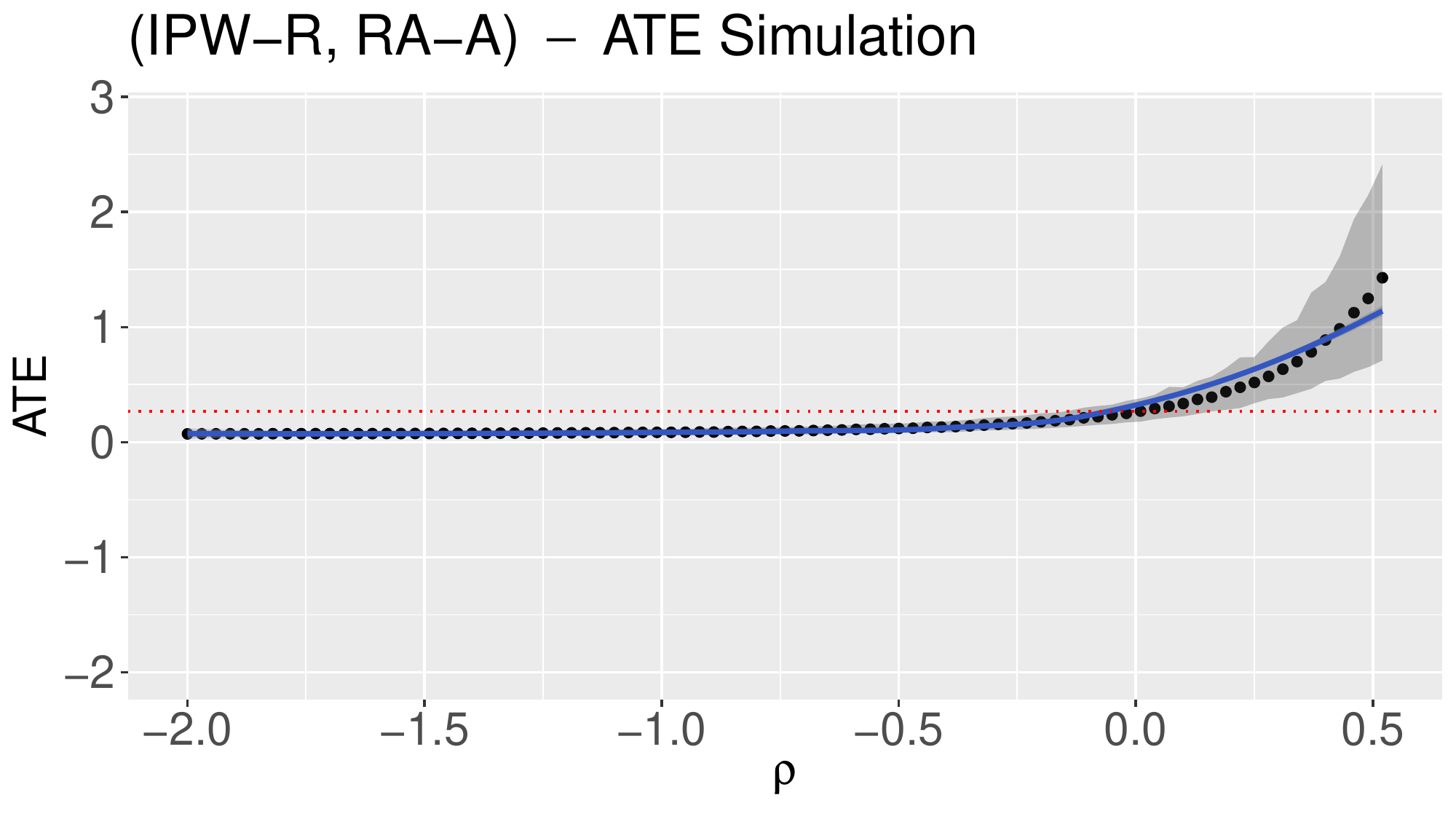}
	\includegraphics[scale=0.3]{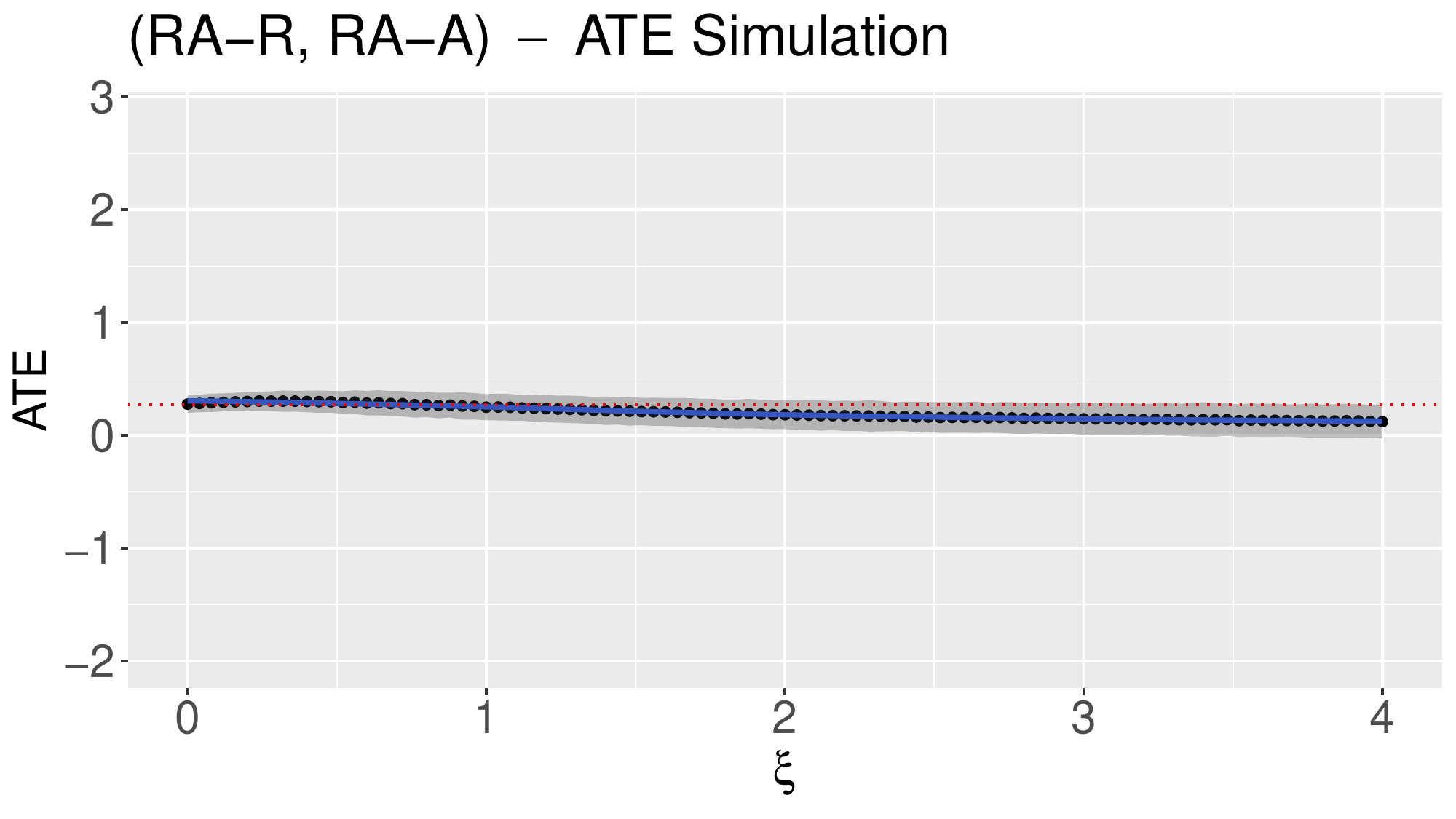}
	\includegraphics[scale=0.3]{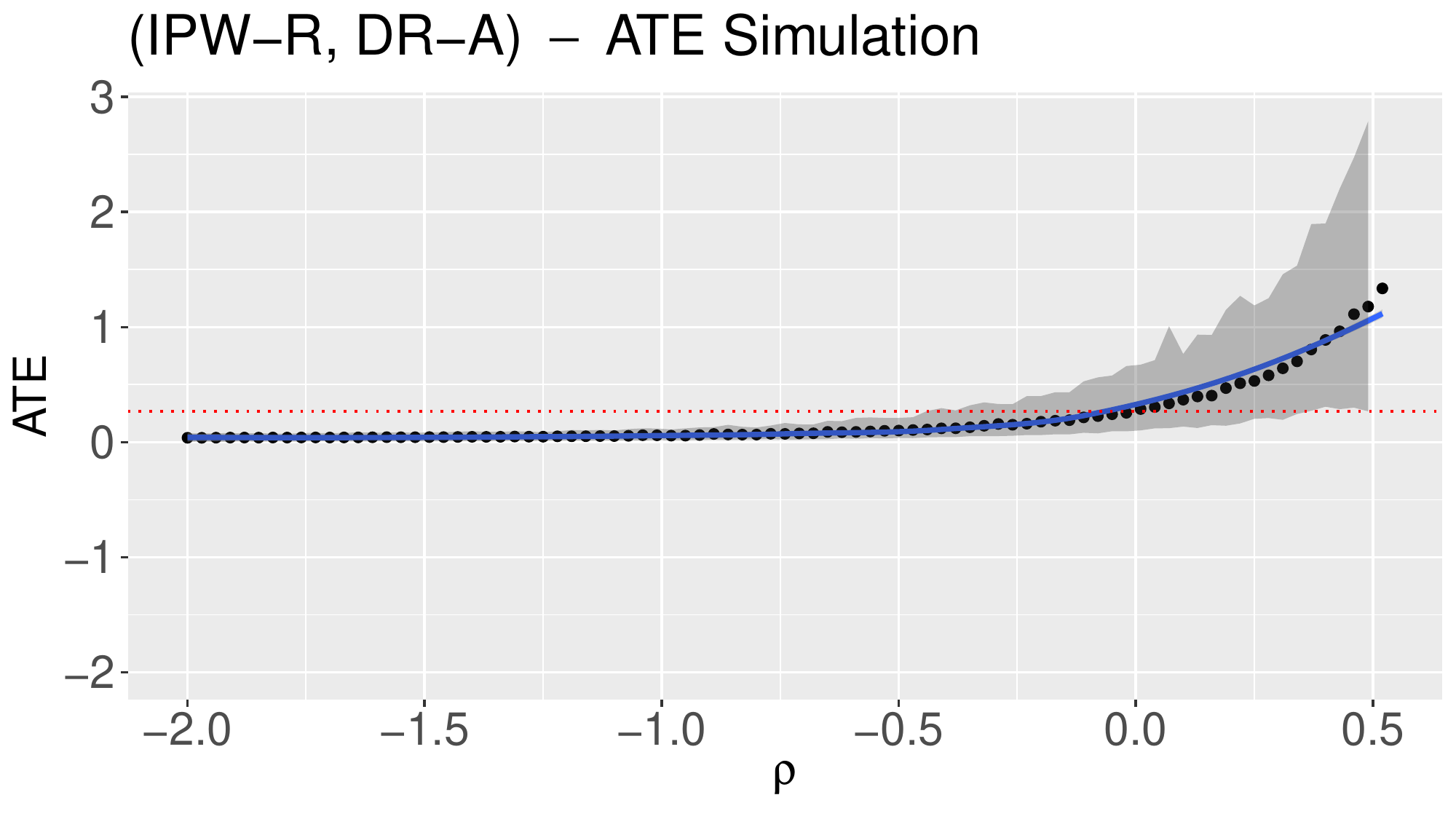}
	\includegraphics[scale=0.3]{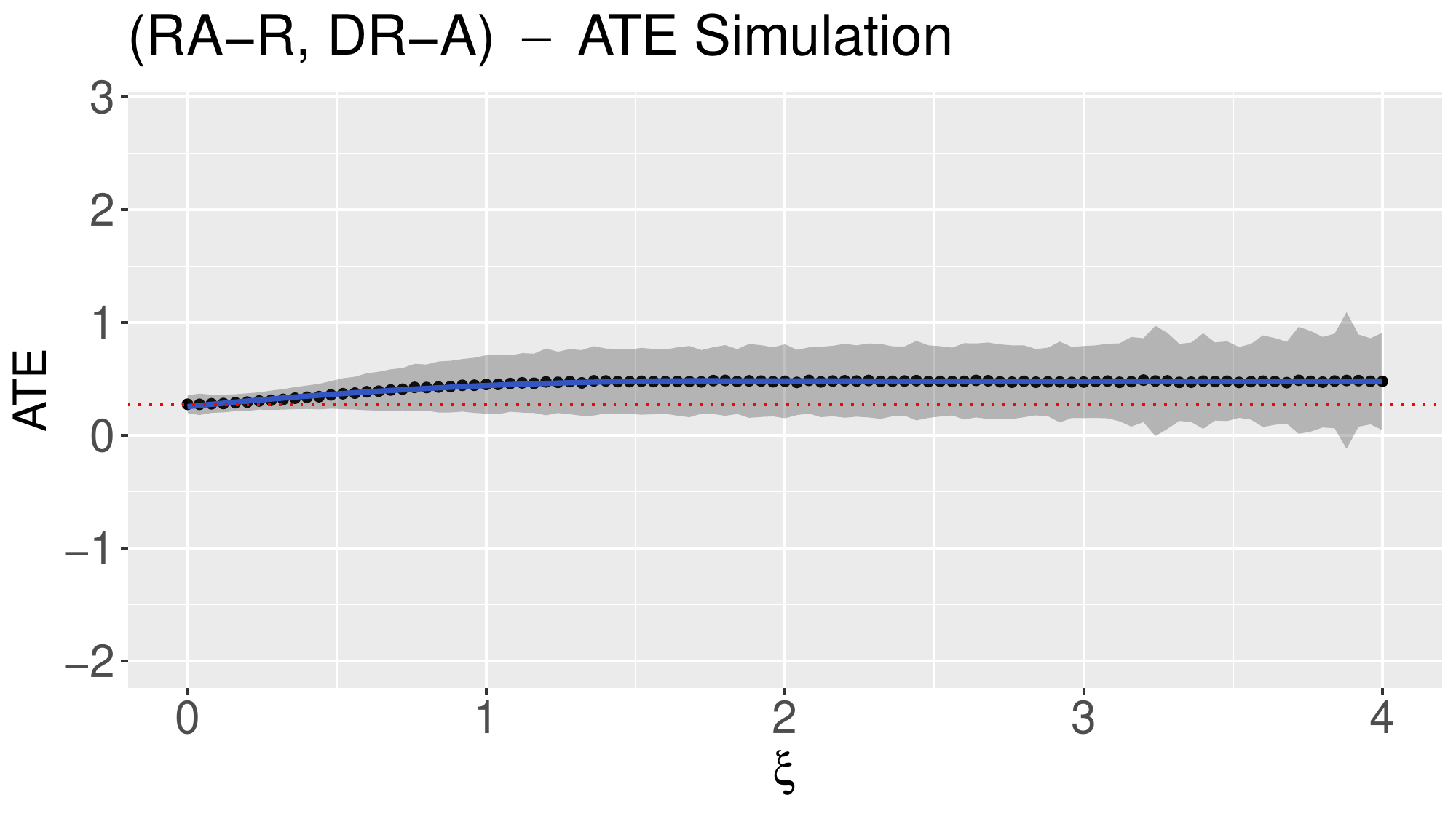}
	\caption{Sensitivity analysis for the Cox model simulation with both the IPW and RA estimators.}
	\label{fig:sensitivity:causal}
\end{figure}

First, we observe that when $\xi$ reaches $4$, we are very close to deterministic imputation, so often in most practical applications, $\xi$ does not need to belong to a large interval.  The estimator where we use RA in both stages (RA-$R$, RA-$A$) performs the best and appears fairly robust to changes in $\xi$.  The (RA-$R$, IPW-$A$) estimator also appears to have minimal bias as we vary $\xi$, but the variance increases.  The IPW estimators are fairly sensitive to changes in $\rho$, and in particular, the variance can severely increase for moderate changes in $\rho$.  This sensitivity analysis suggests that RA tends to provide the most stable estimates when the assumptions are perturbed.  Intuitively, this makes sense because a perturbation in the missing data assumption changes the complete odds, and therefore, can drastically inflate the inverse probability weights in some scenarios.



\section{Additional analysis of the NACC data}

\label{appendix:data}

We performed sensitivity analysis under the IPW and RA settings by perturbing both the IPW and RA estimators with sensitivity parameters $\rho\in[-1,1]$ and $\zeta\in[0,1]$ (since the covariates in the NACC data are binary). 
The details for  sensitivity analysis are outlined Appendices \ref{appendix:sens-odds} and \ref{appendix:sens-binary-imp}.

\begin{figure}
	\centering
	\includegraphics[scale=0.3]{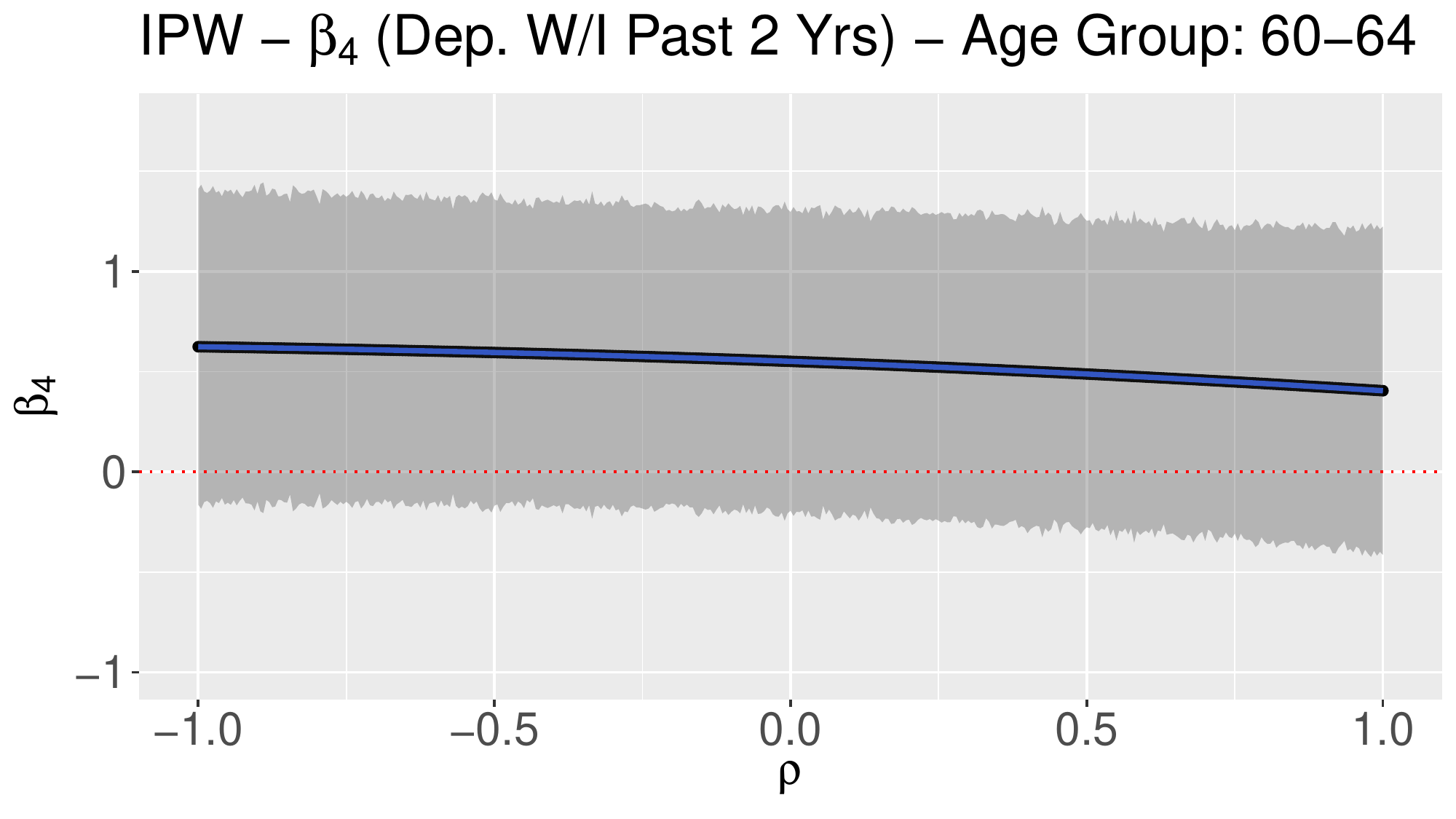}
	\includegraphics[scale=0.3]{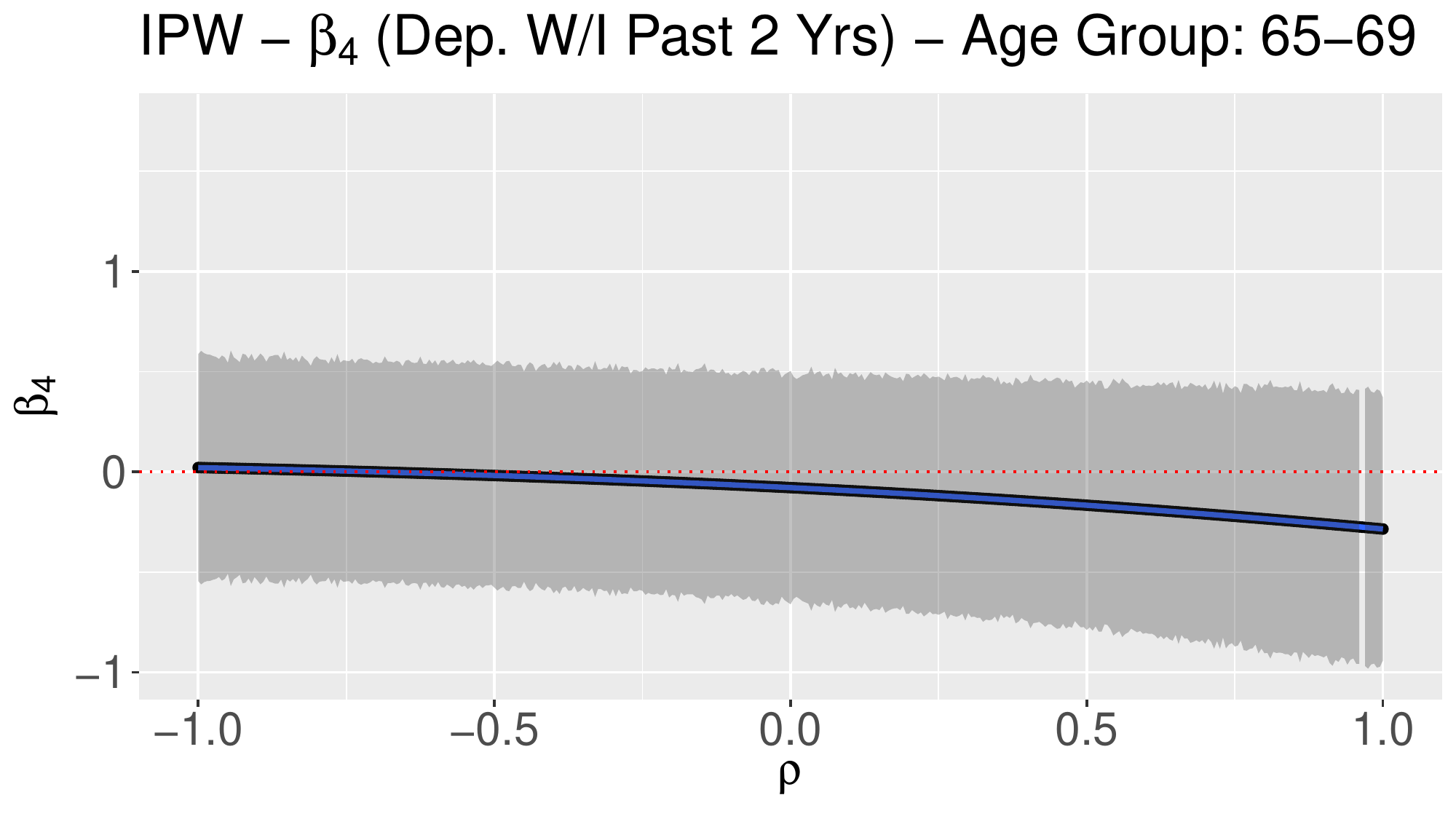}
	\includegraphics[scale=0.3]{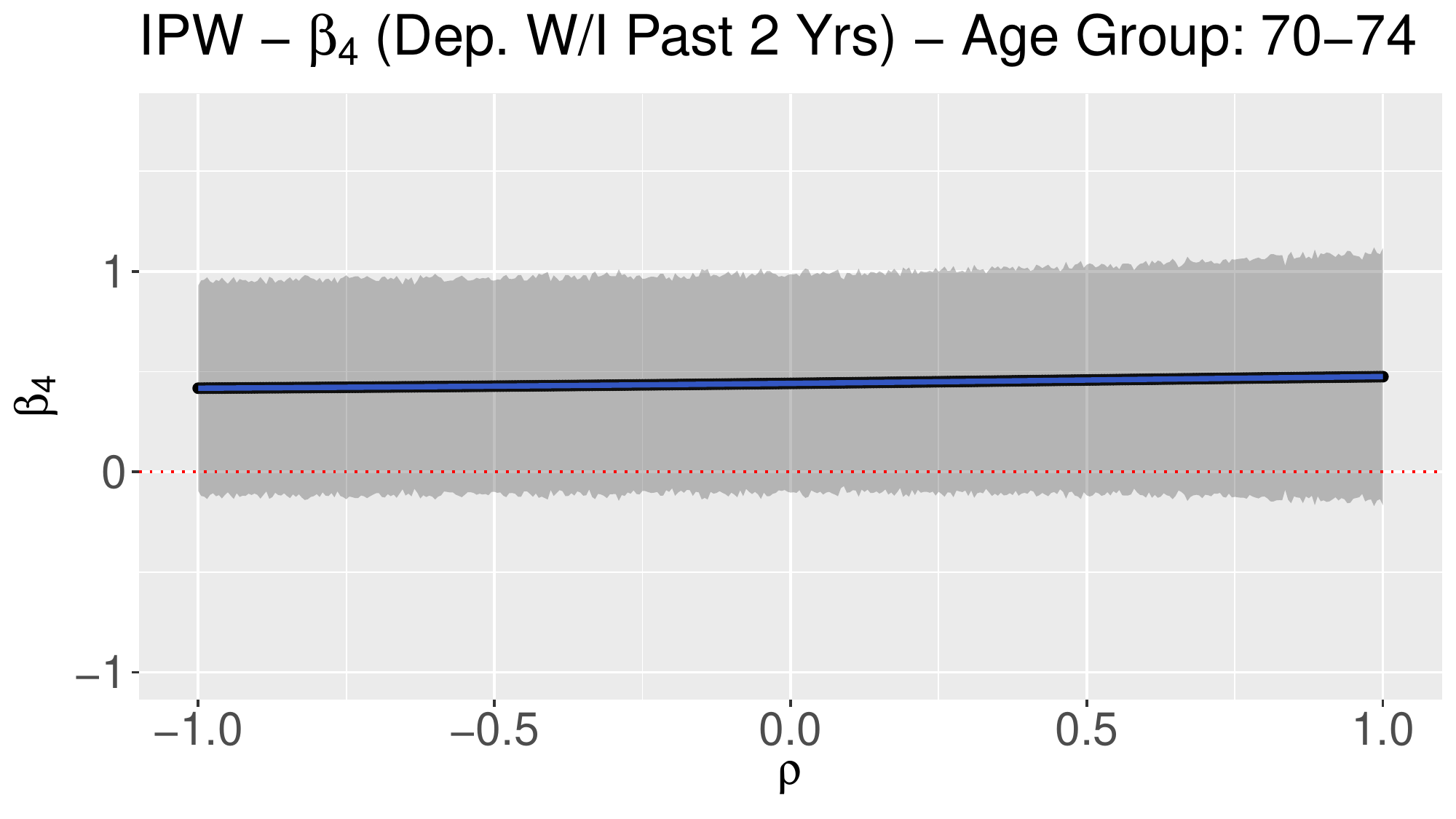}
	\caption{Sensitivity analysis results for the IPW estimator for $\beta_4$.}
	\label{fig:SA:beta4ipw}
\end{figure}

\begin{figure}
	\centering
	\includegraphics[scale=0.3]{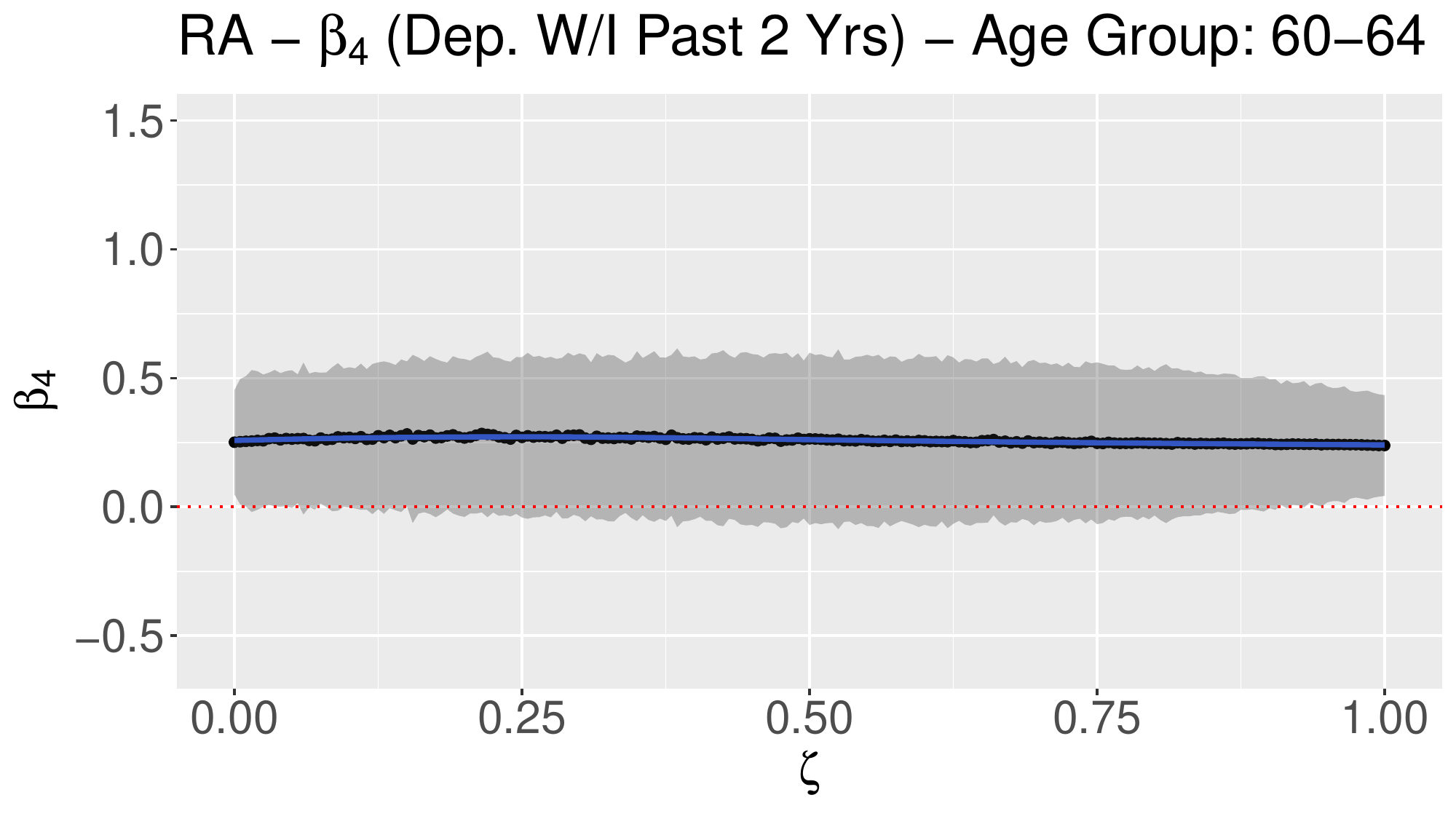}
	\includegraphics[scale=0.3]{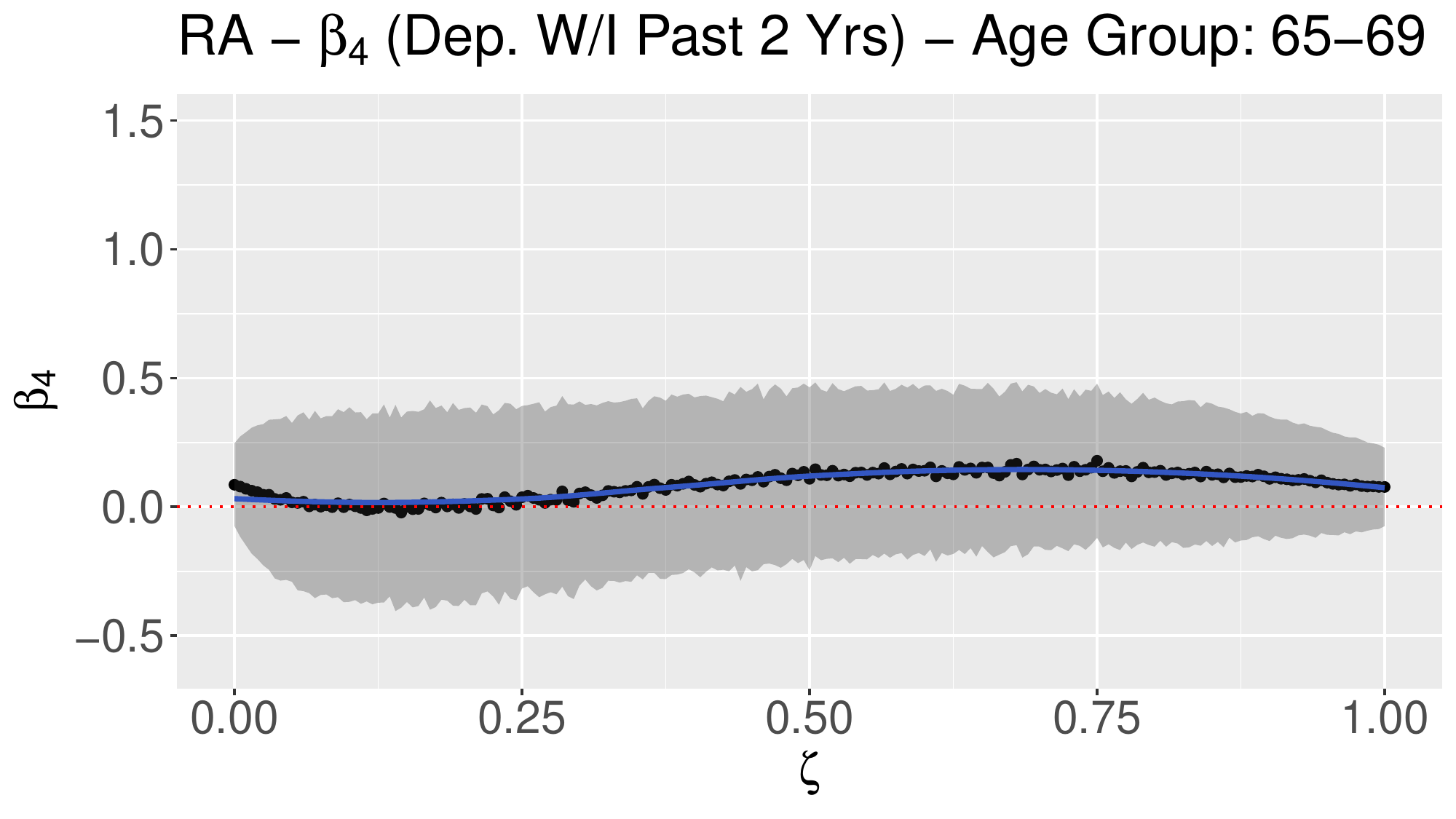}
	\includegraphics[scale=0.3]{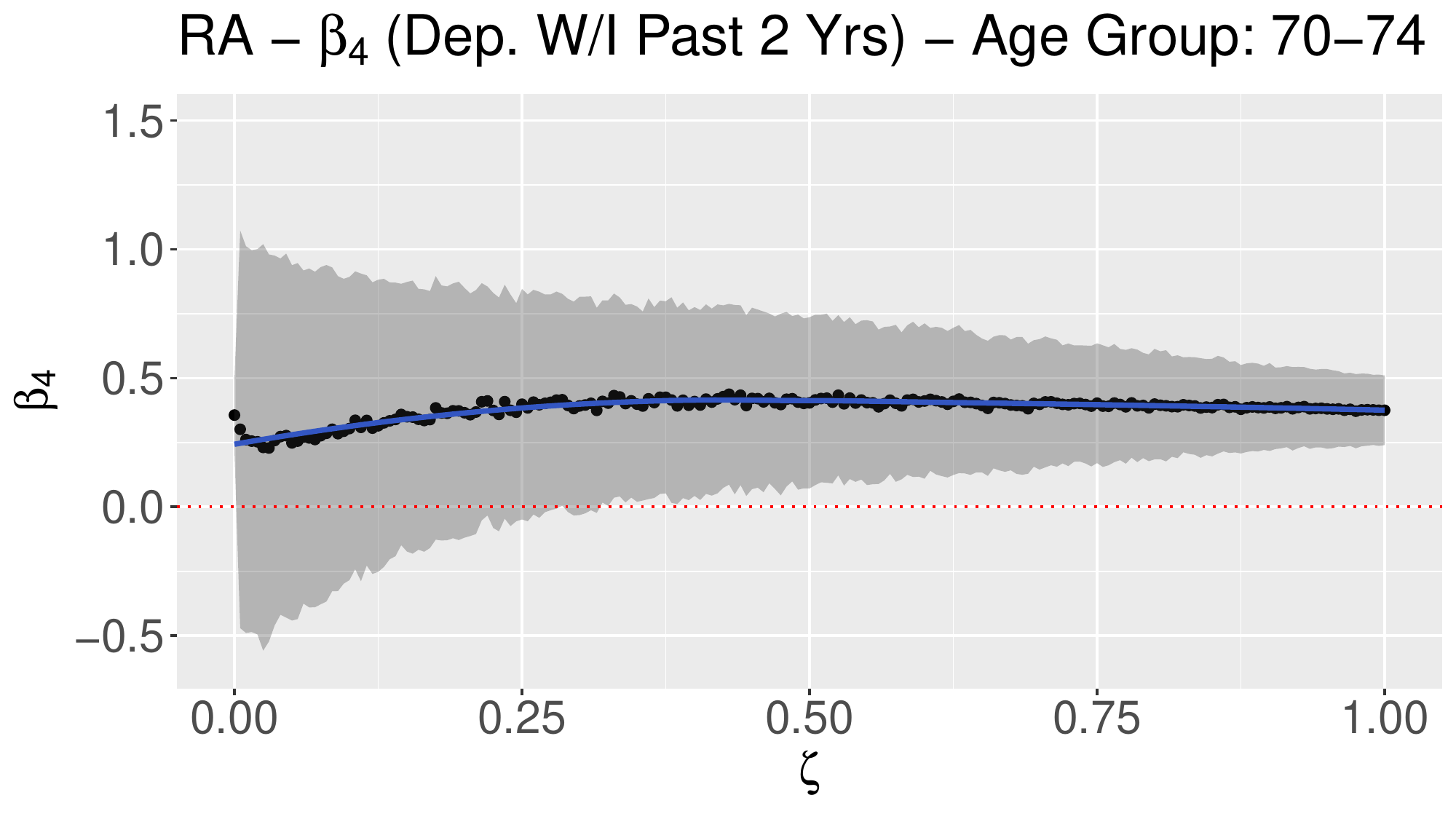}
	\caption{Sensitivity analysis results for the RA estimator for $\beta_4$.}
	\label{fig:SA:beta4ra}
\end{figure}

We include the results of all sensitivity analysis of the depression and anxiety in this section.  
Figures~\ref{fig:SA:beta4ipw} and \ref{fig:SA:beta4ra} show that for three age groups ($60-64$, $65-69$, and $70-74$), the point estimates given by the IPW and RA estimators are generally fairly stable even as we vary the sensitivity parameters.  Moreover, the confidence intervals for the age groups $60-64$ and $70-74$ for both the IPW and RA estimators either exhibit significance or have endpoints close to $0$ across much of the sensitivity parameters' ranges.  This preliminary analysis suggests that at least for some age groups, recent depression is associated with shorter time to dementia.  There is evidence in the literature to support this claim as well \citep{Kitching2015depression}.

\begin{figure}
	\centering
	\includegraphics[scale=0.3]{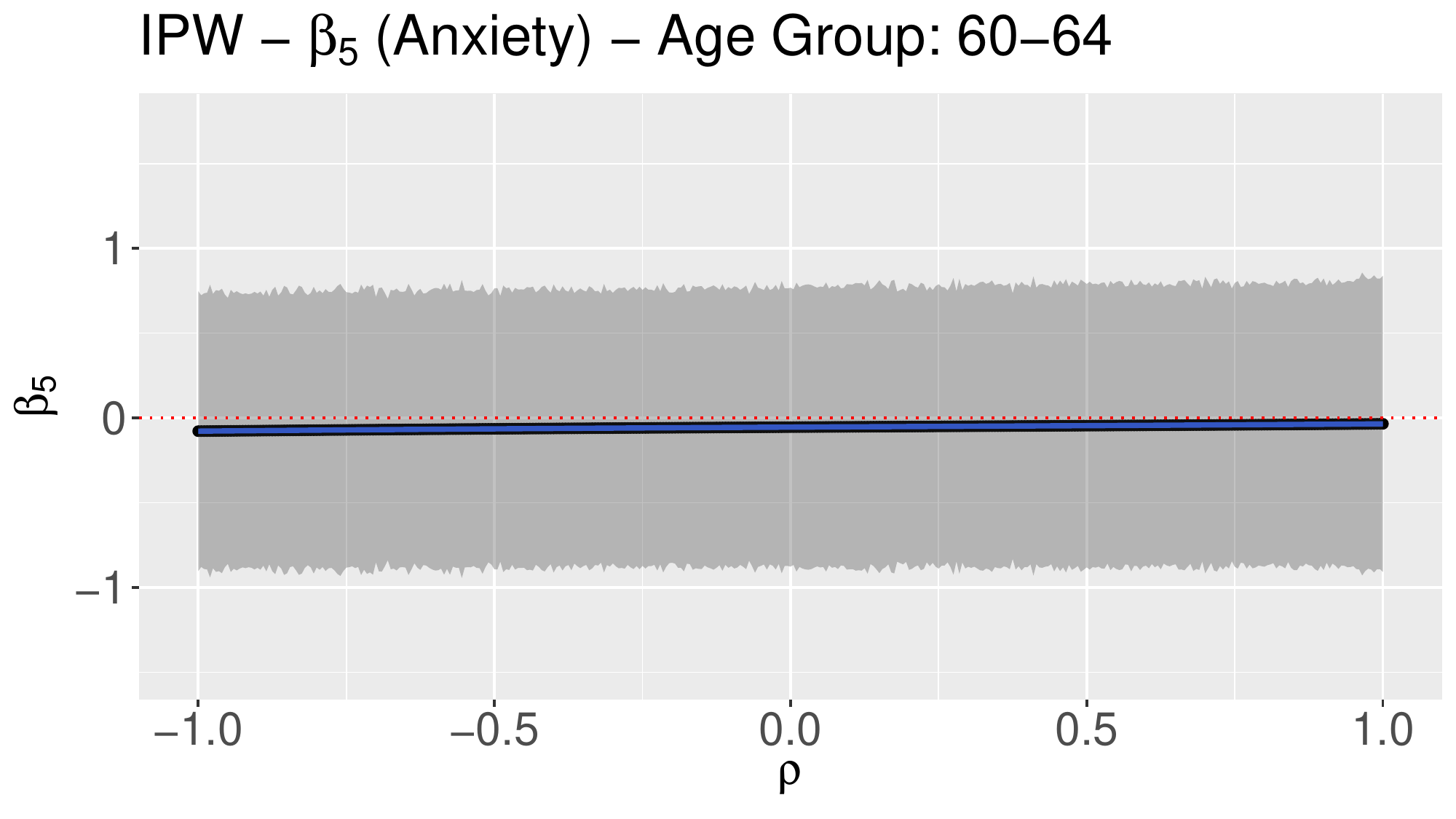}
	\includegraphics[scale=0.3]{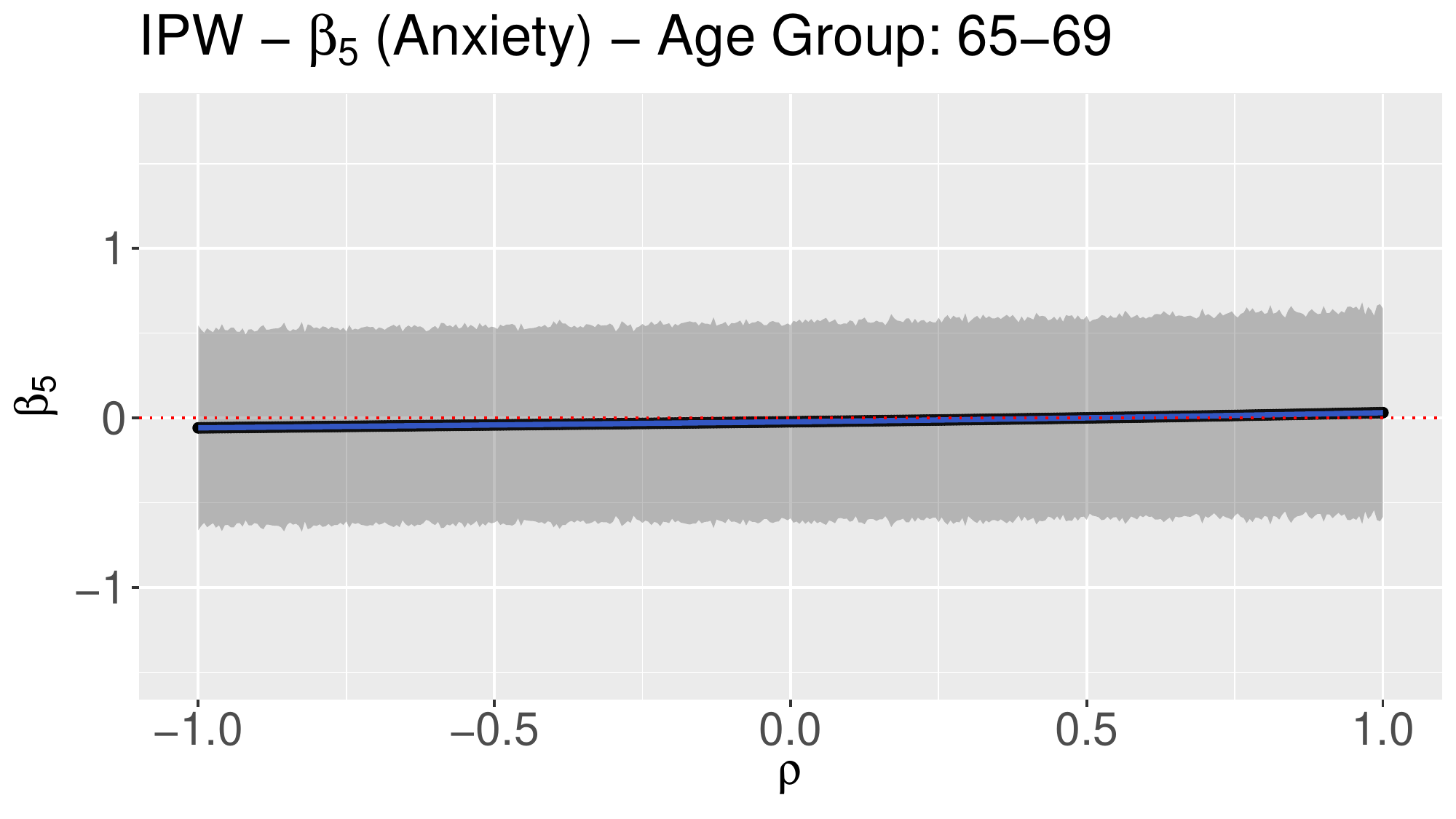}
	\includegraphics[scale=0.3]{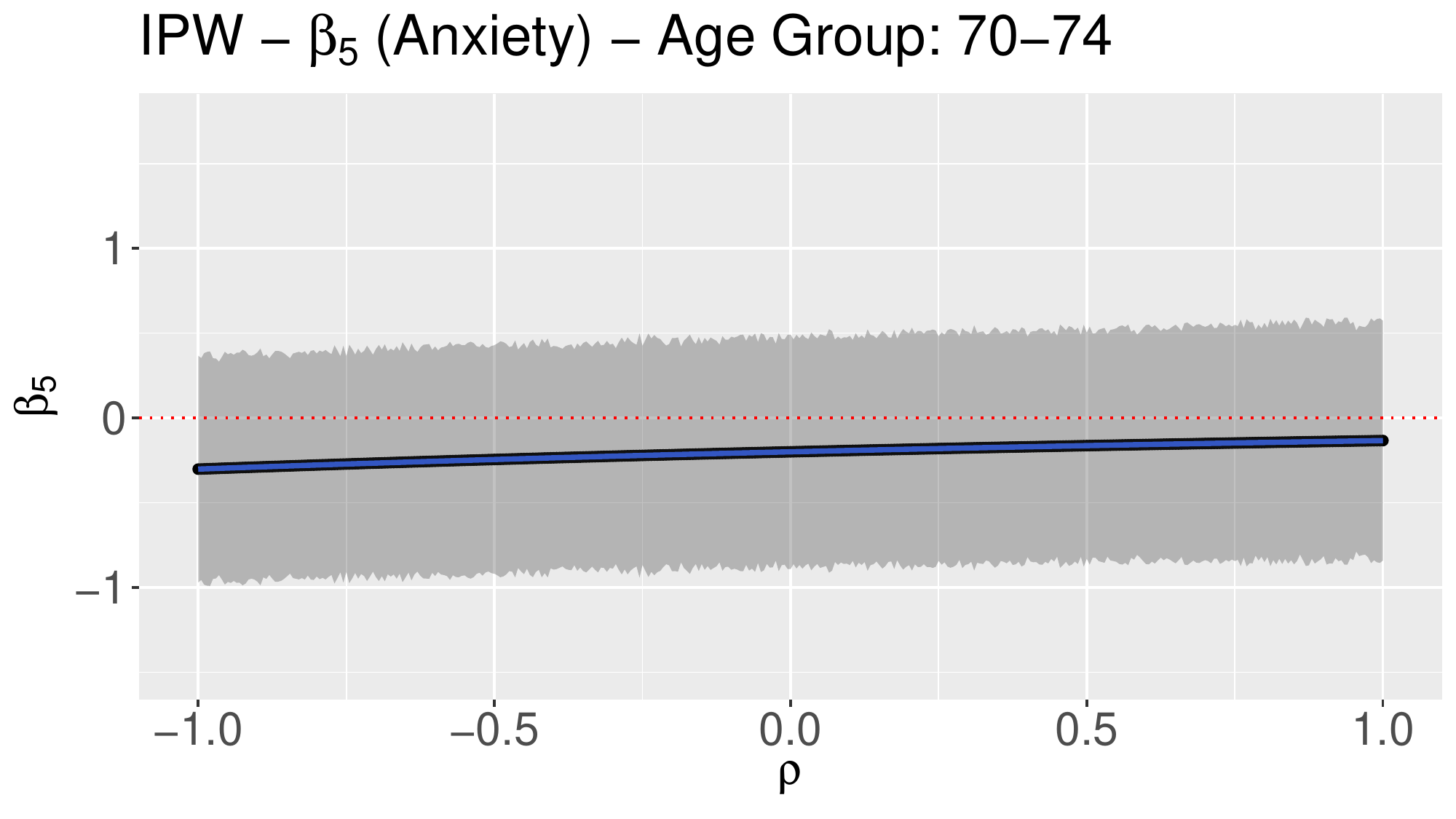}
	\includegraphics[scale=0.3]{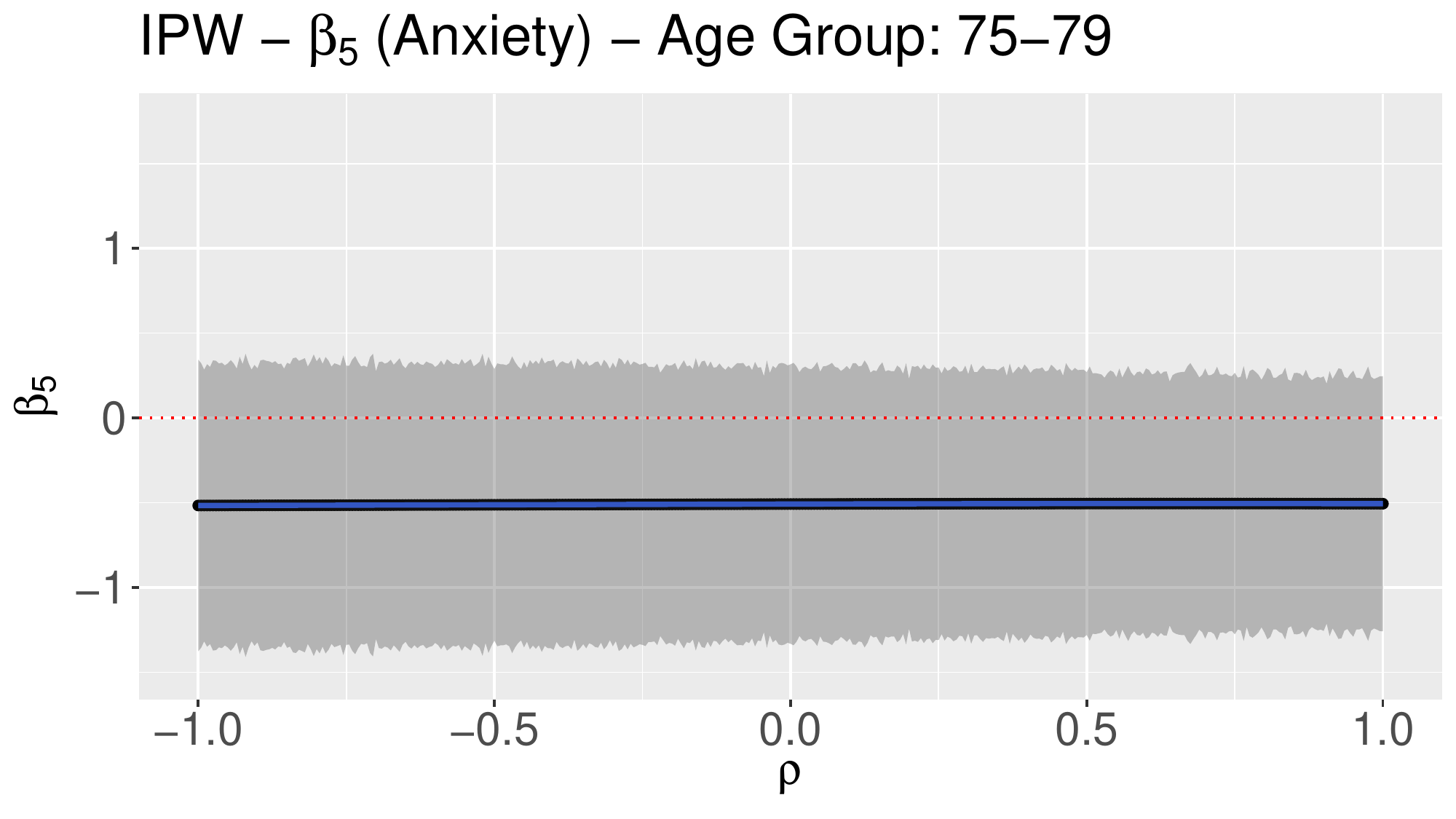}
	\includegraphics[scale=0.3]{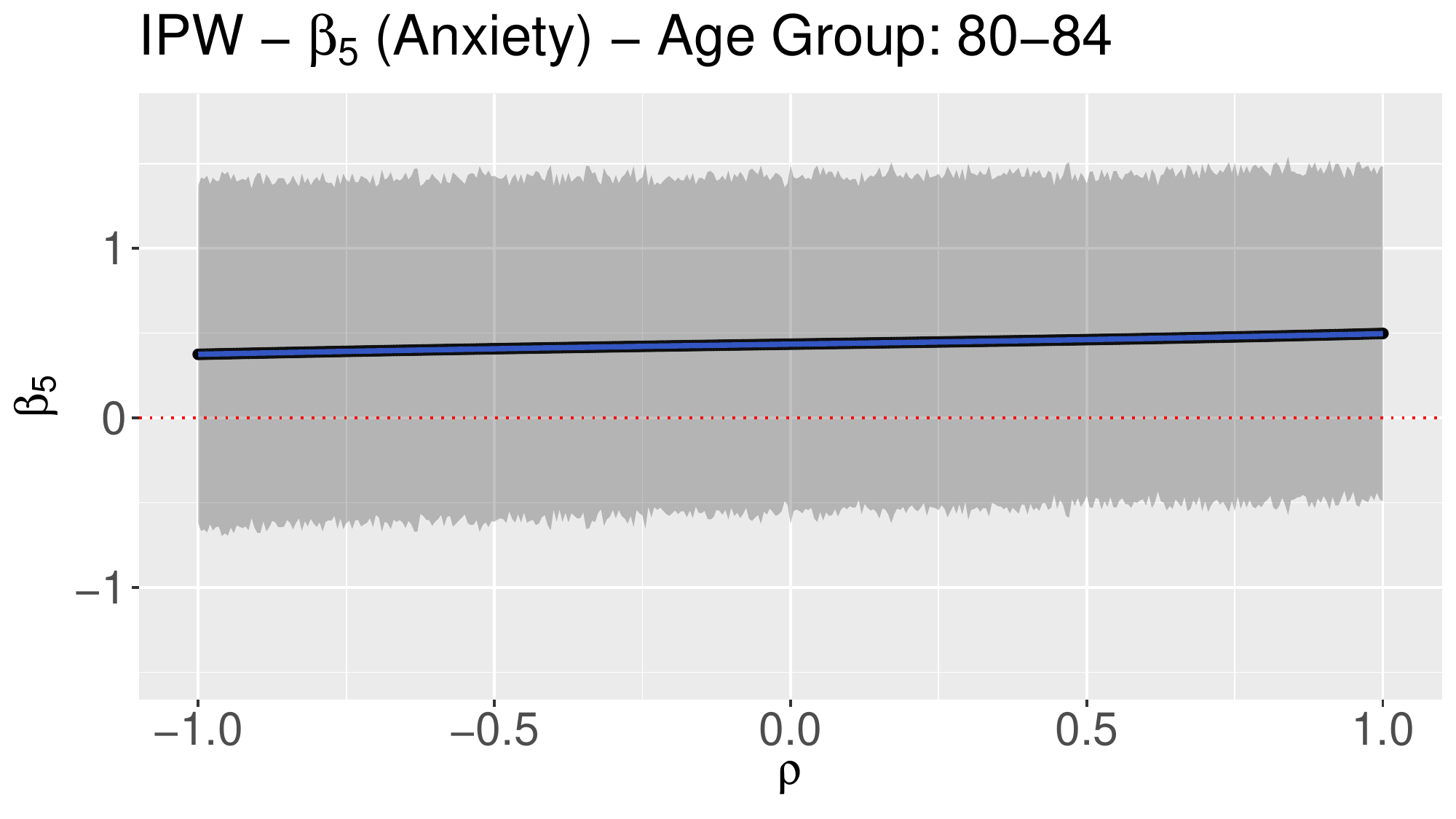}
	\caption{Sensitivity analysis results for the IPW estimator for $\beta_5$.}
	\label{fig:SA:beta5ipw}
\end{figure}

\begin{figure}
	\centering
	\includegraphics[scale=0.3]{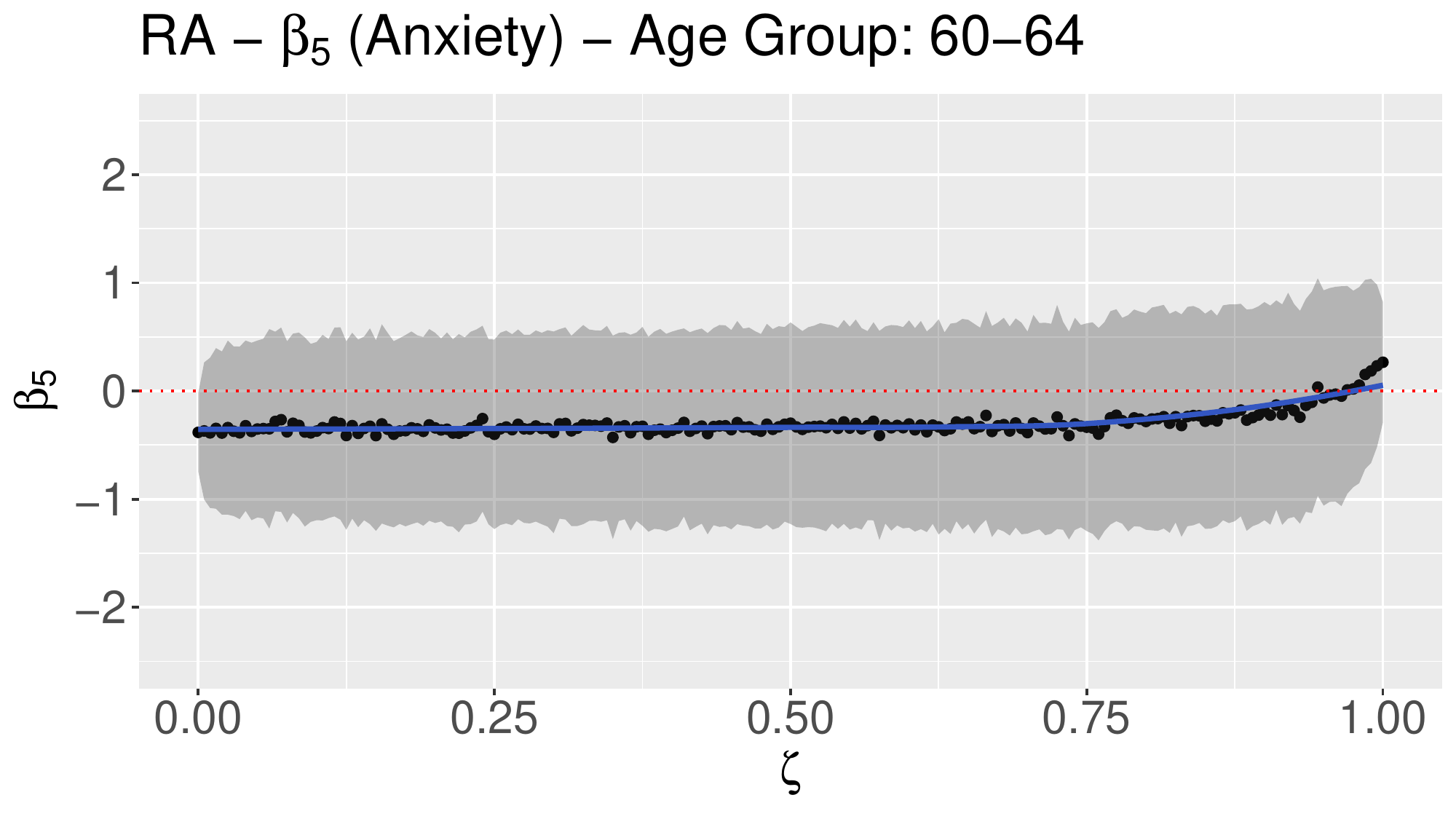}
	\includegraphics[scale=0.3]{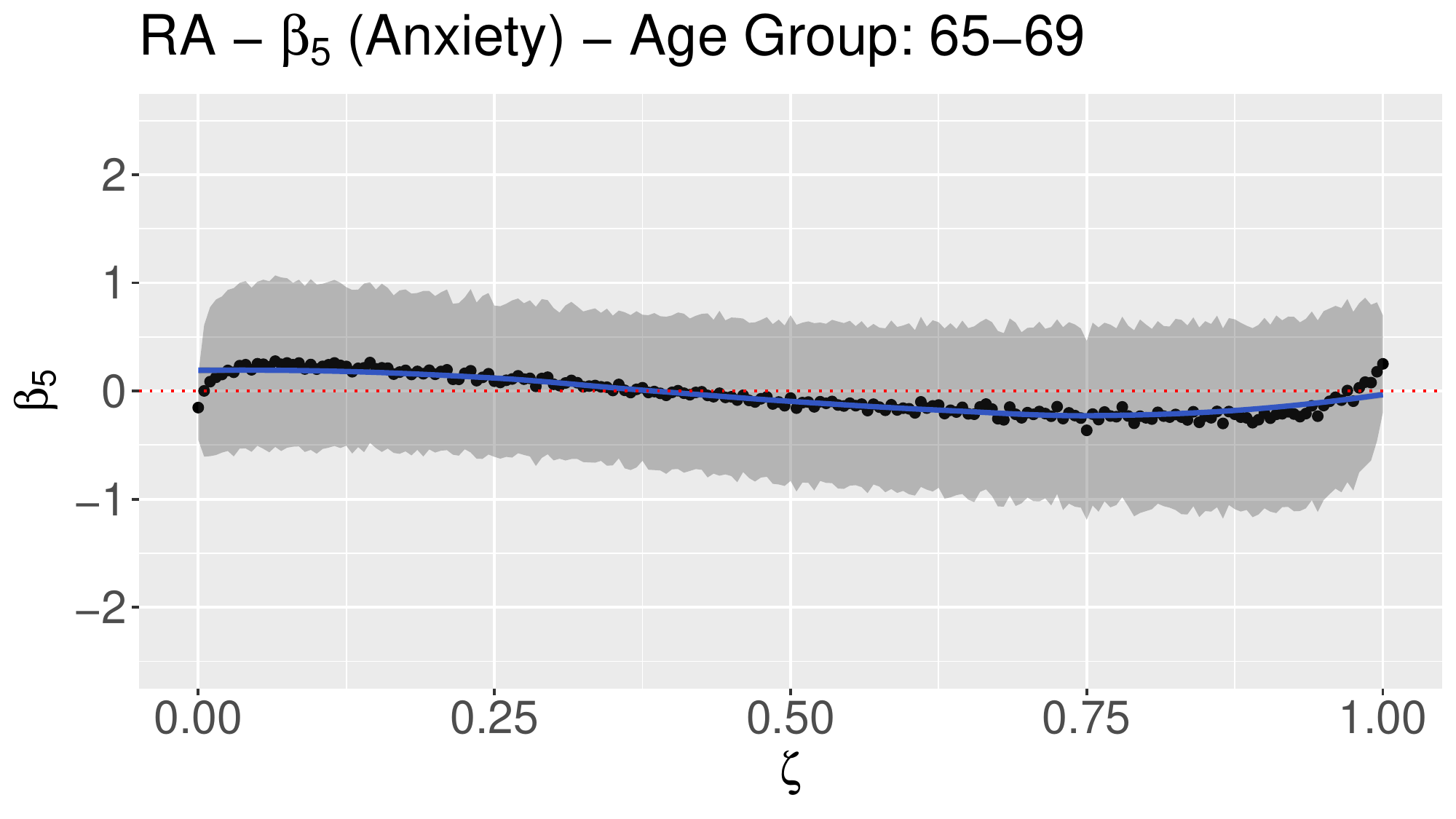}
	\includegraphics[scale=0.3]{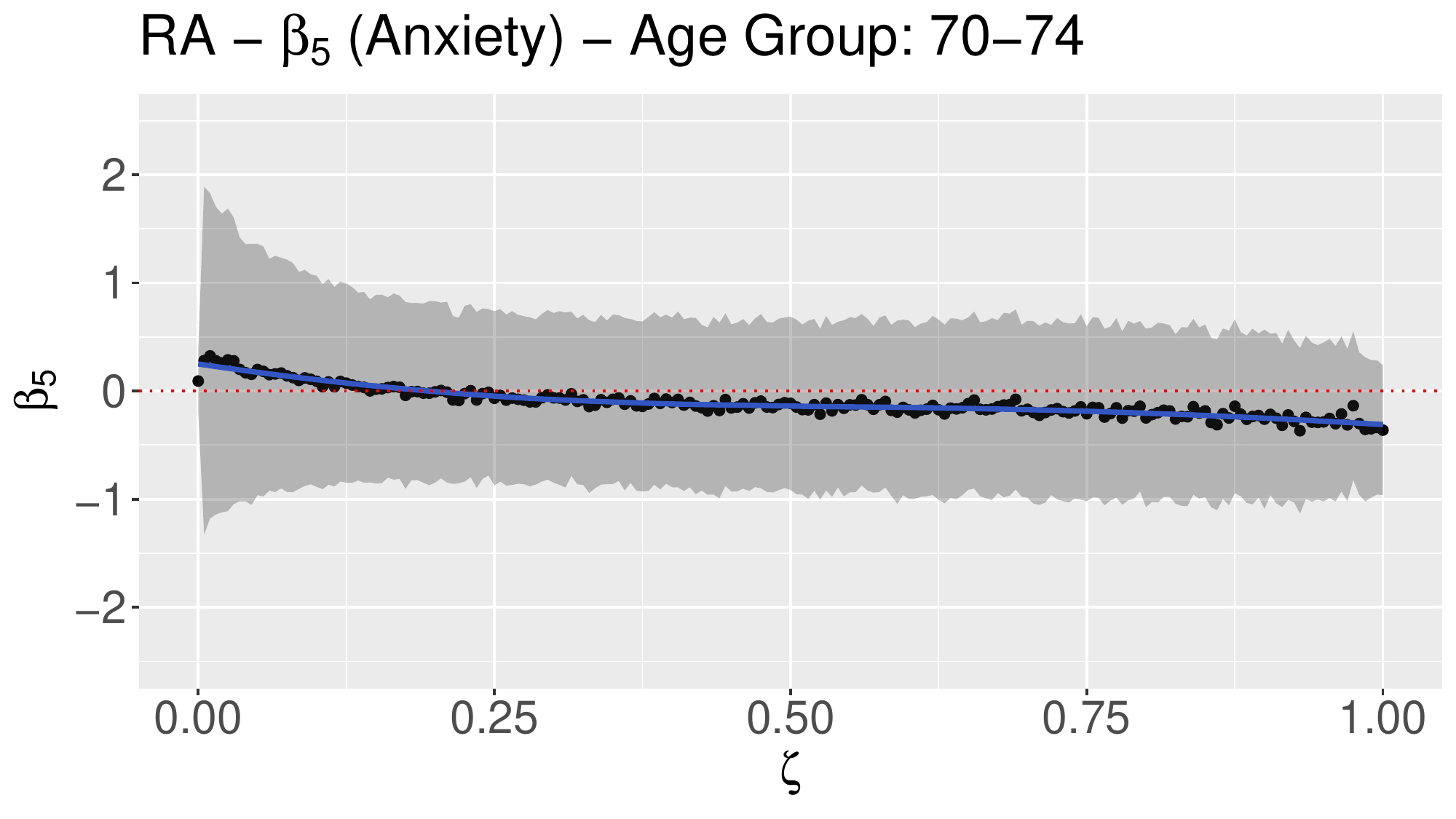}
	\includegraphics[scale=0.3]{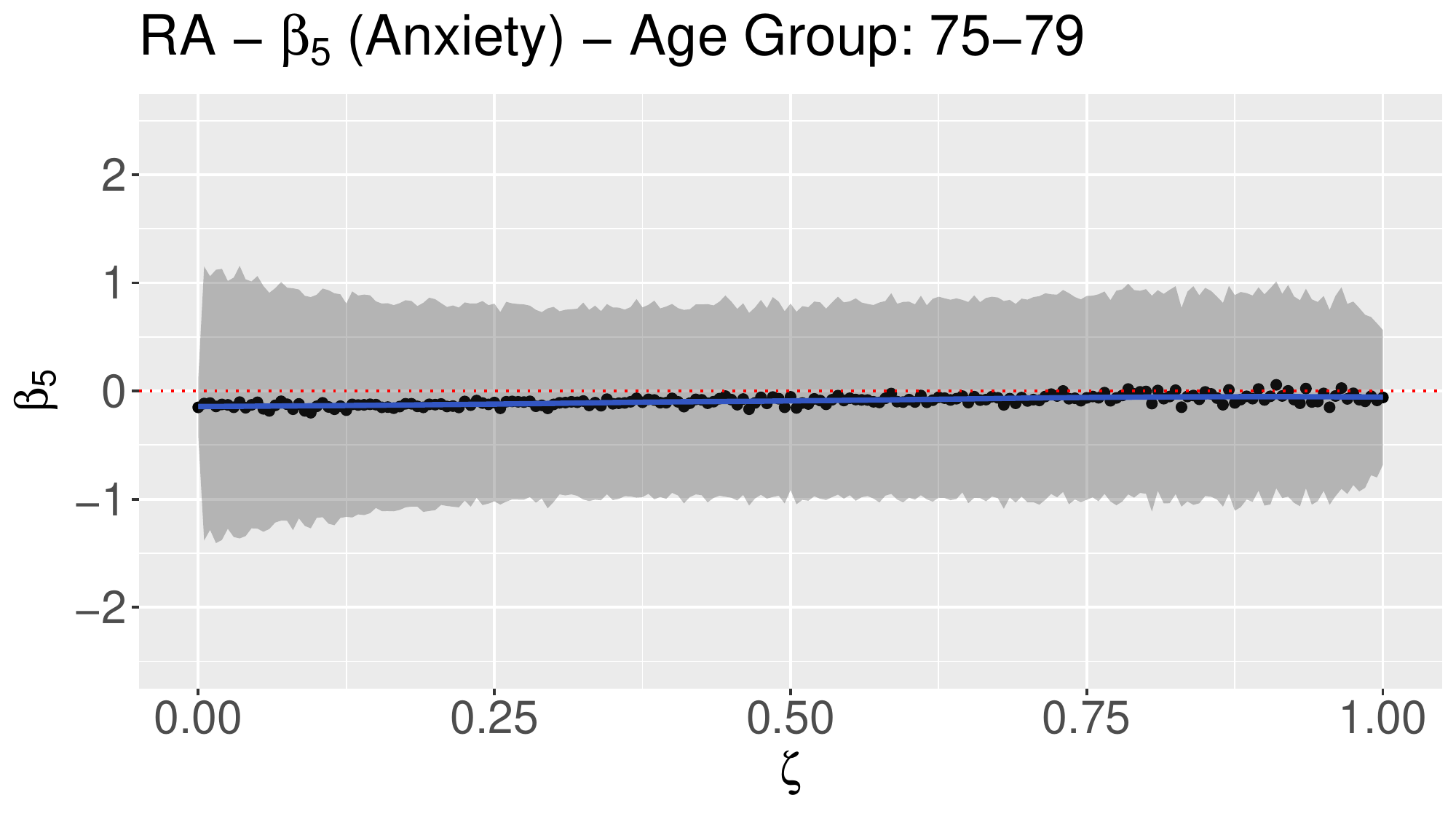}
	\includegraphics[scale=0.3]{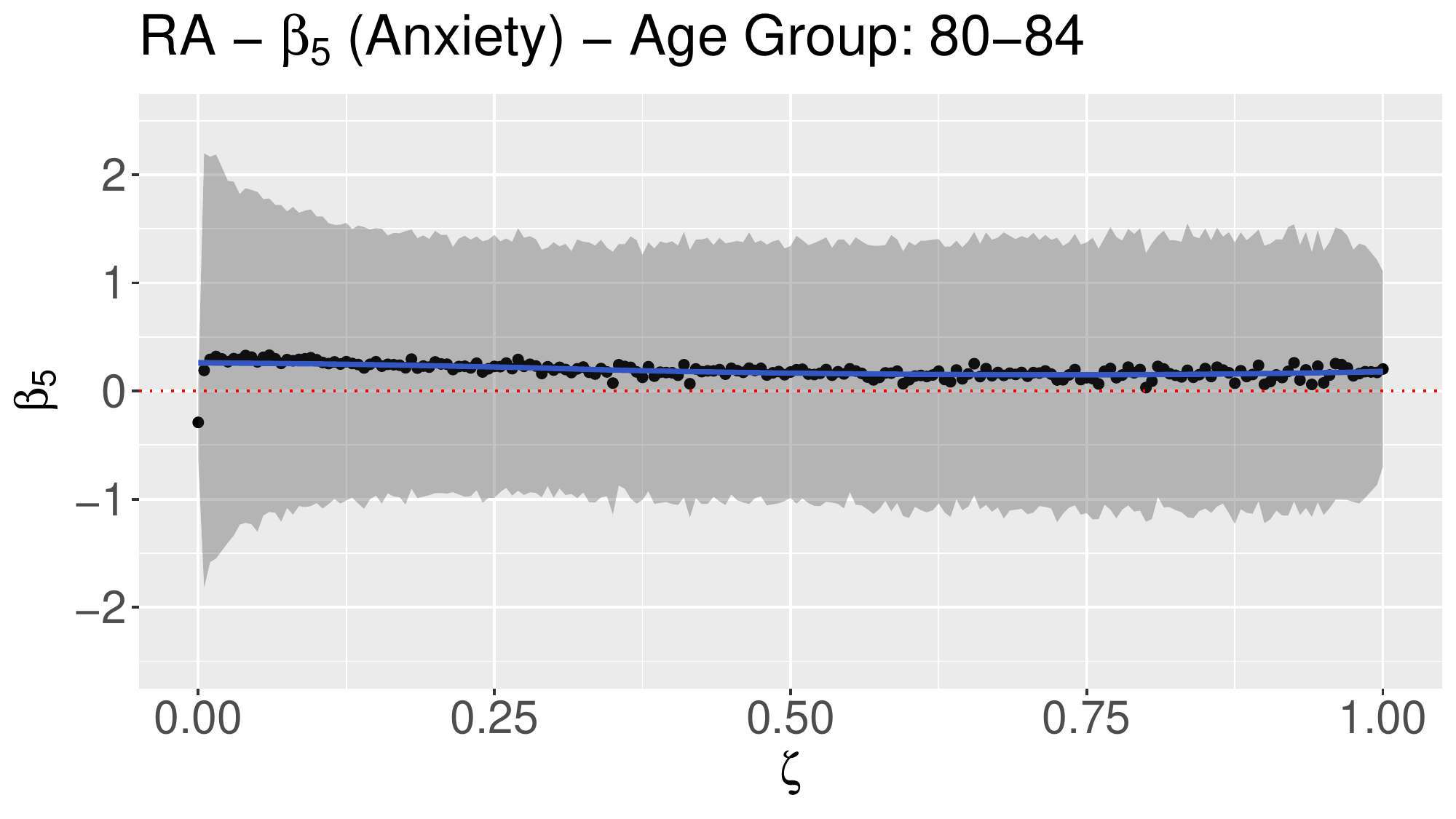}
	\caption{Sensitivity analysis results for the RA estimator for $\beta_5$.}
	\label{fig:SA:beta5ra}
\end{figure}

Figures \ref{fig:SA:beta5ipw} and \ref{fig:SA:beta5ra}
display the results for anxiety. 
We see across all age groups, as we vary the respective sensitive parameters $\rho$ and $\zeta$, the point estimates for $\beta_5$ are all close to $0$ with all of the confidence intervals covering $0$.  This suggests some evidence that the self-reported anxiety is not associated with shorter time to dementia.

\section{Proofs}

\label{appendix:proofs}

The proofs of the results in the paper are provided here.  We separate the proofs into three sections by topic: 1. estimator construction and identification, 2. multiple robustness and efficiency theory, and 3. asymptotic theory.

\subsection{Estimator construction and identification}	\label{app:ID}
This section contains proofs of Lemma \ref{lemma:completeodds}, Proposition \ref{prop:odds-propensity}, Proposition \ref{prop:oddsCCMV}, Theorem \ref{thm:CCMV_ID}, Lemma \ref{lemma:ra}.

\begin{proof}[ of Lemma \ref{lemma:completeodds}]
	Let the observed outcome pattern $a$ be fixed.  We have
	\begin{align*}
		&\E[f_a(X,W_a) 1(A=a)1(R=r)] \\
		&= \int f_a(x,w_a) 1(a'=a) 1(r'=r) \ p(x,r',w_a,a') \ dx dr' dw_a da' \\
		&= \int f_a(x,w_a) 1(a'=a) \ p(x,r,w_a,a') \ dx dw_a da' \\
		&= \int f_a(x,w_a) 1(a'=a) \ \frac{p(x,r,w_a,a')}{p(x,1_d,w_a,a')} \ p(x,1_d,w_a,a') \ dx dw_a da'  \\
		&= \int f_a(x,w_a) 1(a'=a) \frac{P(R=r|x,w_a,a')}{P(R=1_d|x,w_a,a')} \nonumber \\
		&\qquad \times 1(s=1_d) \ p(x,s,w_a,a') \ dx ds dw_a da' \\
		&= \E[f_a(X,W_a)1(A=a) O_{r,a}(X,W_a) 1(R=1_d)],
	\end{align*}	
	as desired.
	
	Finally, note that summing over all patterns $r$ yields the second decomposition
	\begin{equation*}
		\E[f_a(X,W_a)1(A=a)] = \sum_r \E[f_a(X,W_a)1(A=a)O_{r,a}(X,W_a)1(R=1_d)].
	\end{equation*}
\end{proof}

\begin{proof}[ of Proposition \ref{prop:odds-propensity}]
	Fix an observed outcome pattern $a$.  We start with the forward direction.  Suppose the complete odds $O_{r,a}(X,W_a)$ is identifiable for all $r$.  Then,
	\begin{equation*}
		\sum_r O_{r,a}(X,W_a) = \sum_r \frac{P(R=r|X,W_a,A=a)}{P(R=1_d|X,W_a,A=a)} = \frac{1}{P(R=1_d|X,W_a,A=a)}
	\end{equation*}
	is identifiable.  The desired selection probabilities can be recovered because
	\begin{equation*}
		P(R=r|X,W_a,A=a) = P(R=1_d|X,W_a,A=a) \cdot O_{r,a}(X,W_a)
	\end{equation*}
	for any covariate missingness pattern $r$.
	
	For the converse, an identifiable set of selection probabilities $P(R=r|X,W_a,A=a)$ for all $r$ immediately implies that the complete odds can be formed and is identifiable since the complete odds is the exactly the ratio of two selection probabilities.
\end{proof}

\begin{proof}[ of Proposition \ref{prop:oddsCCMV}]
	This result is best seen by showing that the CCMV formulation in \eqref{eq:CCMV} is equivalent to the CCMV formulation in \eqref{eq:CCMV-odds}, so we will proceed with a series of equivalences.  We have
	\begin{align*}
		&p(x_{\bar{r}}|x_r,r,w_a,a) = p(x_{\bar{r}}|x_r,R=1_d,w_a,a) \\
		&\quad \iff \frac{p(x,r,w_a,a)}{p(x_r,r,w_a,a)} = \frac{p(x,1_d,w_a,a)}{p(x_r,1_d,w_a,a)} \\
		&\quad \iff \frac{p(x,r,w_a,a)}{p(x,1_d,w_a,a)} = \frac{p(x_r,r,w_a,)}{p(x_r,1_d,w_a,a)} \\
		&\quad \iff \frac{P(R=r|X,W_a,a)}{P(R=1_d|X,W_a,a)} = \frac{P(R=r|X_r,W_a,a)}{P(R=1_d|X_r,W_a,a)} \\
		&\quad \iff O_{r,a}(X,W_a) = Q_{r,a}(X_r,W_a).
	\end{align*}
	Thus, the CCMV missing data assumption on the covariates is equivalent through the pattern mixture model and complete odds formulations.
\end{proof}

\begin{proof}[ of Theorem \ref{thm:CCMV_ID}]
	We wish to show that under the CCMV assumption, the full-covariate distribution $p(x,r,w_a,a)$ is nonparametrically identified.  First, we factorize the target distribution using a pattern mixture model approach
	\begin{equation*}
		p(x,r,w_a,a) = p(x_{\bar{r}}|x_r,r,w_a,a) p (x_r,w_a,a).
	\end{equation*}
	
	Note that the distribution $p(x_r,r,w_a,a)$ is identifiable from the observed data, so the only part we need to handle is the extrapolation density $p(x_{\bar{r}}|x_r,r,w_a,a)$.  By the CCMV assumption, we have the equality
	\begin{equation*}
		p(x_{\bar{r}}|x_r,r,w_a,a) \stackrel{\text{CCMV}}{=} p(x_{\bar{r}}|x_r,R=1_d,w_a,a).
	\end{equation*}
	Since the distribution $p(x,R=1_d,w_a,a)$ is identifiable from the observed data (consider only the observations with fully-observed covariates and observed outcome pattern $A=a$), the conditional distribution $p(x_{\bar{r}}|x_r,R=1_d,w_a,a)$ is identifiable.  This means that the extrapolation density is equated to an identifiable quantity.
	
	The distribution $p(x,r,w_a,a)$ derived from the CCMV identifying restriction does not conflict with the observed data 
	because after marginalizing out the unobserved variable $x_{\bar r}$, 
	we obtain $p (x_r,w_a,a)$, so we have nonparametric identification.
	
\end{proof}

\begin{proof}[ of Lemma \ref{lemma:ra}]
	The expectation rewrites as
	\begin{align*}
		\E[f_a(X,W_a)1(A=a)1(R=r)] &= \int f_a(x,w_a)1(a'=a)1(r'=r) \ p(x,r,w_a,a') \ dx dr' dw_a da' \\
		&= \int f_a(x,w_a) \ p (x,r,w_a,a) \ dx dw_a \\
		&= \int f_a(x,w_a) p(x_{\bar{r}}|x_r,r,w_a,a) \ dx_{\bar{r}} \ p(x_r,r,w_a,a) \ dx_r dw_a.
	\end{align*}
	Recall that $m_{r,a}(x_r,w_a) = \int f_a(x,w_a) \ p(x_{\bar{r}}|x_r,r,w_a,a) \ dx_{\bar{r}}$ is the regression function. 
	Substituting it into the previous expression yields
	\begin{align*}
		\int m_{r,a}(x_r,w_a) \ p(x_r,r,w_a,a) \ dx_r dw_a &= \E[m_{r,a}(x_r,w_a)1(R=r)1(A=a)].
	\end{align*}
	Finally, summing over all patterns $r$ yields the second result
	\begin{align*}
		\E[f_a(X,W_a)1(A=a)] &= \sum_r \E[m_{r,a}(x_r,w_a)1(R=r)1(A=a)].
	\end{align*}
\end{proof}





\subsection{Multiple robustness and efficiency theory}	\label{app:eff}
This section contains proofs of
Lemma \ref{lemma:DR-form}, Theorem \ref{theorem:multiplyrobust-cox}, Theorem \ref{theorem:multiplyrobust-missingresp}, Corollary \ref{cor:mr-missresp}, Theorem \ref{theorem:eif}.

\begin{proof}[ of Lemma \ref{lemma:DR-form}]
	Fix the observed outcome pattern $a$, and suppose the CCMV assumption holds for $X$.  We will show that $\E[g_r(X,W_a)] = \E[f_a(X,W_a)1(A=a)1(R=r)]$ provided either $O_{r,a}(X,W_a)$ or $m_{r,a}(X_r,W_a)$ is correctly specified.  By Lemmas \ref{lemma:completeodds} and \ref{lemma:ra}, the terms \eqref{eq:odds-exp} and \eqref{eq:ra-exp} are known to have expected value equal to $\E[f_a(X,W_a)1(A=a)1(R=r)]$ when the complete odds $O_{r,a}(X,W_a)$ and the regression functions $m_{r,a}(X_r,W_a)$ are correctly specified, respectively.
	
	Therefore, we focus our attention on term \eqref{eq:dr-exp}.  There are two cases.\\
	
	\textsc{Case 1.} In the first case, we will assume that the complete odds $O_{r,a}(X,W_a)$ is correctly specified in that it equals $P(R=r|X,W_a,A=a)/P(R=1_d|X,W_a,A=a)$.  Since term \eqref{eq:odds-exp} has the desired expectation when the complete odds are correctly specified, it suffices to show that term \eqref{eq:ra-exp} minus term \eqref{eq:dr-exp} has mean 0.  The difference of the terms is
	\begin{align*}
		&m_{r,a}(X_r,W_a)1(R=r)1(A=a) - m_{r,a}(X_r,W_a)1(A=a)1(R=1_d)O_{r,a}(X,W_a).
	\end{align*}
	Taking the expectation yields
	\begin{align*}
		&\int m_{r,a}(x_r,w_a) p(x_r,r,w_a,a) \ dx_r dw_a \nonumber \\
		&\quad - \int m_{r,a}(x_r,w_a)1(a'=a)1(s=1_d)O_{r,a}(x,w_a) \ p(x,s,w_a,a') \ dxds dw_a da' \\
		&= \int m_{r,a}(x_r,w_a) p(x_r,r,w_a,a) \ dx_r dw_a \nonumber \\
		&\qquad - \int m_{r,a}(x_r,w_a)O_{r,a}(x,w_a) \ p(x,1_d,w_a,a) \ dx dw_a.
	\end{align*}
	Since $O_{r,a}(X,W_a)$ is assumed to properly specified, this further simplifies to
	\begin{align*}
		&\int m_{r,a}(x_r,w_a) p(x_r,r,w_a,a) \ dx_r dw_a - \int m_{r,a}(x_r,w_a)p(x,r,w_a,a) \ dx dw_a \\
		&= \int m_{r,a}(x_r,w_a) p(x_r,r,w_a,a) \ dx_r dw_a - \int m_{r,a}(x_r,w_a) p(x_r,r,w_a,a) \ dx_r dw_a \\
		&= 0.
	\end{align*}
	Therefore, $\E[g(X,W_a)] = \E[f_a(X,W_a)1(A=a)]$ when $O_{r,a}(X,W_a)$ is correctly specified.\\
	
	\textsc{Case 2.} We now consider the second case.  Suppose that the regression function $m_{r,a}(X_r,W_a)$ is correctly specified.  Since term \eqref{eq:ra-exp} has the desired expectation under this assumption, it suffices to show that term \eqref{eq:odds-exp} minus term \eqref{eq:dr-exp} has mean 0.
	
	Consider the expectation of \eqref{eq:odds-exp}.  Using the law of total expectation, we have
	\begin{align*}
		&\E[f_a(X,W_a)1(A=a)1(R=1_d)O_{r,a}(X,W_a)]\\
		&\quad= \E[\E[f_a(X,W_a)O_{r,a}(X,W_a)|X_r,R=1_d,W_a,A=a]|R=1_d,A=a] \nonumber \\
		&\qquad \cdot P(R=1_d, A=a) \\
		&\quad \stackrel{\text{CCMV}}{=} \E[\E[f_a(X,W_a)Q_{r,a}(X_r,W_a)|X_r,R=r,W_a,A=a]|R=1_d,A=a] \nonumber \\
		&\qquad \cdot P(R=1_d, A=a) \\
		&\quad = \E[\E[f_a(X,W_a)|X_r,R=r,W_a,A=a]\cdot Q_{r,a}(X_r,W_a)|R=1_d,A=a] \nonumber \\
		&\qquad \cdot P(R=1_d, A=a).
	\end{align*}
	Since the regression function $m_{r,a}(X_r,W_a)$ is assumed to be correctly specified, this further simplifies as
	\begin{align*}
		&\E[m_{r,a}(X_r,W_a)\cdot Q_{r,a}(X_r,W_a)|R=1_d,A=a]\cdot P(R=1_d, A=a)\\
		&\quad = \E[m_{r,a}(X_r,W_a)Q_{r,a}(X_r,W_a)1(A=a)1(R=1_d)] \\
		&\quad  \stackrel{\text{CCMV}}{=} \E[m_{r,a}(X_r,W_a)O_{r,a}(X,W_a)1(A=a)1(R=1_d)].
	\end{align*}
	This expression is exactly equal to the expected value of term \eqref{eq:dr-exp}, so it follows that term \eqref{eq:odds-exp} minus term \eqref{eq:dr-exp} has mean 0.
	
	Since we showed the expectation is correct in these two cases, we can conclude this form is doubly robust.
\end{proof}

\begin{proof}[ of Theorem \ref{theorem:multiplyrobust-cox}]
	The heart of the proof lies with Lemma \ref{lemma:DR-form}.  We will argue that the population estimating equation has mean $0$.  Consider the multiple robust population estimating equation with
	\begin{align*}
		& U_r(\beta) = \E\left[ \Delta \left(X - \frac{s^{(1)}(Y; \beta)}{s^{(0)}(Y;\beta)}\right) Q_{r,1}(X_r,Y)1(R=1_d)\right] \\
		& \qquad + \E\left[\E\left[X - \frac{s^{(1)}(Y;\beta)}{s^{(0)}(Y;\beta)} \biggr |X_r, R,Y, \Delta\right]\cdot 1(R=r) \cdot 1(\Delta=1) \right] \\
		& \qquad - \E\left[\E\left[X - \frac{s^{(1)}(Y;\beta)}{s^{(0)}(Y;\beta)} \biggr |X_r, R,Y, \Delta\right]\cdot 1(\Delta=1) \cdot 1(R=1_d) \cdot Q_{r,1}(X_r,Y) \right],
	\end{align*}
	
	\begin{align*}
		& s^{(1)}_r(\beta) = \sum_{\delta} \E[1(Y\geq t) X e^{\beta^\top X}Q_{r,1}(X_r,Y)1(\Delta=\delta)1(R=1_d)] \\
		& \qquad + \sum_{\delta} \E[m^{(1)}_{r,\delta}(X_r,Y;t,\beta) 1(R=r) 1(\Delta= \delta)] \\
		& \qquad - \sum_{\delta} \E\left[m^{(1)}_{r,\delta}(X_r,Y;t,\beta) \cdot 1(\Delta=1) \cdot 1(R=1_d) \cdot Q_{r,1}(X_r,Y)\right],
	\end{align*}
	
	\begin{align*}
		& s^{(0)}_r(\beta) = \sum_{\delta} \E[1(Y\geq t) e^{\beta^\top X}Q_{r,1}(X_r,Y)1(\Delta=\delta)1(R=1_d)] \\
		& \qquad + \sum_{\delta} \E[m^{(0)}_{r,\delta}(X_r,Y;t,\beta) 1(R=r) 1(\Delta= \delta)] \\
		& \qquad - \sum_{\delta} \E\left[m^{(0)}_{r,\delta}(X_r,Y;t,\beta) \cdot 1(\Delta=1) \cdot 1(R=1_d) \cdot Q_{r,1}(X_r,Y)\right].
	\end{align*}
	
	From Lemma \ref{lemma:DR-form}, the three above terms have expectation equivalent to
	\begin{align*}
		& \E\left[ \Delta \left(X - \frac{s^{(1)}(Y; \beta)}{s^{(0)}(Y;\beta)}\right) 1(R=r) \right], \\
		& \E[1(Y\geq t) X e^{\beta^\top X} 1(R=r)], \\
		& \E[1(Y\geq t) e^{\beta^\top X} 1(R=r)],
	\end{align*}
	respectively, provided at least one of the regression functions or the complete odds $Q_{r,1}(X_r,Y)$ is correctly specified.  Thus, $\sum_r U_r(\beta)$ has mean $0$, and the near-solution $\hat{\beta}_{\text{mr}}$ to the finite-sample estimating equation is consistent under standard regularity conditions.
\end{proof}

\begin{proof}[ of Theorem \ref{theorem:multiplyrobust-missingresp}]
	We will rewrite it in the following form to exploit the results from the previous subsections
	\begin{equation*}
		1(A=1)\cdot\left(\frac{Y-m_1(X)}{\pi_A(X)} + m_1(X)\right) + 1(A=0) \cdot m_1(X).
	\end{equation*}
	Since $\mathcal{A} = \{0,1\}$, we have two functions $f_0$ and $f_1$ to consider.  The previous display suggests taking
	\begin{align}
		&f_0(X) = m_1(X), \label{eq:f0} \\
		&f_1(X,Y) = \frac{Y}{\pi_A(X)} - \frac{m_1(X)}{\pi_A(X)} + m_1(X). \label{eq:f1}
	\end{align}
	We now apply Lemma \ref{lemma:DR-form} to both functions.  For each term of $f_0(X)$ and $f_1(X,Y)$ that is a function of $X$, Lemma \ref{lemma:DR-form} yields $3$ terms.  As such, we obtain a total of 12 terms.  Applying Lemma \ref{lemma:DR-form} to \eqref{eq:f0} yields
	\begin{align*}
		&m_1(X)1(A=0)1(R=1_d)Q_{r,0}(X_r) + M_{m,r,0}(X_r)1(R=r)1(A=0) \\
		&\quad -M_{m,r,0}(X_r)1(A=0)1(R=1_d)Q_{r,0}(X_r).
	\end{align*}
	We obtain the following 9 terms from applying Lemma \ref{lemma:DR-form} to \eqref{eq:f1}:
	\begin{align*}
		&m_1(X)1(A=1)1(R=1_d)Q_{r,1}(X_r,Y) + M_{m,r,1}(X_r,Y)1(R=r)1(A=1) \\
		&\quad - M_{m,r,1}(X_r,Y)1(A=1)1(R=1_d)Q_{r,1}(X_r,Y) \\
		&\quad + \frac{Y}{\pi_A(X)} 1(A=1) 1(R=1_d) Q_{r,1}(X_r,Y) + Y M_{1/\pi_A, r,1}(X_r,Y) 1(R=r) 1(A=1) \\
		&\quad - Y M_{1/\pi_A,r,1}(X_r,Y)1(A=1)1(R=1_d)Q_{r,1}(X_r,Y) \\
		&\quad - \frac{m_1(X)}{\pi_A(X)} 1(A=1) 1(R=1_d) Q_{r,1}(X_r,Y) - M_{m/\pi_A,r,1}(X_r,Y)1(R=r) 1(A=1) \\
		&\quad + M_{m/\pi_A,r,1}(X_r,Y)1(A=1) 1(R=1_d) Q_{r,1}(X_r,Y).
	\end{align*}
	The double robustness follows from the proof of Lemma \ref{lemma:DR-form}.
\end{proof}

\begin{proof}[ of Corollary \ref{cor:mr-missresp}]
	Compute the uncentered doubly robust influence function $q_r(X,R,Y,A)$ for each missing pattern $r$ using Theorem \ref{theorem:multiplyrobust-missingresp}.  By construction, each influence function is consistent for $\theta_r$ provided either $Q_{r,a}(X_r,W_a)$ and $p(x_{\bar{r}} | x_r, w_a, a)$ are correctly specified.  Each estimator for a given missing covariate pattern $r$ is doubly robust from Theorem \ref{theorem:multiplyrobust-missingresp}.  The doubly robustness for each pattern implies that the final estimator is $2^{|\mathcal{R}|-1}$-robust.
\end{proof}

\begin{proof}[ of Theorem \ref{theorem:eif}]
	Let $P_0$ denotes the true distribution.
	We solve for efficient influence function using the semiparametric theory outlined in \cite{van1998asymptotic}.  By Lemma \ref{lemma:completeodds} and under CCMV, the parameter of interest $\psi(P_0)$ can be written as the following statistical functional
	\begin{align*}
		\theta^{(0)} &= \int y \ P_0(dx, dr, dy, da) \\
		&= \int \frac{1(a=1)y}{P_0(A=1|x)} \sum_r Q_{r,1}^{(0)}(x_r,y) 1(r'=1_d) \ P_0(dx,dr',dy,da),
	\end{align*}
	where $Q_{r,1}^{(0)}(x_r,y)$ denotes the complete odds $Q_{r,1}(x_r,y)$ under $P_0$.

	For ease of typesetting, in our abbreviations, we will adopt the convention that $R$ comes before $A$.  In other words, the expression $p(x,1_d,y,1)$ should be taken to be equivalent to $p(x,R=1_d,y,A=1)$.
	
	Define $p_0(x,r',y,1)$ and $p_0(x,r',0)$ to be the true model, and consider a first order perturbation of the model
	\begin{align*}
		p_t(x,r',y,1) = p_0(x,r',y,1)[1 + t\cdot g_1(x,r',y,1)], \\
		p_t(x,r',0) = p_0(x,r',0)[1 + t\cdot g_0(x,r',0)]
	\end{align*}
	for functions $g_1$ and $g_0$ satisfying $\E_{p_0}[1(A=1)g_1(X,R',Y,A)]=\E_{p_0}[1(A=0)g_0(X,R',A)]=0$.  Then, the efficient influence function is the mean-zero function $\text{EIF}(x,r',y,a) = 1(A=1)\text{EIF}_1(x,r',y) + 1(A=0)\text{EIF}_0(x,r')$ satisfying
	\begin{align*}
		\lim_{t\to 0} \frac{\theta^{(t)}-\theta^{(0)}}{t} &= \int \text{EIF}_1(x,r',y) p_0(x,r',y,1) g(x,r',y,1) \ dx dr' dy \\
		&\quad + \int \text{EIF}_0(x,r') p_0(x,r',0) g(x,r',0) \ dx dr'.
	\end{align*}
	
	There are three terms to be perturbed: $p(x,r',y,1)$, $Q_{r,1}(x_r,y)$, and $P(A=1|x)$.  It suffices to perturb each one up to first order and hold the remaining two unchanged.  Thus, we divide our task into 3 parts and proceed as follows:
	
	\textbf{Part (I).}  From perturbing $p(x,r',y,1)$, a direct calculation shows that
	\begin{align}
		\frac{1(A=1)Y}{P(A=1|X)} \sum_r Q_{r,1}(X_r,Y) 1(R=1_d)
	\end{align}
	contributes to the EIF.
	
	\textbf{Part (II).}  We now perturb the odds $Q_{r,1}(x_r,y)$.  Up to first order, we have
	\begin{align*}
		Q_{r,1}^{(t)}(x_r,y) &= \frac{p_t(x_r,r,y,a)}{p_t(x_r,1_d,y,a)} \\
		&= \frac{p_0(x_r,r,y,a) + t\int p_0(x_{\bar{r}}', x_r, r, y,a) g_1(x_{\bar{r}}', x_r, r, y,a) \ dx_{\bar{r}}'}{p_0(x_r,1_d,y,a) + t\int p_0(x_{\bar{r}}', x_r, 1_d, y,a) g_1(x_{\bar{r}}', x_r, 1_d, y,a) \ dx_{\bar{r}}'} + O(t^2) \\
		&= Q_{r,1}^{(0)}(x_r,y) + t \biggr( \frac{\int p_0(x_{\bar{r}}', x_r,r,y,a) g_1(x_{\bar{r}}', x_r, r, y,a) \ dx_{\bar{r}}'}{p_0(x_r,1_d,y,a)} \\
		&\qquad - \frac{p_0(x_r,r,y,a) \int p_0(x_{\bar{r}}', x_r,1_d,y,a) g_1(x_{\bar{r}}',x_r,1_d,y,a) \ dx_{\bar{r}}'}{p(x_r,1_d,y,a)p(x_r,1_d,y,a)} \biggr) + O(t^2).
	\end{align*}
	
	Thus, there are 2 additional terms that arise from this perturbation.  The first term can be found by simplifying the following expression
	\begin{align*}
		&\int \frac{1(a=1)y}{P_0(A=1|x)} \cdot \sum_r \frac{\int p_0(x_{\bar{r}}', x_r,r,y,a) g_1(x_{\bar{r}}', x_r, r, y,a) \ dx_{\bar{r}}'}{p_0(x_r,1_d,y,a)} \cdot 1(r'=1_d) \ P_0(dx,dr',dy,da) \\
		&= \int \frac{1(a=1)y}{P_0(A=1|x)} \cdot \sum_r 1(s=r) \cdot p_0(x_{\bar{r}}|x_r,1_d,y,a) \ dx_{\bar{r}} \\
		&\qquad p_0(x_{\bar{r}}', x_r,s,y,a) g_1(x_{\bar{r}}', x_r, s, y,a) \  dx_{\bar{r}}' dx_r ds dy da.
	\end{align*}
	So, the first term is
	\begin{equation}
		1(A=1)Y \sum_r 1(R=r) M_{1/\pi_A,r,1}(X_r,Y),
	\end{equation}
	where the notations $M_{f,r,a}$ is defined in equations \eqref{regf1}-\eqref{regf4}.
	
	The second term can be found by simplifying 
	\begin{align*}
		&-\int \frac{1(a=1)y}{P_0(A=1|x)} \cdot \sum_r \frac{p_0(x_r,r,y,a) \int p_0(x_{\bar{r}}', x_r,1_d,y,a) g_1(x_{\bar{r}}',x_r,1_d,y,a) \ dx_{\bar{r}}'}{p(x_r,1_d,y,a)p(x_r,1_d,y,a)} \\
		&\qquad \cdot 1(r'=1_d) \ P_0(dx,dr',dy,da) \\
		&=-\int \frac{1(a=1)y}{P_0(A=1|x)} \cdot \sum_r \frac{Q_{r,1}^{(0)}(x_r,y) \int 1(s=1_d) p_0(x_{\bar{r}}', x_r,s,y,a) g_1(x_{\bar{r}}',x_r,s,y,a) \ dx_{\bar{r}}' ds}{p(x_r,1_d,y,a)} \\
		&\qquad \cdot 1(r'=1_d) \ P_0(dx,dr',dy,da) \\
		&=-\int \frac{1(a=1)y}{P_0(A=1|x)} \cdot \sum_r  Q_{r,1}^{(0)}(x_r,y) \int 1(s=1_d) p_0(x_{\bar{r}}', x_r,s,y,a) g_1(x_{\bar{r}}',x_r,s,y,a) \ dx_{\bar{r}}' ds \\
		&\qquad \cdot p(x_{\bar{r}} | x_r, 1_d, y, a) \ dx_{\bar{r}} dx_r dy da \\
		&=-\int \frac{1(a=1)y}{P_0(A=1|x)} \cdot \sum_r  Q_{r,1}^{(0)}(x_r,y) \cdot 1(s=1_d) \cdot p(x_{\bar{r}} | x_r, 1_d, y, a) \ dx_{\bar{r}}\\
		&\qquad \cdot p_0(x_{\bar{r}}', x_r,s,y,a) g_1(x_{\bar{r}}',x_r,s,y,a) \ dx_{\bar{r}}' dx_r ds dy da.
	\end{align*}
	So, the second term is
	\begin{equation}
		-1(A=1)Y 1(R=1_d)\sum_r Q_{r,1}(X_r,Y)M_{1/\pi_A,r,1}(X_r,Y).
	\end{equation}

	\textbf{Part (III).}  This part requires the most tedious calculations.  We start by rewriting $P(A=1|x)$ as an identifiable expression under CCMV
	\begin{equation}
		\begin{aligned}
			&\frac{1}{P(A=1|x)} \\
			& = \frac{p(x)}{p(x,A=1)}\\
			&= \frac{\displaystyle\sum_{s\neq 1_d} \biggr[\int \overbrace{Q_{s,1}(x_s,y')}^{(i)} \overbrace{p(x,1_d,y',1)}^{(ii)} \ dy' + \overbrace{Q_{s,0}(x_s) }^{(iii)}\overbrace{p(x,1_d,0)}^{(iv)} \biggr] + \overbrace{p(x,1_d)}^{(vi)}}{\displaystyle\sum_{s\neq 1_d}\biggr[\int \underbrace{Q_{s,1}(x_s,y')}_{(v)} p(x,1_d,y',1) \ dy'\biggr] + p(x,1_d,1)}.
		\end{aligned}
		\label{eq:eff:III}
	\end{equation}
	
	Our goal is to perturb the above expression and find the first order approximation.  We consider each term in the above expression, and we expand up to first order:
	
	\begin{align*}
		(i):\quad Q_{s,1}^{(t)}(x_s,y') &= \frac{p_t(x_s,s,y',1)}{p_t(x_s,1_d,y',1)} \\
		&= \frac{p_0(x_s,s,y',1) + t\int p_0(x_{\bar{s}}', x_s, s, y',1) g_1(x_{\bar{s}}', x_s, s, y',1) \ dx_{\bar{s}}'}{p_0(x_s,1_d,y',1) + t\int p_0(x_{\bar{s}}', x_s, 1_d, y',1) g_1(x_{\bar{s}}', x_s, 1_d, y',1) \ dx_{\bar{s}}'} + O(t^2) \\
		&= Q_{s,1}^{(0)}(x_s,y') + t \biggr( \frac{\int p_0(x_{\bar{s}}', x_s,s,y',1) g_1(x_{\bar{s}}', x_s, s, y',1) \ dx_{\bar{s}}'}{p_0(x_s,1_d,y',1)} \\
		&\qquad - \frac{p_0(x_s,s,y',1) \int p_0(x_{\bar{s}}', x_s,1_d,y',1) g_1(x_{\bar{s}}',x_s,1_d,y',1) \ dx_{\bar{s}}'}{p(x_s,1_d,y',1)p(x_s,1_d,y',1)} \biggr) + O(t^2).
	\end{align*}
	
	\begin{align*}
		(ii):\quad p_t(x,1_d,y',1)	&= p_0(x,1_d,y',1) + t p_0(x,1_d,y',1) g_1(x,1_d,y',1) + O(t^2).
	\end{align*}
	
	\begin{align*}
		(iii):\quad Q_{s,0}^{(t)}(x_s) &= \frac{p_t(x_s,s,0)}{p_t(x_s,1_d,0)} \\
		&= \frac{p_0(x_s,s,0) + t\int p_0(x_{\bar{s}}', x_s, s, 0) g_0(x_{\bar{s}}', x_s, s, 0) \ dx_{\bar{s}}'}{p_0(x_s,1_d,0) + t\int p_0(x_{\bar{s}}', x_s, 1_d, 0) g_0(x_{\bar{s}}', x_s, 1_d, 0) \ dx_{\bar{s}}'} + O(t^2) \\
		&= Q_{s,0}^{(0)}(x_s) + t \biggr( \frac{\int p_0(x_{\bar{s}}', x_s,s,0) g_0(x_{\bar{s}}', x_s, s,0) \ dx_{\bar{s}}'}{p_0(x_s,1_d,0)} \\
		&\qquad - \frac{p_0(x_s,s,0) \int p_0(x_{\bar{s}}', x_s,1_d,0) g_0(x_{\bar{s}}',x_s,1_d,0) \ dx_{\bar{s}}'}{p(x_s,1_d,0)p(x_s,1_d,0)} \biggr) + O(t^2).
	\end{align*}
	
	\begin{align*}
		(iv):\quad p_t(x,1_d,0) &= \int p_t(x,1_d,0) \ dy'\\
		&= p_0(x,1_d,0) + t\int p_0(x,1_d,0) g_0(x,1_d,0) \ dy' + O(t^2).
	\end{align*}
	
	\begin{align*}
		(v):\quad p_t(x,1_d,1) &= \int p_t(x,1_d,y',1) \ dy'\\
		&= p_0(x,1_d,1) + t\int p_0(x,1_d,y',1) g_1(x,1_d,y',1) \ dy' + O(t^2).
	\end{align*}
	
	\begin{align*}
		(vi):\quad p_t(x,1_d) &= \int p_t(x,1_d,y',1) \ dy' + p_t(x,1_d,0) \\
		&= p_0(x,1_d) + t\int p_0(x,1_d,y',1) g_1(x,1_d,y',1) \ dy' da' + t p_0(x,1_d,0) g_0(x,1_d,0) + O(t^2).
	\end{align*}
	
	Thus, the denominator of equation \eqref{eq:eff:III}, $p(x,A=1)$, takes the following form (up to first order):
	\begin{align*}
		A_0+A_1 t &= p_0(x,1) + t \int p_0(x,1_d,y',1)g_1(x,1_d,y',1) \ dy' \\
		&\quad + t \sum_{s\neq 1_d} \int Q_{s,1}(x_s,y') p_0(x,1_d,y',1)g_1(x,1_d,y',1) \ dy' \\
		&\quad + t \sum_{s\neq 1_d}\int \frac{p_0(x,1_d,y',1)}{p(x_s,1_d,y',1)} p(x_{\bar{s}}',x_s,y',1) g_1(x_{\bar{s}}', x_s, s, y', 1) \ dx_{\bar{s}}' dy' \\
		&\quad - t \sum_{s\neq 1_d}\int \frac{p_0(x,1_d,y',1) p_0(x_s,s,y',1)}{p(x_s,1_d,y',1) p(x_s,1_d,y',1)} p(x_{\bar{s}}',x_s,y',1) g_1(x_{\bar{s}}', x_s, s, y', 1) \ dx_{\bar{s}}' dy' \\
		&= p_0(x,1) + \underbrace{t \int p_0(x,1_d,y',1)g_1(x,1_d,y',1) \ dy'}_{(a)} \\
		&\quad + \underbrace{t \sum_{s\neq 1_d} \int Q_{s,1}(x_s,y') p_0(x,1_d,y',1)g_1(x,1_d,y',1) \ dy'}_{(b)} \\
		&\quad + \underbrace{t \sum_{s\neq 1_d}\int p_0(x_{\bar{s}}|x_s,1_d,y',1) p_0(x_{\bar{s}}',x_s,s,y',1) g_1(x_{\bar{s}}', x_s, s, y', 1) \ dx_{\bar{s}}' dy'}_{(c)} \\
		&\quad \underbrace{- t \sum_{s\neq 1_d}\int p_0(x_{\bar{s}}|x_s,1_d,y',1) Q_{s,1}^{(0)}(x_s,y') p(x_{\bar{s}}',x_s,1_d,y',1) g_1(x_{\bar{s}}', x_s, 1_d, y', 1) \ dx_{\bar{s}}' dy'}_{(d)}
	\end{align*}
	
	The numerator of equation \eqref{eq:eff:III}, $p(x)$, takes the following form (up to first order):
	\begin{align*}
		a_0+a_1 t &= p_0(x) + t \int p_0(x,1_d,y',1)g_1(x,1_d,y',1) \ dy' da'\\
		&\quad + t p_0(x,1_d,0) g_0(x,1_d,0) \\
		&\quad + t \sum_{s\neq 1_d} \int Q_{s,1}(x_s,y') p_0(x,1_d,y',1)g_1(x,1_d,y',1) \ dy' \\
		&\quad + t \sum_{s\neq 1_d}\int \frac{p_0(x,1_d,y',1)}{p(x_s,1_d,y',1)} p(x_{\bar{s}}',x_s,y',1) g_1(x_{\bar{s}}', x_s, s, y', 1) \ dx_{\bar{s}}' dy' \\
		&\quad - t \sum_{s\neq 1_d}\int \frac{p_0(x,1_d,y',1) p_0(x_s,s,y',1)}{p_0(x_s,1_d,y',1) p_0(x_s,1_d,y',1)} p_0(x_{\bar{s}}',x_s,y',1) g_1(x_{\bar{s}}', x_s, s, y', 1) \ dx_{\bar{s}}' dy' \\
		&\quad + t \sum_{s\neq 1_d} Q_{s,0}(x_s) p_0(x,1_d,0)g_0(x,1_d,0) \\
		&\quad + t \sum_{s\neq 1_d}\int \frac{p_0(x,1_d,0)}{p_0(x_s,1_d,0)} p_0(x_{\bar{s}}',x_s,0) g_0(x_{\bar{s}}', x_s, s, 0) \ dx_{\bar{s}}' \\
		&\quad - t \sum_{s\neq 1_d}\int \frac{p_0(x,1_d,0) p_0(x_s,s,0)}{p_0(x_s,1_d,0) p_0(x_s,1_d,0)} p_0(x_{\bar{s}}',x_s,0) g_0(x_{\bar{s}}', x_s, s, 0) \ dx_{\bar{s}}' \\
		&= p_0(x) + \underbrace{t \int p_0(x,1_d,y',1)g_1(x,1_d,y',1) \ dy'}_{(e)} \\
		&\quad + \underbrace{t p_0(x,1_d,0)g_0(x,1_d,0) }_{(f)} \\
		&\quad + \underbrace{t \sum_{s\neq 1_d} \int Q_{s,1}(x_s,y') p_0(x,1_d,y',1)g_1(x,1_d,y',1) \ dy'}_{(g)} \\
		&\quad + \underbrace{t \sum_{s\neq 1_d}\int p_0(x_{\bar{s}}|x_s,1_d,y',1) p_0(x_{\bar{s}}',x_s,s,y',1) g_1(x_{\bar{s}}', x_s, s, y', 1) \ dx_{\bar{s}}' dy'}_{(h)} \\
		&\quad \underbrace{- t \sum_{s\neq 1_d}\int p_0(x_{\bar{s}}|x_s,1_d,y',1) Q_{s,1}^{(0)}(x_s,y') p(x_{\bar{s}}',x_s,1_d,y',1) g_1(x_{\bar{s}}', x_s, 1_d, y', 1) \ dx_{\bar{s}}' dy'}_{(i)} \\
		&\quad + \underbrace{t \sum_{s\neq 1_d} Q_{s,0}(x_s) p_0(x,1_d,0)g_0(x,1_d,0)}_{(j)} \\
		&\quad + \underbrace{t \sum_{s\neq 1_d}\int p_0(x_{\bar{s}}|x_s,1_d,0) p_0(x_{\bar{s}}',x_s,0) g_0(x_{\bar{s}}', x_s, s, 0) \ dx_{\bar{s}}'}_{(k)} \\
		&\quad \underbrace{- t \sum_{s\neq 1_d}\int p_0(x_{\bar{s}}|x_s,1_d,0) Q_{s,0}^{(0)}(x_s) p_0(x_{\bar{s}}',x_s,1_d,0) g_0(x_{\bar{s}}', x_s, 1_d, 0) \ dx_{\bar{s}}'}_{(l)} \\
	\end{align*}
	
	A fraction of the form $\dfrac{a_0+a_1t + O(t^2)}{A_0 + A_1t + O(t^2)}$ has the following first order expansion
	\begin{equation*}
		\frac{a_0}{A_0} + \left(\frac{a_1}{A_0} - \frac{a_0A_1}{A_0^2}\right) t + O(t^2).
	\end{equation*}
	
	Thus, the perturbation of $1/P(A=1|x)$ yields 12 terms, based on terms (a)-(l).  We now derive each individually.
	
	\textbf{Term (a).} The contribution from (a) gives
	\begin{align*}
		&-t\int \frac{1(r'=1_d)1(a=1)y p_0(x,r',y,a) \sum_r Q_{r,1}(x_r,y)}{P_0(A=1|x)p_0(x,1)} \\
		&\qquad \times p_0(x,1_d,y',1) g_1(x,1_d,y',1) \ dy' \ dxdr'dyda  \\
		&= -t\int \frac{1(r'=1_d)1(a=1)y p_0(x,1_d,y,1) \sum_r Q_{r,1}(x_r,y)}{P_0(A=1|x)p_0(x,1)} \\
		&\qquad \times p_0(x,r',y',a) g_1(x,r',y',a) \ dy \ dxdr'dy'da  \\
		&= -t\int \frac{1(r'=1_d)1(a=1)y p(y|x,1)}{P_0(A=1|x)} \\
		&\qquad \times p_0(x,r',y',a) g_1(x,r',y',a) \ dy \ dxdr'dy'da  \\
		&= -t\int \frac{1(r'=1_d)1(a=1) m_1(X)}{P_0(A=1|x)} \\
		&\qquad \times p_0(x,r',y',a) g_1(x,r',y',a) \ dxdr'dy'da.
	\end{align*}
	So, (a) contributes
	$$-1(R=1_d)1(A=1)\frac{m_1(X)}{\pi_A(X)}.$$

	\textbf{Term (b).} The contribution from (b) gives
	
	\begin{align*}
		&-t\int \frac{1(r'=1_d)1(a=1)y p_0(x,r',y,a) \sum_r Q_{r,1}(x_r,y)}{P_0(A=1|x)p_0(x,1)} \\
		&\qquad \times \sum_{s\neq 1_d} Q_{s,1}(x_s,y') p_0(x,1_d,y',1) g_1(x,1_d,y',1) \ dy' \ dxdr'dyda \\
		&= -t\int \frac{1(r'=1_d)1(a=1)y p(y|x,1)}{P_0(A=1|x)} \\
		&\qquad \times \sum_{s\neq 1_d} Q_{s,1}(x_s,y') p_0(x,r',y',a) g_1(x,r',y',a) \ dy \ dxdr'dy'da \\
	\end{align*}
	So, (b) contributes
	$$-1(R=1_d)1(A=1)\frac{m_1(X)}{\pi_A(X)} \sum_{r\neq 1_d} Q_{r,1}(X_r, Y).$$
	
	\textbf{Term (c).} The contribution from (c) gives
	
	\begin{align*}
		&-t\int \frac{1(r'=1_d)1(a=1)y p_0(x,r',y,a) \sum_r Q_{r,1}(x_r,y)}{P_0(A=1|x)p_0(x,1)} \\
		&\qquad \times \sum_{s\neq 1_d}p_0(x_{\bar{s}}| x_s, 1_d, y', 1) p_0(x_{\bar{s}}', x_s, s, y', 1) g_1(x_{\bar{s}}', x_s, s, y', 1) \ dx_{\bar{s}}' dy' \ dx_{\bar{s}} dx_s dr' dyda \\
		&= -t\int \frac{1(a=1) m_1(X)}{P_0(A=1|x)} \\
		&\qquad \times \sum_{s\neq 1_d}p_0(x_{\bar{s}}| x_s, 1_d, y', 1) p_0(x_{\bar{s}}', x_s, s, y', a) g_1(x_{\bar{s}}', x_s, s, y', a) \ dx_{\bar{s}}  \ dx_{\bar{s}}' dx_s dy'da \\
		&= -t\int 1(a=1) \\
		&\qquad \times \sum_{s\neq 1_d} M_{m/\pi_A,s,1}(x_s,y') p_0(x_{\bar{s}}', x_s, s, y', a) g_1(x_{\bar{s}}', x_s, s, y', a) \ dx_{\bar{s}}  \ dx_{\bar{s}}' dx_s dy'da \\
		&= -t\int 1(a=1) \\
		&\qquad \times \sum_{s\neq 1_d} 1(r''=s) M_{m/\pi_A,s,1}(x_s,y') p_0(x_{\bar{s}}', x_s, r'', y', a) g_1(x_{\bar{s}}', x_s, r'', y', a) \ dx_{\bar{s}}  \ dx_{\bar{s}}' dx_s dr'' dy'da 
	\end{align*}
	So, (c) contributes
	$$-1(A=1) \sum_{r\neq 1_d} 1(R=r) M_{m/\pi_A,r,1}(X_r,Y).$$
	
	\textbf{Term (d).} The contribution from (d) gives
	\begin{align*}
		&t\int \frac{1(r'=1_d)1(a=1)y p_0(x,r',y,a) \sum_r Q_{r,1}(x_r,y)}{P_0(A=1|x)p_0(x,1)} \\
		&\qquad \times \sum_{s\neq1_d} Q_{s,1}(x_s,y') p_0(x_{\bar{s}}| x_s, 1_d, y', 1) p_0(x_{\bar{s}}', x_s, 1_d, y', 1) g_1(x_{\bar{s}}', x_s, 1_d, y', 1) \ dx_{\bar{s}}' dy' \ dx_{\bar{s}} dx_s dr' dyda \\
		&= t\int \frac{1(a=1)m_1(X)}{P_0(A=1|x)} \\
		&\qquad \times \sum_{s\neq1_d} Q_{s,1}(x_s,y') p_0(x_{\bar{s}}| x_s, 1_d, y', 1) p_0(x_{\bar{s}}', x_s, 1_d, y', 1) g_1(x_{\bar{s}}', x_s, 1_d, y', 1) \ dx_{\bar{s}} \ dx_{\bar{s}}' dx_s dy'da \\
		&= t\int 1(a=1) \\
		&\qquad \times \sum_{s\neq1_d} Q_{s,1}(x_s,y') M_{m/\pi_A,s,1}(x_s,y') p_0(x_{\bar{s}}', x_s, 1_d, y', 1) g_1(x_{\bar{s}}', x_s, 1_d, y', 1) \ dx_{\bar{s}}' dx_s dy'da \\
		&= t\int 1(a=1) \\
		&\qquad \times \sum_{s\neq1_d} 1(r''=s) Q_{s,1}(x_s,y') M_{m/\pi_A,s,1}(x_s,y') p_0(x_{\bar{s}}', x_s, r'', y', 1) g_1(x_{\bar{s}}', x_s, r'', y', 1) \ dx_{\bar{s}}' dx_s dr'' dy'da 
	\end{align*}
	So, (d) contributes
	$$1(A=1) \sum_{r\neq 1_d} 1(R = 1_d) M_{m/\pi_A,r,1}(X_r,Y) Q_{r,1}(X_r,Y).$$

	\textbf{Term (e).} The contribution from (e) gives
	\begin{align*}
		&t\int \frac{1(r'=1_d)1(a=1)y p_0(x,r',y,a) \sum_r Q_{r,1}(x_r,y)}{p_0(x,1)} \\
		&\qquad \times p_0(x,1_d,y',1) g_1(x,1_d,y',1) \ dy' \ dxdr'dyda  \\
		&= t\int \frac{1(r'=1_d)1(a=1)y p_0(x,1_d,y,1) \sum_r Q_{r,1}(x_r,y)}{p_0(x,1)} \\
		&\qquad \times p_0(x,r',y',a) g_1(x,r',y',a) \ dy \ dxdr'dy'da  \\
		&= t\int 1(r'=1_d)1(a=1)y p(y|x,1) \\
		&\qquad \times p_0(x,r',y',a) g_1(x,r',y',a) \ dy \ dxdr'dy'da  \\
		&= t\int 1(r'=1_d)1(a=1) m_1(X) \\
		&\qquad \times p_0(x,r',y',a) g_1(x,r',y',a) \ dxdr'dy'da.
	\end{align*}
	So, (e) contributes
	$$1(R=1_d)1(A=1)m_1(X).$$

	\textbf{Term (f).} The contribution from (f) gives
	\begin{align*}
		&t\int \frac{1(r'=1_d)1(a=1)y p_0(x,r',y,a) \sum_r Q_{r,1}(x_r,y)}{p_0(x,1)} \\
		&\qquad \times p_0(x,1_d,0)g_0(x,1_d,0) \ dxdr'dyda  \\
		&= t\int \frac{1(r'=1_d)1(a=1)1(a'=0)y p_0(x,r',y,a) \sum_r Q_{r,1}(x_r,y)}{p_0(x,1)} \\
		&\qquad \times p_0(x,r',a')g_0(x,r',a') \ da dy \ dxdr'da'  \\
		&= t\int \frac{1(r'=1_d)1(a=1)1(a'=0)y p_0(x,y,a) }{p_0(x,1)} \\
		&\qquad \times p_0(x,r',a')g_0(x,r',a') \ da dy \ dxdr'da'  \\
		&= t\int 1(r'=1_d)1(a'=0)m_1(X) \\
		&\qquad \times p_0(x,r',a')g_0(x,r',a') \ da dy \ dxdr'da'  \\
	\end{align*}
	So, (f) contributes
	$$1(R=1_d)1(A=0)m_1(X).$$
	
	\textbf{Term (g).} The contribution from (g) gives
	\begin{align*}
		&t\int \frac{1(r'=1_d)1(a=1)y p_0(x,r',y,a) \sum_{r} Q_{r,1}(x_r,y) \cdot \sum_{s\neq 1_d} Q_{s,1}(x_s,y') }{p_0(x,1)} \\
		&\qquad \times p_0(x,1_d,y',1)g_1(x,1_d,y',1) \ dy' \ dxdr'dyda  \\
		&= t\int 1(r'=1_d)1(a=1)m_1(X) \cdot \sum_{s\neq 1_d} Q_{s,1}(x_s,y') \\
		&\qquad \times p_0(x,r',y',a)g_1(x,r',y',a) \ dxdr'dy'da  \\
	\end{align*}
	So, (g) contributes
	$$1(R=1_d)1(A=1)m_1(X)\sum_{r\neq 1_d} Q_{r,1}(X_r,Y).$$
	
	\textbf{Term (h).} The contribution from (h) gives
	
	\begin{align*}
		&t\int \frac{1(r'=1_d)1(a=1)y p_0(x,r',y,a) \sum_r Q_{r,1}(x_r,y)}{p_0(x,1)} \\
		&\qquad \times \sum_{s\neq 1_d}p_0(x_{\bar{s}}| x_s, 1_d, y', 1) p_0(x_{\bar{s}}', x_s, s, y', 1) g_1(x_{\bar{s}}', x_s, s, y', 1) \ dx_{\bar{s}}' dy' \ dx_{\bar{s}} dx_s dr' dyda \\
		&= t\int 1(a=1) m_1(X) \\
		&\qquad \times \sum_{s\neq 1_d}p_0(x_{\bar{s}}| x_s, 1_d, y', 1) p_0(x_{\bar{s}}', x_s, s, y', a) g_1(x_{\bar{s}}', x_s, s, y', a) \ dx_{\bar{s}}  \ dx_{\bar{s}}' dx_s dy'da \\
		&= t\int 1(a=1) \\
		&\qquad \times \sum_{s\neq 1_d} M_{m,s,1}(x_s,y') p_0(x_{\bar{s}}', x_s, s, y', a) g_1(x_{\bar{s}}', x_s, s, y', a) \ dx_{\bar{s}}  \ dx_{\bar{s}}' dx_s dy'da \\
		&= t\int 1(a=1) \\
		&\qquad \times \sum_{s\neq 1_d} 1(r''=s) M_{m,s,1}(x_s,y') p_0(x_{\bar{s}}', x_s, r'', y', a) g_1(x_{\bar{s}}', x_s, r'', y', a) \ dx_{\bar{s}}  \ dx_{\bar{s}}' dx_s dr'' dy'da 
	\end{align*}
	So, (h) contributes
	$$1(A=1) \sum_{r\neq 1_d} 1(R=r) M_{m,r,1}(X_r,Y).$$

	\textbf{Term (i).} The contribution from (i) gives
	\begin{align*}
		&-t\int \frac{1(r'=1_d)1(a=1)y p_0(x,r',y,a) \sum_r Q_{r,1}(x_r,y)}{p_0(x,1)} \\
		&\qquad \times \sum_{s\neq1_d} Q_{s,1}(x_s,y') p_0(x_{\bar{s}}| x_s, 1_d, y', 1) p_0(x_{\bar{s}}', x_s, 1_d, y', 1) g_1(x_{\bar{s}}', x_s, 1_d, y', 1) \ dx_{\bar{s}}' dy' \ dx_{\bar{s}} dx_s dr' dyda \\
		&= -t\int 1(a=1)m_1(X) \\
		&\qquad \times \sum_{s\neq1_d} Q_{s,1}(x_s,y') p_0(x_{\bar{s}}| x_s, 1_d, y', 1) p_0(x_{\bar{s}}', x_s, 1_d, y', 1) g_1(x_{\bar{s}}', x_s, 1_d, y', 1) \ dx_{\bar{s}} \ dx_{\bar{s}}' dx_s dy'da \\
		&= -t\int 1(a=1) \\
		&\qquad \times \sum_{s\neq1_d} Q_{s,1}(x_s,y') M_{m,s,1}(x_s,y') p_0(x_{\bar{s}}', x_s, 1_d, y', 1) g_1(x_{\bar{s}}', x_s, 1_d, y', 1) \ dx_{\bar{s}}' dx_s dy'da \\
		&= -t\int 1(a=1) \\
		&\qquad \times \sum_{s\neq1_d} 1(r''=s) Q_{s,1}(x_s,y') M_{m,s,1}(x_s,y') p_0(x_{\bar{s}}', x_s, r'', y', 1) g_1(x_{\bar{s}}', x_s, r'', y', 1) \ dx_{\bar{s}}' dx_s dr'' dy'da 
	\end{align*}
	So, (i) contributes
	$$-1(A=1) \sum_{r\neq 1_d} 1(R = 1_d) M_{m,r,1}(X_r,Y) Q_{r,1}(X_r,Y).$$
	
	\textbf{Term (j).} The contribution from (j) gives
	\begin{align*}
		&t\int \frac{1(r'=1_d)1(a=1)y p_0(x,r',y,a) \sum_{r} Q_{r,1}(x_r,y) \cdot \sum_{s\neq 1_d} Q_{s,0}(x_s) }{p_0(x,1)} \\
		&\qquad \times p_0(x,1_d,0)g_0(x,1_d,0) \ dxdr'dyda  \\
		&= t\int 1(r'=1_d)1(a'=0)m_1(X) \cdot \sum_{s\neq 1_d} Q_{s,0}(x_s) \\
		&\qquad \times p_0(x,r',a')g_0(x,r',a') \ dxdr'da' \\
	\end{align*}
	So, (j) contributes
	$$1(R=1_d)1(A=0)m_1(X)\sum_{r\neq 1_d} Q_{r,0}(X_r).$$
	
	\textbf{Term (k).} The contribution from (k) gives
	
	\begin{align*}
		&t\int \frac{1(r'=1_d)1(a=1)y p_0(x,r',y,a) \sum_r Q_{r,1}(x_r,y)}{p_0(x,1)} \\
		&\qquad \times \sum_{s\neq 1_d}\int p_0(x_{\bar{s}}|x_s,1_d,0) p_0(x_{\bar{s}}',x_s,0) g_0(x_{\bar{s}}', x_s, s, 0) \ dx_{\bar{s}}' \ dx_{\bar{s}} dx_s dr' dyda \\
		&= t\int 1(a'=-) m_1(X) \\
		&\qquad \times \sum_{s\neq 1_d}p_0(x_{\bar{s}}|x_s,1_d,a') p_0(x_{\bar{s}}',x_s,a') g_0(x_{\bar{s}}', x_s, s, a') \ dx_{\bar{s}}  \ dx_{\bar{s}}' dx_s da' \\
		&= t\int 1(a'=0) \\
		&\qquad \times \sum_{s\neq 1_d} 1(r''=s) M_{m,s,0}(x_s) p_0(x_{\bar{s}}',x_s,r'',a') g_0(x_{\bar{s}}', x_s, r'', a') \ dx_{\bar{s}}' dx_s dr'' da' \\
	\end{align*}
	So, (k) contributes
	$$1(A=0) \sum_{r\neq 1_d} 1(R=r) M_{m,r,0}(X_r).$$
	
	\textbf{Term (l).} The contribution from (l) gives
	\begin{align*}
		&-t\int \frac{1(r'=1_d)1(a=1)y p_0(x,r',y,a) \sum_r Q_{r,1}(x_r,y)}{p_0(x,1)} \\
		&\qquad \times t \sum_{s\neq 1_d}\int p_0(x_{\bar{s}}|x_s,1_d,0) Q_{s,0}^{(0)}(x_s) p_0(x_{\bar{s}}',x_s,1_d,0) g_0(x_{\bar{s}}', x_s, 1_d, 0) \ dx_{\bar{s}}' \ dx_{\bar{s}} dx_s dr' dyda \\
		&-t\int 1(a'=0)1(r'=1_d)m_1(X) \\
		&\qquad \times t \sum_{s\neq 1_d}\int p_0(x_{\bar{s}}|x_s,1_d,0) Q_{s,0}^{(0)}(x_s) p_0(x_{\bar{s}}',x_s,r',a') g_0(x_{\bar{s}}', x_s, r', a') \ dx_{\bar{s}}' \ dx_{\bar{s}} dx_s dr' da' \\
		&-t\int 1(a'=0)1(r'=1_d) \\
		&\qquad \times t \sum_{s\neq 1_d}\int M_{m,s,0}(x_s) Q_{s,0}^{(0)}(x_s) p_0(x_{\bar{s}}',x_s,r',a') g_0(x_{\bar{s}}', x_s, r', a') \ dx_{\bar{s}}' \ dx_{\bar{s}} dx_s dr' da' \\
	\end{align*}
	So, (l) contributes
	$$-1(A=0) \sum_{r\neq 1_d} 1(R = 1_d) M_{m,r,0}(X_r) Q_{r,0}(X_r).$$
	
	\textbf{Merging Terms.}  Adding all of the terms together from all three parts yields the following:
	\begin{align*}
		&\frac{1(A=1)Y}{P(A=1|X)} \sum_r Q_{r,1}(X_r,Y) 1(R=1_d) + 1(A=1)Y \sum_r 1(R=r) M_{1/\pi_A,r,1}(X_r,Y) \\
		&\quad-1(A=1)Y 1(R=1_d)\sum_r Q_{r,1}(X_r,Y)M_{1/\pi_A,r,1}(X_r,Y) \\
		&\quad-1(R=1_d)1(A=1)\frac{m_1(X)}{\pi_A(X)} -1(R=1_d)1(A=1)\frac{m_1(X)}{\pi_A(X)} \sum_{r\neq 1_d} Q_{r,1}(X_r, Y) \\
		&\quad-1(A=1) \sum_{r\neq 1_d} 1(R=r) M_{m/\pi_A,1}(X_r,Y) +1(A=1) \sum_{r\neq 1_d} 1(R = 1_d) M_{m/\pi_A,1}(X_r,Y) Q_{r,1}(X_r,Y) \\
		&\quad+1(R=1_d)1(A=1)m_1(X) +1(R=1_d)1(A=0)m_1(X) \\
		&\quad+1(R=1_d)1(A=1)m_1(X)\sum_{r\neq 1_d} Q_{r,1}(X_r,Y) \\
		&\quad+1(A=1) \sum_{r\neq 1_d} 1(R=r) M_{m,r,1}(X_r,Y) -1(A=1) \sum_{r\neq 1_d} 1(R = 1_d) M_{m,r,1}(X_r,Y) Q_{r,1}(X_r,Y) \\
		&\quad+1(R=1_d)1(A=0)m_1(X)\sum_{r\neq 1_d} Q_{r,0}(X_r) \\
		&\quad+1(A=0) \sum_{r\neq 1_d} 1(R=r) M_{m,r,0}(X_r) -1(A=0) \sum_{r\neq 1_d} 1(R = 1_d) M_{m,r,0}(X_r) Q_{r,0}(X_r) \\
		&= \frac{1(A=1)Y}{P(A=1|X)} \sum_r Q_{r,1}(X_r,Y) 1(R=1_d) + 1(A=1)Y \sum_r 1(R=r) M_{1/\pi_A,r,1}(X_r,Y) \\
		&\quad-1(A=1)Y 1(R=1_d)\sum_r Q_{r,1}(X_r,Y)M_{1/\pi_A,r,1}(X_r,Y) \\
		&\quad-1(R=1_d)1(A=1)\frac{m_1(X)}{\pi_A(X)} \sum_{r} Q_{r,1}(X_r, Y) \\
		&\quad-1(A=1) \sum_{r\neq 1_d} 1(R=r) M_{m/\pi_A,1}(X_r,Y) +1(A=1) \sum_{r\neq 1_d} 1(R = 1_d) M_{m/\pi_A,1}(X_r,Y) Q_{r,1}(X_r,Y) \\
		&\quad +1(R=1_d)1(A=0)m_1(X)\sum_{r} Q_{r,0}(X_r) +1(R=1_d)1(A=1)m_1(X)\sum_{r} Q_{r,1}(X_r,Y) \\
		&\quad+1(A=1) \sum_{r\neq 1_d} 1(R=r) M_{m,r,1}(X_r,Y) -1(A=1) \sum_{r\neq 1_d} 1(R = 1_d) M_{m,r,1}(X_r,Y) Q_{r,1}(X_r,Y) \\
		&\quad+1(A=0) \sum_{r\neq 1_d} 1(R=r) M_{m,r,0}(X_r) -1(A=0) \sum_{r\neq 1_d} 1(R = 1_d) M_{m,r,0}(X_r) Q_{r,0}(X_r) \\
		&= q(X,R,Y,A).
	\end{align*}
	This has mean equivalent to $\theta$, so the efficient influence function is $\text{EIF}(X,R,Y,A) = q(X,R,Y,A) - \theta$.  This shows that the multiply-robust estimator for the missing response problem achieves the semiparametric efficiency bound.
	
\end{proof}





\subsection{Asymptotic theory}	\label{app:asymp}
This section contains proof of 
Theorem \ref{theorem:ipw} and Theorem \ref{theorem:ra}.
For the ease of derivation and referencing the literature (e.g. Chapter 19 of \citealt{van1998asymptotic}), 
we introduce notations from the empirical process theory. 
We denote $P$ as the true probability measure that generates our data
and $P_n$ be the empirical measure. 
Namely, in the case of observing IID random elements $Z_1,\cdots, Z_n \in \mathcal{Z}$, $P(z)$ is the underlying distribution
that generates $Z_1,\cdots, Z_n$ and $P_n(z) = \frac{1}{n}\sum_{i=1}^n I(Z_i\leq z)$
is the corresponding empirical distribution. 
For any function $f: \mathcal{Z}\rightarrow \mathbb{R}$, 
we denote 
$$
P f = \int f(z) P(dz) = \E(f(Z)),\quad P_n f = \int f(z) P_n(dz) = \frac{1}{n}\sum_{i=1}^n f(Z_i).
$$


\begin{proof}[ of Theorem~\ref{theorem:ipw}]
	Following Lemma \ref{lemma:completeodds}, we rewrite the population quantity of interest as
	\begin{align*}
		\theta &= \E[h(X,W,A)] \\
		&= \sum_{a\in\mathcal{A}} \E[f_a(X,W_a)1(A=a)] \\
		&= \sum_{a\in\mathcal{A}} \sum_r \E[f_a(X,W_a)1(A=a)O_{r,a}(X,W_a)1(R=1_d)] \\
		&= \sum_{a\in\mathcal{A}} \sum_r \theta_{\text{IPW},r,a}.
	\end{align*}
	Since we need to estimate $f_a$ and $O_{r,a}$, our estimator is
	\begin{align*}
		\hat{\theta}_{\text{IPW}} &= \sum_{a\in\mathcal{A}} \sum_r \left(\frac{1}{n}\sum_{i=1}^n \hat{f}_a(X_i, W_{i,a}) 1(A_i = a) \hat{O}_{r,a}(X_i, W_{i,a}) 1(R_i = 1_d)\right) \\
		&= \sum_{a\in\mathcal{A}} \sum_r \hat{\theta}_{\text{IPW}, r,a}.
	\end{align*}
	We will show that $\hat{\theta}_{\text{IPW}, r,a}$, as defined above, is an asymptotically linear estimator for $\theta_{\text{IPW},r,a}$.  For ease of typesetting, define
	\begin{align*}
		&q_{r,a}(X,R,W_a,A) := f_a(X,W_a)1(A=a)O_{r,a}(X,W_a)1(R=1_d), \\
		&\hat{q}_{r,a}(X,R,W_a,A) := \hat{f}_a(X,W_a)1(A=a)\hat{O}_{r,a}(X,W_a)1(R=1_d).
	\end{align*}
	
	Using the empirical process notations, we have
	\begin{align*}
		\hat{\theta}_{\text{IPW},r,a} - \theta_{\text{IPW},r,a} &= P_n \hat{q}_{r,a} - P_0 q_{r,a} \\
		&= \underbrace{(P_n-P_0)q_{r,a}}_{(I)} + \underbrace{P_0(\hat{q}_{r,a}- q_{r,a})}_{(II)} + \underbrace{(P_n-P_0)(\hat{q}_{r,a}- q_{r,a})}_{(III)}.
	\end{align*}
	Therefore, there are 3 terms we need to consider to show asymptotic normality.
	
	\textbf{Term (I)}.  Expanding this term, we obtain
	\begin{align*}
		(P_n - P_0) q_{r,a} &= \frac{1}{n}\sum_{i=1}^n (q(X,R,W_a,A) - \E[q(X,R,W_a,A)]),
	\end{align*}
	which is immediately asymptotically linear.
	
	\textbf{Term (II).}  Collect the parameters $\tau_a$ and $\eta_{r,a}$ into a single parameter $\kappa_{r,a}$ for simplicity.  As the first derivative of $q_{r,a}(x,r,w_a,a;\kappa_{r,a})$ exists and is nonzero (assumptions \ref{A4} and \ref{B4}), Term (II) equals
	\begin{align*}
		P_0(\hat{q}_{r,a} - q_{r,a}) &= \int [q_{r,a}(x,r,w_a,a;\hat{\kappa}_{r,a}) - q_{r,a}(x,r,w_a,a;\kappa^*_{r,a})] \ P(dx,dr,dw_a,da) \\
		&= (\hat{\kappa}_{r,a}-\kappa^*_{r,a})^\top \left (\int \nabla_{\kappa_{r,a}} q_{r,a}(x,r,w_a,a;\kappa_{r,a}) \ P(dx,dr,dw_a,da) \right) \nonumber\\
		&\qquad + o_p(1/\sqrt{n}).
	\end{align*}
	The interchange of the derivative and the integral follows from the Dominated Convergence Theorem.  Since we assume the $\hat{\kappa}_{r,a}$ admits an asymptotically linear expansion (assumptions \ref{A1} and \ref{B2}), this further simplifies to
	\begin{equation*}
		\frac{1}{n}\sum_{i=1}^n \varphi_{r,a}(X_i,R_i,W_{i,a},A_i)^\top \left(\int \nabla_{\kappa_{r,a}} q_{r,a}(x,r,w_a,a;\kappa_{r,a}) \ P(dx,dr,dw_a,da) \right) + o_p(1/\sqrt{n}),
	\end{equation*}
	where $\varphi_{r,a}$ is the influence function for $\hat{\kappa}_{r,a}$.
	
	\textbf{Term (III).}  For the last term, we show that is negligible up to first order.  We apply Lemma 19.24 from \citep{van1998asymptotic}.  We verify the following two conditions:
	\begin{enumerate}
		\item $\mathcal{Q} := \{q_{r,a}(x,r,w_a;\tau,\eta) : \tau,\eta\}$ is a $P$-Donsker class: \\
		Since $\{f_a(x,w_a;\tau) : \tau\}$ and $\{O_{r,a}(x,w_a;\eta) : \eta\}$ are uniformly bounded $P$-Donsker classes, it follows that $\mathcal{Q}$ is also $P$-Donsker.
		
		The class of functions formed by the pairwise products of functions from two uniformly bounded $P$-Donsker classes is $P$-Donsker (this permanence property can be seen using a Lipschitz transformation argument; see Example 2.10.8 from \citep{van1996weak}).
		
		\item $q_{r,a}(x,r,w_a,a;\hat{\tau}_a,\hat{\eta}_{r,a})$ converges to some $q_{r,a}(x,r,w_a,a;\tau^*_a,\eta^*_{r,a})$ in $L_2(P)$:\\  
		We have
		\begin{align*}
			&\int [q_{r,a}(x,s,w_a,a';\hat{\tau}_a,\hat{\eta}_{r,a}) - q_{r,a}(x,s,w_a,a';\tau^*_a,\eta^*_{r,a})]^2 \ P(dx,ds,dw_a,da') \\
			&\quad = \int 1(a'=a)1(s=1_d) \biggr[f_a(x,w_a;\hat{\tau}_a)O_{r,a}(x,w_a;\hat{\eta}_{r,a}) \\
			&\qquad - f_a(x,w_a;\tau^*_a)O_{r,a}(x,w_a;\eta^*_{r,a})\biggr]^2 \ P(dx,ds,dw_a,da') \\
			&\quad \leq \int [f_a(x,w_a;\hat{\tau}_a)O_{r,a}(x,w_a;\hat{\eta}_{r,a}) - f_a(x,w_a;\tau^*_a)O_{r,a}(x,w_a;\eta^*_{r,a})]^2 \nonumber \\
			&\qquad \qquad P(dx,ds,dw_a,da').
		\end{align*}
		We can upper bound the $L_2(P)$ norm via
		\begin{align*}
			&\lVert f_a(x,w_a;\hat{\tau}_a)O_{r,a}(x,w_a;\hat{\eta}_{r,a}) - f_a(x,w_a;\tau^*_a)O_{r,a}(x,w_a;\eta^*_{r,a}) \rVert_2  \\
			&\quad \leq \lVert f_a(x,w_a;\hat{\tau}_a) [O_{r,a}(x,w_a;\hat{\eta}_{r,a}) - O_{r,a}(x,w_a;\eta^*_{r,a})] \rVert_2 \nonumber \\
			&\quad\quad + \lVert [f_a(x,w_a;\hat{\tau}_a) - f_a(x,w_a;\tau^*)]O_{r,a}(x,w_a;\eta^*_{r,a}) \rVert_2 \\
			&\quad \leq \lVert f_a(x,w_a;\hat{\tau}_a) \rVert_\infty \cdot \lVert O_{r,a}(x,w_a;\hat{\eta}_{r,a}) - O_{r,a}(x,w_a;\eta^*_{r,a}) \rVert_2 \nonumber \\
			&\quad \quad + \lVert f_a(x,w_a;\hat{\tau}_a) - f_a(x,w_a;\tau^*_a) \rVert_2 \cdot \lVert O_{r,a}(x,w_a;\eta^*_{r,a}) \rVert_\infty
		\end{align*}
		by Minkowski's and Holder's inequalities.  By assumptions \ref{A2}, \ref{A3}, \ref{B1}, and \ref{B4}, we have $L_2(P)$ convergence of $f_a(x,w_a;\hat{\tau}_a)O_{r,a}(x,w_a;\hat{\eta}_{r,a})$ to the function $f_a(x,w_a;\tau^*_a)O_{r,a}(x,w_a;\eta^*_{r,a})$.
	\end{enumerate}
	Therefore, the sufficient conditions of the Lemma 19.24 of \citep{van1998asymptotic} are satisfied, and it follows that Term (III) is $o_p(1/\sqrt{n})$.
	
	\textbf{Merging terms.}  Finally, putting everything together, we have
	\begin{align*}
		&\sqrt{n}(\hat{\theta}_{r,a} - \theta_{r,a}) \\
		&\quad= \frac{1}{\sqrt{n}}\sum_{i=1}^n (q(X,R,W_a,A) - \E[q(X,R,W_a,A)]) \\
		&\qquad +  \frac{1}{\sqrt{n}}\sum_{i=1}^n \varphi_{r,a}(X_i,R_i,W_{i,a},A_i)^\top \left(\int \nabla_{\kappa_{r,a}} q_{r,a}(x,r,w_a,a;\kappa_{r,a}) \ P(dx,dr,dw_a,da) \right) \\
		&\qquad + o_p(1)
	\end{align*}
	for some influence function $\varphi_{r,a}$ for $\hat{\kappa}_{r,a}$ (this follows from the asymptotic linearity assumptions \ref{A1} and \ref{B2}.  Next, we note that $\hat{\theta}_{\text{IPW},r,a}$ is generally correlated with this $\hat{\theta}_{\text{IPW},r',a'}$ for some $r\neq r'$ and $a\neq a'$.  Therefore, they do not typically converge jointly to a degenerate distribution.  As we have provided the asymptotic linear forms for each of the estimators marginally, we have joint convergence of the $\{\sqrt{n}(\hat{\theta}_{\text{IPW},r,a} - \theta_{\text{IPW},r,a}) \}_{r,a}$ to a zero-mean multivariate Gaussian by the Central Limit Theorem.  The asymptotic normality of $\hat{\theta}_{\text{IPW}}$ follows from the Continuous Mapping Theorem.  Consistency follows from asymptotic normality since we have convergence at the standard $\sqrt{n}$-rate.
\end{proof}

\begin{proof}[ of Theorem~\ref{theorem:ra}] Using Lemma \ref{lemma:ra}, we can decompose $\theta$ using a regression adjustment method as follows
	\begin{align*}
		\theta &= \E[h(X,W,A)] \\ 
		&= \sum_{a}\E[f_a(X,W_a)1(A=a)] \\
		&= \sum_a \sum_r \E[ m_{r,a}(X_r,W_a) 1(R=r) 1(A=a)] \\
		&= \sum_a \sum_r \theta_{\text{RA},r,a},
	\end{align*}
	where $m_{r,a}(X_r,W_a) = \ E[f_a(X,W_a) | X_r, R=r,W_a, A=a]$ is a regression function.  Here $\theta_{\text{RA},r,a} = \int \E[f_a(X,W_a) |X_r=x_r, R=r, W_a=w_a, A=a] \ P_{obs}(dx_r,r,dw_a,a)$.
	
	To start, we fix a covariate pattern $r$ and an observed outcome pattern $a$.  Define
	\begin{align*}
		&\hat{\theta}_{\text{RA},r,a} = \int f_a(x,w_a; \hat{\tau}_a) \  \hat{P}_{ex}(dx_{\bar{r}}|x_r,r,w_a,a)  \ \hat{P}_{obs}(dx_r,r,dw_a,a),\\
		&\tilde{\theta}_{\text{RA},r,a} = \int f_a(x,w_a; \tau_a^*) \  \hat{P}_{ex}(dx_{\bar{r}}|x_r,r,w_a,a)  \ \hat{P}_{obs}(dx_r,r,dw_a,a),\\
		&\theta_{\text{RA},r,a} = \int f_a(x,w_a; \tau_a^*) \  P_{ex}(dx_{\bar{r}}|x_r,r,w_a,a)  \ P_{obs}(dx_r,r,dw_a,a).
	\end{align*}
	
	Then, we can decompose the difference via
	\begin{equation*}
		\sqrt{n}(\hat{\theta}_{\text{RA},r,a} - \theta_{RA,r,a}) = \underbrace{\sqrt{n}[\hat{\theta}_{RA,r,a} - \tilde{\theta}_{\text{RA},r,a}]}_{(I)} + \underbrace{\sqrt{n}[\tilde{\theta}_{RA,r,a} - \theta_{RA,r,a}]}_{(II)}.
	\end{equation*}
	
	\textbf{Term (I).} First, we start with Term (I), which rewrites as
	\begin{align*}
		&\quad\sqrt{n} \int [f_a(x,w_a;\hat{\tau}_a) - f_a(x,w_a;\tau_a^*)]  \ \hat{P}_{ex}(dx_{\bar{r}}|x_r,w_a,a)  \ \hat{P}_{obs}(dx_r,r,dw_a,a) \\
		&\quad = \sqrt{n}(\hat{\tau}_a - \tau_a^*)^\top  \left( \int \nabla_{\tau_a} \ f_a(x,w_a;\tau_a) \ \hat{P}(dx,r,dw_a,a)\right) + o_p(1) \\
		&\quad = \frac{1}{\sqrt{n}}\sum_{i=1}^n \phi_{a}(X_i, R_i, W_{i,A_i}, A_i) \left( \int \nabla_{\tau_a} \ f_a(x,w_a;\tau_a) \ \hat{P}(dx,r,dw_a,a)\right) + o_p(1)
	\end{align*}
	by the parametric assumption of \ref{A1}.
	
	\textbf{Term (II).} Now, we consider Term (II), which requires slightly more work.  Expanding it and grouping terms, we obtain the following decomposition.
	
	\begin{align*}
		&\sqrt{n}(\tilde{\theta}_{\text{RA},r,a} - \theta_{\text{RA},r,a})\\
		&= \sqrt{n}\int \int f_a(x,w_a; \tau_a^*) \ \hat{P}_{ex}(dx_{\bar{r}}|x_r,r,w_a,a) \hat{P}_{obs}(dx_r,r,dw_a,a) \\
		&\quad - \sqrt{n} \int \int f_a(x,w_a; \tau_a^*) \ P_{ex}(dx_{\bar{r}}|x_r,r,w_a,a) P_{obs}(dx_r,r,dw_a,a) \\
		&=  \underbrace{\sqrt{n} \int \int f_a(x,w_a; \tau_a^*) \ \hat{P}_{ex}(dx_{\bar{r}}|x_r,r,w_a,a) (\hat{P}_{obs}(dx_r,r,dw_a,a)-P_{obs}(dx_r,r,dw_a,a))}_\text{(A)} \\
		&\quad - \underbrace{\sqrt{n}\int \int f_a(x,w_a; \tau_a^*) \  (\hat{P}_{ex}(dx_{\bar{r}}|x_r,r,w_a,a)-P_{ex}(dx_{\bar{r}}|x_r,r,w_a,a)) P_{obs}(dx_r,r,dw_a,a)}_\text{(B)}
	\end{align*}
	
	\textbf{Term A of (I).} Term A rewrites as
	\begin{equation*}
		\sqrt{n}\int m_{r,a}(x_r,w_a;\hat{\eta}_{r,a}) \ (\hat{P}_{obs}(dx_r,r,dw_a,a)-P_{obs}(dx_r,r,dw_a,a)),
	\end{equation*}
	
	This is equal to
	\begin{align*}
		& \sqrt{n}\int (m_{r,a}(x_r,w_a;\hat{\eta}_{r,a})-m_{r,a}(x_r,w_a;\eta^*_{r,a})) \ (\hat{P}_{obs}(dx_r,r,dw_a,a)-P_{obs}(dx_r,r,dw_a,a)) \nonumber \\
		& \quad + \sqrt{n}\int m_{r,a}(x_r,w_a;\eta^*_{r,a}) \ (\hat{P}_{obs}(dx_r,r,dw_a,a)-P_{obs}(dx_r,r,dw_a,a)).
	\end{align*}
	
	Observe that the conditions of Lemma 19.24 of \citep{van1998asymptotic} are satisfied (namely, $\hat{m}_{r,a}$ belongs to a Donsker class by \ref{C2} and we have $L_2(P)$ convergence to the true regression function by \ref{C3}).  Therefore, it follows that the first term in the previous display is $o_p(1)$.  Thus, for asymptotic analysis, we can replace $\hat{\eta}_{r,a}$ in Term A with the population version $\eta_{r,a}^*$.  This means that Term A
	\begin{equation*}
		\sqrt{n}\int m_{r,a}(x_r,w_a;\hat{\eta}_{r,a}) \ (\hat{P}_{obs}(dx_r,r,dw_a,a)-P_{obs}(dx_r,r,dw_a,a))
	\end{equation*}
	is asymptotically equivalent to (up to an $o_p(1)$ term)
	\begin{equation*}
		\sqrt{n}\int m_{r,a}(x_r,w_a;\eta^*_{r,a}) \ (\hat{P}_{obs}(dx_r,r,dw_a,a)-P_{obs}(dx_r,r,dw_a,a)).
	\end{equation*}
	
	\textbf{Term B of (II).} We now consider the term B.  It rewrites as
	\begin{align*}
		&\sqrt{n} \int (m_{r,a}(x_r, w_a; \hat{\eta}_{r,a}) - m_{r,a}(x_r,w_a; \eta_{r,a}^*)) \ P_{obs}(dx_r,r,dw_a,a) \\
		&\quad= \sqrt{n}(\hat{\eta}_{r,a}-\eta_{r,a}^*)^\top \left( \int \nabla_{\eta_{r,a}} \ m_{r,a}(x_r,w_a;\eta_{r,a}) \ P_{obs}(dx_r,r,dw_a,a)\right) + o_p(1).
	\end{align*}
	Since the first derivative of $m_{r,a}(x_r,w_a;\eta_{r,a})$ is bounded, the interchange of the derivative and the integral follows from the Dominated Convergence Theorem.
	
	\textbf{Collecting Terms (I) and (II).}  We combine Terms A and B of Term (II) and add them to Term (I), and it follows that
	\begin{align*}
		&\sqrt{n}(\hat{\theta}_{\text{RA},r,a} - \theta_{\text{RA},r,a}) \\
		&= \frac{1}{\sqrt{n}}\sum_{i=1}^n  \phi_{a}(X_i, R_i, W_{i,A_i}, A_i) \left( \int \nabla_{\tau_a} \ f_a(x,w_a;\tau_a) \ \hat{P}(dx,r,dw_a,a)\right) \\
		&\quad + \sqrt{n}\int m_{r,a}(x_r,w_a;\eta^*_{r,a}) \ (\hat{P}_{obs}(dx_r,r,dw_a,a)-P_{obs}(dx_r,r,dw_a,a)) \\
		&\quad + \sqrt{n}(\hat{\eta}_{r,a}-\eta_{r,a}^*)^\top \left( \int \nabla_{\eta_{r,a}} \ m_{r,a}(x_r,w_a;\eta_{r,a}) \ P_{obs}(dx_r,r,dw_a,a)\right) \\
		&\quad + o_p(1) \\
		&= \frac{1}{\sqrt{n}}\sum_{i=1}^n  \phi_{a}(X_i, R_i, W_{i,A_i}, A_i) \left( \int \nabla_{\tau_a} \ f_a(x,w_a;\tau_a) \ \hat{P}(dx,r,dw_a,a)\right) \\
		&\quad + \sqrt{n}(\hat{\eta}_{r,a}-\eta_{r,a}^*)^\top \left( \int \nabla_{\eta_{r,a}} \ m_{r,a}(x_r,w_a;\eta_{r,a}) \ P_{obs}(dx_r,r,dw_a,a)\right) \\
		&\quad + \frac{1}{\sqrt{n}}\sum_{i=1}^n \biggr[ m_{r,a}(X_{i,R_i}, W_{i,A_i}; \eta_{r,a}^*) \\
		&\quad \qquad - \E[ 1(R=r)1(A=a)m_{r,a}(X_R, W_A; \beta_{r,a}^*)]\biggr] \\
		&\quad + o_p(1).
	\end{align*}
	By Slutsky's lemma and the WLLN, this is equivalent up to an $o_p(1)$ term to
	\begin{align*}
		&\frac{1}{\sqrt{n}}\sum_{i=1}^n \phi_{a}(X_i, R_i, W_{i,A_i}, A_i) \cdot \E [ 1(R=r)1(A=a)\nabla_{\tau_a} \ f_a(X,W_a;\tau_a)] \nonumber \\
		&\quad + \frac{1}{\sqrt{n}}\sum_{i=1}^n \psi_{r,a}(X_{i}, R_i, W_{i,A_i}, A_i) \cdot \E[1(R=r)1(A=a) \nabla_{\beta_{r,a}} \ m_{r,a}(X_r,W_a;\eta_{r,a})]  \nonumber \\
		&\quad + \frac{1}{\sqrt{n}}\sum_{i=1}^n \biggr[ m_{r,a}(X_{i,R_i}, R_i, W_{i,A_i}, A_i; \beta_{r,a}^*) \nonumber \\
		&\quad \qquad \qquad - \E[ 1(R=r)1(A=a)m_{r,a}(X_R, R, W_A, A; \beta_{r,a}^*)]\biggr],
	\end{align*}
	where $\psi_{r,a}$ is the influence function of $\hat{\eta}_{r,a}$ since it is assumed to be an asymptotic linear estimator by Assumption \ref{C1}.
	
	Thus, we have showed that $\hat{\theta}_{\text{RA},r,a}$ is an asymptotically linear estimator.  Next, note that in standard missing data problems, different subsets of the observed data set are used to estimate the extrapolation distribution $P_{ex}(x_{\bar{r}}|x_r,r,w_a,a)$ for different values of $r$ and $a$, as this relies on the identifying restrictions used.  Therefore, $|\Corr(\hat{\theta}_{\text{RA},r,a}, \hat{\theta}_{\text{RA},r',a'})| \neq 1$ when $r\neq r'$ or $a\neq a'$, so the joint convergence of all the $\hat{\theta}_{\text{RA},r,a}$s to a multivariate normal distribution follows since marginally, they all admit an asymptotic linear expansion.  Finally, applying the Continuous Mapping Theorem yields the asymptotic normality of $\hat{\theta}_{\text{RA}}$, and consistency follows.
\end{proof}

---
\end{document}